\documentclass[12pt]{article}

\usepackage{latexsym,amsmath,amssymb,amsthm,amsfonts,graphicx,natbib}
\usepackage{bbold}
\usepackage{epsfig}
\usepackage{bm}
\usepackage[euler]{textgreek}
\usepackage{xcolor}
\usepackage{algorithm,algpseudocode}
\usepackage{hyperref}
\usepackage{multirow}
\usepackage{algcompatible}
\usepackage[normalem]{ulem} 
\usepackage[pagewise,mathlines]{lineno}

\include{def}
\def\dispmuskip{\thinmuskip= 3mu plus 0mu minus 2mu \medmuskip=  4mu plus 2mu minus 2mu \thickmuskip=5mu plus 5mu minus 2mu}
\def\textmuskip{\thinmuskip= 0mu                    \medmuskip=  1mu plus 1mu minus 1mu \thickmuskip=2mu plus 3mu minus 1mu}
\def\beq{\dispmuskip\begin{equation}}    \def\eeq{\end{equation}\textmuskip}
\def\beqn{\dispmuskip\begin{displaymath}}\def\eeqn{\end{displaymath}\textmuskip}
\def\bea{\dispmuskip\begin{eqnarray}}    \def\eea{\end{eqnarray}\textmuskip}
\def\bean{\dispmuskip\begin{eqnarray*}}  \def\eean{\end{eqnarray*}\textmuskip}

\newcommand{\eps}{\epsilon}
\newcommand{\wh}{\widehat}

\newcommand{\bol}[1]{\textbf{#1}}

\def\E{{\rm E}}                         

\def\F{{\cal F}}

\def\N{{\cal N}}

\def\Var{\text{\rm Var}}

\begin{document}
	
	\title{ A Statistical Recurrent Stochastic Volatility Model for Stock Markets}
\author{T.-N. Nguyen,\;\; M.-N. Tran,\;\;D. Gunawan,\;\; R. Kohn\thanks{{\bf Nguyen} and {\bf Tran}: Discipline of Business Analytics, University of Sydney Business School and ACEMS. {\bf Gunawan}: School of Mathematics and Applied Statistics, University of Wollongong and ACEMS. {\bf Kohn}: School of Economics, UNSW Business School and ACEMS.}
}

	\maketitle
	\begin{abstract}
The Stochastic Volatility (SV) model and its variants are widely used in the financial sector while recurrent neural network (RNN) models are successfully used in many large-scale industrial applications of Deep Learning. Our article combines these two methods in a non-trivial way and proposes a model, which we call the Statistical Recurrent Stochastic Volatility (SR-SV) model, to capture the dynamics of stochastic volatility. 
The proposed model is able to capture complex volatility effects (e.g., non-linearity and long-memory auto-dependence) 
overlooked by the conventional SV models, 
is statistically interpretable and has an impressive out-of-sample forecast performance. 
These properties are carefully discussed and illustrated through extensive simulation studies and applications to five international stock index datasets: The German stock index DAX30, the Hong Kong stock index HSI50, the France market index CAC40, the US stock market index SP500 and the Canada market index TSX250. 
An user-friendly software package together with the examples reported in the paper are available at \url{https://github.com/vbayeslab}.   
\end{abstract}
\section{Introduction}
The volatility of a financial time series, such as stock returns, 
is defined as the variance of the returns and serves as a measure of the uncertainty about the returns. The volatility, which is of great interest to financial econometricians, is unobserved and therefore often modelled statistically in order to estimate it. The two model classes most frequently used in volatility modelling are the Generalized Autoregressive Conditional Heteroscedastic (GARCH) models and the Stochastic Volatility (SV) models. 
The GARCH model \citep{Bollerslev1986} expresses the current volatility, conditional on the previous returns and volatilities, as a {\it deterministic} and linear function of the squared returns and the conditional volatilities in the previous time period. The SV model \citep{Taylor:1982,Taylor:1986}, on the other hand, uses a latent stochastic process to model the volatility, which is usually taken as a first order autoregressive process.
It is well documented that the GARCH and SV models are able to capture important effects exhibited in the variance of financial returns. For example, the volatilities in financial returns are observed to be highly autocorrelated in certain time periods and exhibit periods of both low and high volatility \citep{Benoit1967}. This so-called volatility clustering phenomenon can be modeled by the volatility processes introduced in the GARCH and SV models, making these volatility models widely employed in financial time series modelling.  
				
Although the GARCH and SV models were independently and almost concurrently introduced, the GARCH models were initially more widely adopted as it is much easier to estimate GARCH models than SV models. This is because the likelihood of a GARCH model can be obtained explicitly, while the likelihood of a SV model is intractable as it is an integral over the latent process.
However, the conditional variance process of GARCH models is deterministic and hence GARCH models might not capture efficiently the random oscillatory behavior of financial volatility \citep{Nelson:1991}.
SV models are considered as an attractive alternative to GARCH models because they overcome this limitation  \citep{Kim:1998,Yu:2002}.
Recent advances in Bayesian computation such as particle Markov chain Monte Carlo (PMCMC) \citep{Andrieu:2010} allow straightforward estimation and inference for SV models. 
	
Standard SV models still cannot appropriately capture some important features arising in financial volatility. For example, a large amount of both theoretical and empirical evidence indicates that there exists long-range persistence in the volatility process of many financial returns, see, e.g, \cite{Lo:1991}, \cite{Ding:1993}, \cite{Crato:1994} and \cite{Bollerslev:1996}. 
The long-memory property of a time series implies that the decay of the autocorrelations of the series is slower than exponential. 
The standard SV model of \citep{Taylor:1982} uses an AR(1) process to model the log of the volatility and hence might fail to capture this type of persistence \citep{Breidt:1998}. Another line of the literature shows strong evidence of non-linear auto-dependence in the volatility process of some stock and currency exchange returns \citep{KILIC2011368} and that the simple linear AR(1) process cannot effectively capture the underlying non-linear volatility dynamics.     
	
\cite{Breidt:1998} proposed the Long Memory Stochastic Volatility (LMSV) model to overcome the short-memory limitation of the standard SV model.  
LMSV uses an ARFIMA process \citep{Granger:1980} as an alternative to the AR(1) process to capture the long-memory dependence in the volatility. 
The empirical evidence in \cite{Breidt:1998} suggests that the LMSV model is able to capture the long-memory volatility behaviour in some stock return datasets.
However, the literature is unclear about whether the LMSV model can capture non-linear dynamics within the volatility process, because the ARFIMA model is linear.
Additionally, it is challenging to estimate the LMSV model as its likelihood is intractable. 
We are unaware of any available software package that implements the LMSV methodology. In another approach, \cite{YuJun:2006} introduced a family of non-linear SV (N-SV) models to capture the possible departure from the log transform commonly used in SV models. In the standard SV model, the logarithm of volatility is assumed to follow an AR(1) process; N-SV uses other non-linear transformations, such as the Box-Cox power function, rather than the logarithm. The simulation studies and empirical results on currency exchange and option pricing data in \cite{YuJun:2006} show that the N-SV model using the Box-Cox transformation is able to detect some interesting effects in the underlying volatility process. The general use of N-SV models requires the user to select an appropriate non-linear transformation for the dataset under consideration, and this might lead to a challenging model selection problem.
Neither \cite{Breidt:1998} nor \cite{YuJun:2006} clearly discussed the out-of-sample forecast performance of their LMSV and N-SV models.
	
Recurrent neural networks (RNN) in the Deep Learning literature have impressive prediction performance and have been successfully deployed in a large number of industrial-level applications (language translation, image captioning, speech synthesis, etc.).  
The RNN models are well-known for their ability to efficiently capture the long-range memory and non-linear dependence existing within various types of sequential data, and are considered as the state-of-the-art models for many sequence learning problems \citep{Lipton:2018}. Many researchers and practitioners have used RNN for mean modelling in financial time series analysis, but the general consensus is that these machine learning models do not clearly outperform the traditional time series models such as ARMA and ARIMA (see, e.g., \cite{Makridakis:2018} and \cite{ZHANG2003159}). 
\cite{Makridakis:2018} note that without careful modifications, Machine Learning models are usually less accurate than the statistical approaches that have been extensively investigated in the financial time series literature. Recently, the idea of using the RNN models to improve the predictive performance of GARCH-type models has also been proposed for volatility modelling. For example, \cite{Kim:2018} use the volatility estimates from several GARCH-type models as inputs to a RNN model, which then non-linearly transforms these inputs to output the final estimate of the volatility. 
The empirical results on the Korean stock market KOPSI 200 index show a significant improvement of forecast performance of the proposed hybrid model over several GARCH-type benchmark models. 
However, similar to many engineering-oriented Machine Learning models, \citeauthor{Kim:2018}'s model overlooks the interpretation aspect in volatility modelling, which is often of main interest to econometricians. 
One of the main motivations of our article is to develop deep learning based volatility models that are not only able to produce accurate prediction, but also interpretable and have meaningful in-sample analysis.
These models should not overlook the well-established features of traditional econometric models, that are motivated by the well-known stylized facts in financial time series such as volatility clustering and fat tails.   

In the SV literature, there is still lack of research using RNN structures to model the stochastic volatility dynamics of financial time series, perhaps because of two reasons.
First, it is non-trivial to sensibly incorporate RNN into the statistical volatility models. Simple adaptations of RNN to volatility models easily overlook the important stylized facts exhibited in financial volatility, which are well captured by the AR(1) process in the SV model. It is important to select appropriate RNN structures that are not only able to 
produce accurate out-of-sample volatility forecast, but also explain well the volatility dynamics.  
Second, a stochastic volatility model that incorporates a RNN structure into its latent stochastic process is highly sophisticated and thus challenging to estimate. 
	
This paper combines the SV and RNN models in a non-trivial way, and proposes a new model, called the Statistical Recurrent Stochastic Volatility (SR-SV) model. In particular, we use the Statistical Recurrent Unit (SRU) structure of \cite{Oliva:2017}, which is a special type of RNN models, to capture complex volatility effects overlooked by an AR(1) process in the standard SV model but still retain the essential components of the SV model. This combination allows the SR-SV model to enjoy much of advances from both worlds of deep learning (e.g., flexibility and excellent predictive performance) and econometric volatility modelling (e.g., excellent interpretability of volatility effects). 
The SR-SV model belongs to the class of parametric state space models whose Bayesian inference can be performed using recent advances in the Sequential Monte Carlo (SMC) and particle MCMC literature \citep{Andrieu:2009,Andrieu:2010,Duan:2015,Deligiannidis:2018}. 
The simulation studies and empirical results on the five stock index datasets demonstrate that the SR-SV model can efficiently capture the potential non-linear and long-memory effects in the underlying volatility dynamics, and provide better out-of-sample forecasts than the standard SV, N-SV and LMSV models. 
We note that we have tested SR-SV on a wider range of stock returns but only report in the paper the results for five of them,
as we constantly observed a similar improvement of the model compared to the other three counterpart. 
A Matlab software package implementing Bayesian estimation and inference for SR-SV together with the examples reported in this paper are available on github\footnote{The link is provided in the unblinded version.}.
	
The article is organized as follows. Section \ref{sec:lstm_sv} briefly reviews the SV and SRU models, and presents the SR-SV model. Section \ref{Sec:Bayes} discusses in detail Bayesian estimation and inference for the SR-SV model. Section \ref{sec:simulation and applications} presents the simulation study and applies the SR-SV model to analyze the five stock index datasets. Section \ref{Sec:conclusion} concludes. The Appendix gives details of the implementation and further empirical results.

\section{The SR-SV model}\label{sec:lstm_sv}
\subsection{The SV model and its possible weaknesses}\label{sec:SVmodel}
Let $y=\{y_t,\ t=1,...,T\}$ be a series of financial returns.
We consider a basic version of SV models \citep{Taylor:1982}
\bea
z_t &=& \mu+\phi (z_{t-1}-\mu)+\eps^z_t,\;\;\eps^z_t\sim\N(0,\sigma^2),\;\;t=2,...,T,\;\;\;z_1\sim\N\Big(\mu,\frac{\sigma^2}{1-\phi^2}\Big),\label{SVmodel1}\\
y_t &=& e^{\frac12z_t}\eps^y_t,\;\;\eps^y_t\sim\N(0,1),\;\;t=1,2,...,T.\label{SVmodel2}
\eea
The persistence parameter $\phi$ is assumed to be in $(-1,1)$ to enforce stationarity of both the $z$ and $y$ processes.  
In this SV model, the log volatility process $z$ is assumed to follow an AR(1) model.
It is well documented in the financial econometrics literature
that financial time series data often exhibit a long-term auto-dependence, which forces  
the persistence parameter $\phi$ to be close to 1 \citep{Jacquier:1994,Kim:1998}.
Write $p(z|\theta)$ for the density of $z$ given the model parameters $\theta=(\mu,\phi,\sigma^2)$ and $p(y|z)$ for the density of the data $y$ conditional on $z$.
We can view $p(z|\theta)$ as the prior with $\theta$ being the hyper-parameters and $p(y|z)$ as the likelihood \citep{Jacquier:1994}.
Under this perspective, the SV model \eqref{SVmodel1}-\eqref{SVmodel2} puts non-zero prior mass on AR(1) stochastic processes, and zero or almost-zero mass on stochastic processes that are far from being well approximated by an AR(1).
This means that the SV model in \eqref{SVmodel1}-\eqref{SVmodel2} might not be able to capture more complex dynamics in the posterior behavior of the log volatility process $z$, 
such as long-term memory or non-linear auto-dependence, and that a more flexible prior distribution should be put on $z$.
We will design such a flexible prior by combining the attractive features from both SV and RNN time series modeling techniques.  

\cite{YuJun:2006} propose a class of non-linearity N-SV models as a variant of SV which allows a more flexible link between the variance $\Var(y_t|z_t)$ and the AR(1) process $z$. Their N-SV model, using the Box-Cox transformation for $\Var(y_t|z_t)$, is written as 
\bea
z_t &=& \mu+\phi (z_{t-1}-\mu)+\eps^z_t,\;\;\eps^z_t\sim\N(0,\sigma^2),\;\;t=2,...,T,\;\;\;z_1\sim\N\Big(\mu,\frac{\sigma^2}{1-\phi^2}\Big),\label{NSVmodel2} \\
y_t &=& (1+\delta z_t)^{1/2\delta}\eps^y_t,\;\;\eps^y_t\sim\N(0,1),\;\;t=1,2,...,T,\label{NSVmodel1}
\eea
where $\delta$ is the auxiliary parameter that measures the degree of non-linearity rather than the log transform. As $\delta \rightarrow 0$, $(1+\delta z_t)^{1/2\delta} \rightarrow e^{\frac12z_t}$ and hence the N-SV model includes the SV model as a special case. 
The term non-linearity here might cause some confusion, as it does not refer to the non-linear auto-dependence within the log volatility process $z$, but the non-linearity 
between $\Var(y_t|z_t)$ and $z_t$. 

\cite{Breidt:1998} suggest to use an ARFIMA$(p,d,q)$ process \citep{Granger:1980,Hosking:1981} for the log volatility $z_t$ to capture the long-memory auto-dependence exhibited in financial time series. Their LMSV model is written as  
\bea
(1-B)^d \Phi(B) z_t &=& \Theta (B)\eta_t,\;\;\eta_t\sim\N(0,\sigma^2_\eta),\label{eq:LMSVmodel2} \\
y_t &=& \sigma_t \eps_t,\;\;\sigma_t = \kappa e^{\frac12z_t},\;\;\eps_t\sim\N(0,1),\;\;t=1,2,...,T,\label{eq:LMSVmodel1}
\eea
where $\Phi(B) = 1-\phi_1 B - \phi_2 B^2-...- \phi_p B^p$, $\Theta(B) = 1+\theta_1 B +\theta_2 B^2+ ... + \theta_q B^q$,
and $B$ is a backshift operator, i.e., $B^s X_t = X_{t-s}$.
To ensure the stationarity and invertability of the log volatility process $z_t$, the fractional integration parameter $d$ is assumed to be in $(-0.5,0.5)$ and the roots of $\Phi(B)$ and $\Theta(B)$ have to lie outside the unit circle.  

Another notable line of research in the SV literature is the class of semi-parametric stochastic volatility models that incorporate 
non-parametric techniques into modelling the conditional distribution of financial returns. For example, the stochastic volatility, Dirichlet process mixture (SV-DPM) model of \cite{Jensen:2010} uses a Dirichlet process prior \citep{Ferguson:1973} to characterize the conditional distribution of $y_t$. 
Semi-parametric models are different from the SR-SV model in two important aspects. 
First, the SV-DPM model is proposed to capture the asymmetries and leptokurtotic behaviors of financial returns, while the SR-SV model focuses on modeling the non-linearity and long-memory auto-dependence in the log-volatility dynamics. Second, the SV-DPM model is a semi-parametric model in the sense that the model cannot be described using a finite number of parameters as it uses a non-parametric prior, e.g. Dirichlet process, to simulate the conditional return and retains the parametric structure, e.g. AR(1), of the log-volatility in the standard SV model. The SR-SV model, on the other hand, is a parametric model with eleven parameters whose mathematical representation will be discussed in Section \ref{sec:lstm-sv}. Our article therefore uses parametric models including the standard SV, N-SV and LMSV models as the benchmarks to evaluate the SR-SV model.

\subsection{The SRU model}
There are at least two approaches to modeling time series data.
One approach is to represent time effects {\it explicitly} via some simple function, often a linear function, of the lagged values of the time series.
This is the mainstream time series data analysis approach in the statistics literature with the well-known models such as AR or ARMA.
The alternative approach is to represent time effects {\it implicitly} via latent variables,
which are designed to store the memory of the dynamics in the data.
These latent variables, also called hidden states, are updated in a recurrent and deterministic manner using the information carried over by their values from the previous time steps
and the information from the data at the current time step.
Recurrent neural networks (RNN), belong to the second category, were first developed in cognitive science and successfully used in computer science and other fields.
Another class of models that represent time implicitly is state space models, albeit the recurrent update is stochastic, which are widely used in econometrics and statistics. The SV model discussed in Section \ref{sec:SVmodel} is an example of state space models. 

For the purpose of this section, we denote the time series data as $\{D_t=(x_t,z_t),\ t=1,2,...\}$ where $x_t$ is the vector of inputs and $z_t$ the scalar output. 
In our article, it is useful to think of $x_t$ as scalar; however, the RNN approach is often efficiently used to model multivariate time series. 
If the time series of interest has the form $\{z_t,\ t=1,2,...\}$, it can be written as $\{(x_t,z_t),\ t=2,...\}$ with $x_t=z_{t-1}$.
Our goal is to model the conditional distribution $p(z_t|x_t,D_{1:t-1})$.
If the serial dependence structure is ignored, then a feedforward neural network (FNN) can be used to 
transform the raw input data $x_t$ into a set of hidden units $h_t$, also called {\it learned features} or {\it summary statistics}, for the purpose of explaining or predicting $z_t$.
However, this approach is unsuitable for time series data as the time effects or the serial dependence are totally ignored.
The main idea behind RNN is to let the set of hidden units $h_t$ to feed itself using its lagged value $h_{t-1}$ from the previous time step $t-1$.
Hence, RNN can be best thought of as a FNN that allows a connection of the hidden units to their value from the previous time step, enabling the network to possess memory.
Mathematically, this RNN model \citep{Elman:1990} is written as
\bea
h_t&=&\Psi(w_xx_t+w_hh_{t-1}+b),\label{eq:simpleRNN1}\\
\eta_t&=&\beta_0+\beta_1 h_t,\label{eq:simpleRNN2}\\
z_t|\eta_t&\sim&p(z_t|\eta_t). \label{eq:simpleRNN3}
\eea
The model parameters include $w_x$, $w_h$, $b$, $\beta_0$ and $\beta_1$, $\Psi(\cdot)$ is a non-linear activation function, e.g., common choices are the {\it sigmoid} $\Psi(z)=1/(1+e^{-z})$ and the {\it tanh} $\Psi(z) = (e^{z} - e^{-z})/(e^{z} +e^{-z})$, and $p(z_t|\eta_t)$ is a probability density depending on the learning task.
For example, if $z_t$ is continuous, then typically $p(z_t|\eta_t)$ is a Gaussian density with mean $\eta_t$;
if $z_t$ is binary, then $z_t|\eta_t$ follows a Bernoulli distribution with probability $\Psi(\eta_t)=1/(1+e^{-\eta_t})$.
Usually one sets $h_1=0$, i.e. the neural network initially does not have any memory. 

\begin{figure}[ht]
	\centering
	\includegraphics[width=1\columnwidth]{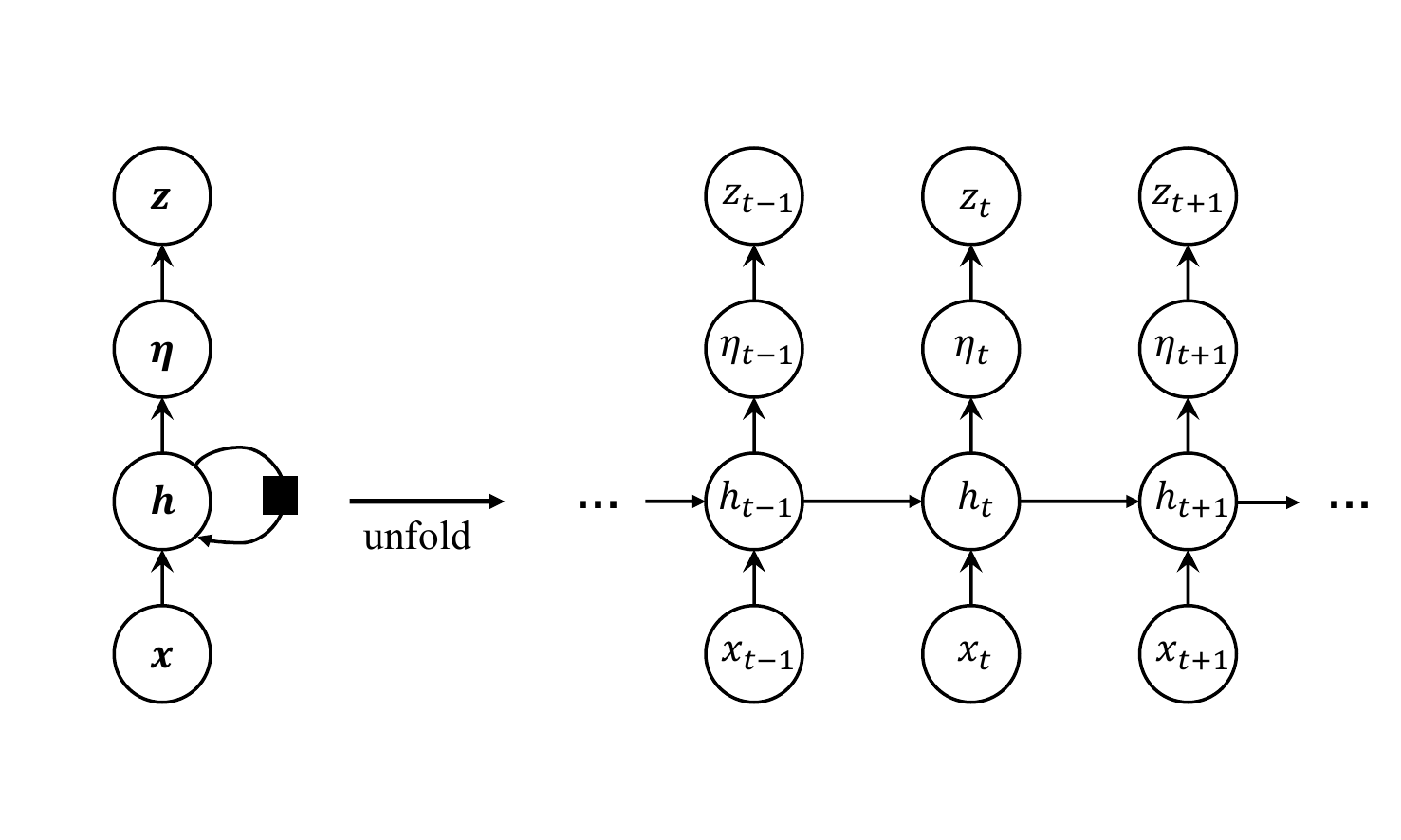}
	\caption{Graphical representation of the RNN model in \eqref{eq:simpleRNN1}-\eqref{eq:simpleRNN3}. }
	\label{f:RNN}
\end{figure}

Figure \ref{f:RNN} illustrates graphically the RNN model \eqref{eq:simpleRNN1}-\eqref{eq:simpleRNN3}. We follow \cite{Goodfellow:2016} and use a black square to indicate the delay of a single time step in the circuit diagram (\textit{left}). The circuit diagram can be interpreted as an unfolded computational graph (\textit{right}), where each node is associated with a particular time step. 

The unfolded graph in Figure \ref{f:RNN} suggests that the hidden state at time $t$ is the output of a composite function 
\begin{equation}
\label{eq:composite}
h_t = f\Big(x_t,f(x_{t-1},...,f(x_1,h_0))\Big),\;\;\;\text{where}\;\;\; f(x_t,h_{t-1}):=\Psi(w_xx_t+w_hh_{t-1}+b),
\end{equation}
which somewhat resembles a multiplication structure in terms of the weight $w_h$.
Consequently, the gradient of $h_t$ with respect to the model parameters might either explode or vanish if $t$ is sufficiently large and $w_h$ is not equal to 1, and hence making it inefficient for the Simple RNN model to learn in long time series. See \cite{Goodfellow:2016} for further explanation.

\begin{figure}[h]
	\centering
	\includegraphics[width=1\columnwidth]{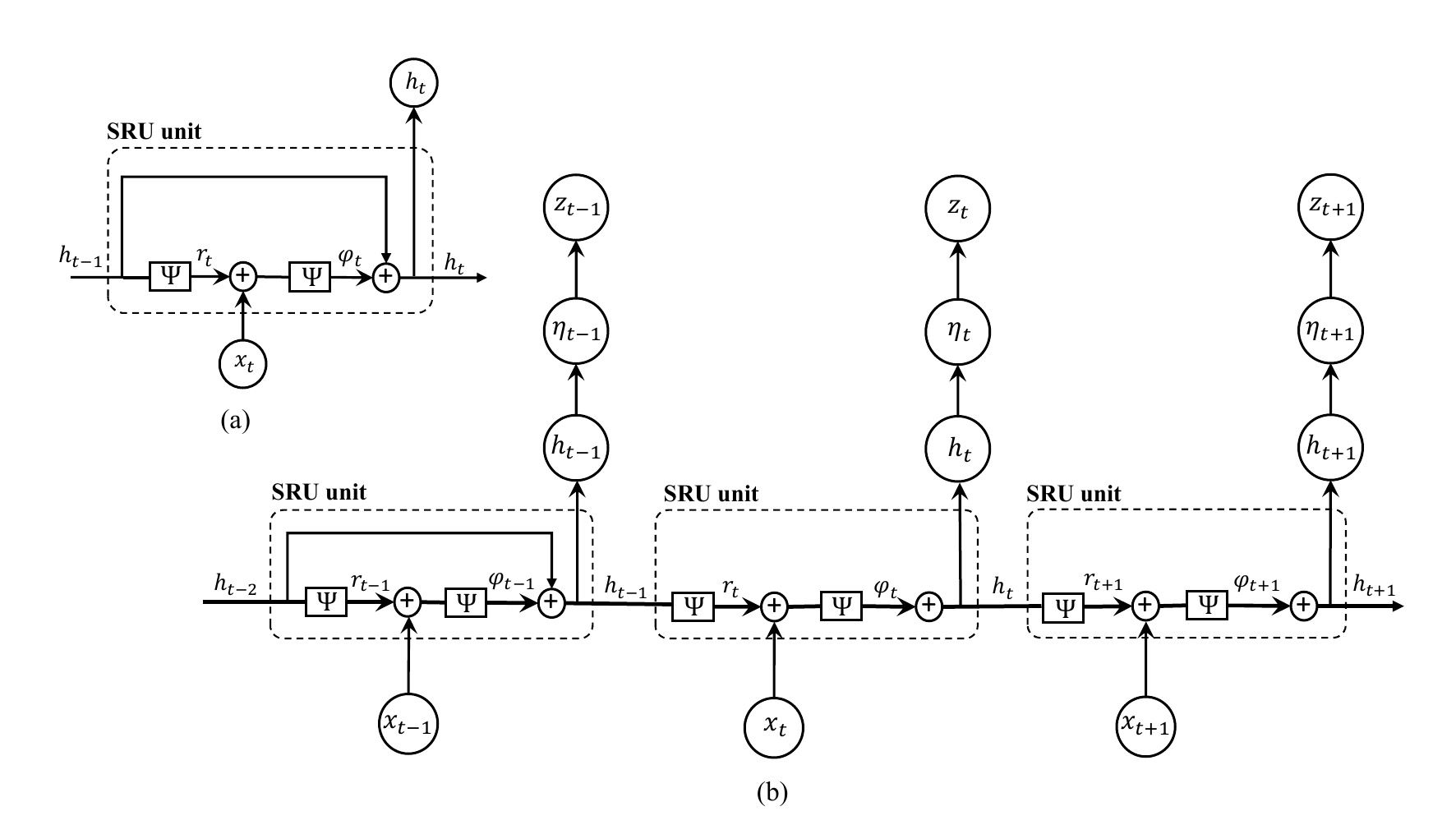}
	\caption{The structure of the SRU unit (\textit{a}) and the graphical representation of the SRU model (\textit{b}), which uses the SRU unit to compute the latent state $h_t$.}
	\label{f:rnn_sru_unit}
\end{figure}

Many sophisticated RNN structures have been proposed to overcome the aforementioned problem in the Simple RNN model; for example, the Long Short-term Memory model of \cite{Hochreiter1997},  
the Gated Recurrent Unit of \cite{cho-etal-2014-learning} and the Statistical Recurrent Unit (SRU) of  \cite{Oliva:2017}.
The SRU allows the vector of summary statistics $h_t$ to traverse through the network using a moving average.
We will use the SRU in this paper as its structure and some of its main parameters carry statistical meaning; see Section \ref{sec:lstm-sv}.
A general SRU structure is mathematically written as
\begin{subequations}
\begin{align}
r_t        &=\Psi(W_h h_{t-1}+b_r),\label{eq:SRU1}\\
\varphi_t  &=\Psi(W_{r}r_t+W_x x_t +b_{\varphi}),\label{eq:SRU2}\\
h_t^{(\alpha_j)} &= \alpha_j h_{t-1}^{(\alpha_j)} + (1-\alpha_j)\varphi_t,\;j=1,...,m;\;\;\;h_t=\big(h_t^{(\alpha_1)},\cdots,h_t^{(\alpha_m)}\big)^\top,\label{eq:SRU3}
\end{align}
\end{subequations}
where $\alpha=(\alpha_1,...,\alpha_m) \in (0,1)$ is a vector of moving average weights, 
and $W_h$, $b_r$, $W_{r}$, $W_x$ and $b_{\varphi}$ are the model parameters. We denote the functional learning structure in \eqref{eq:SRU1}-\eqref{eq:SRU3} as $h_t=\text{SRU}(x_t,h_{t-1})$, which takes $x_t$ - the input data at current time $t$ - and $h_{t-1}$ - the previous output of the SRU - as the input arguments.  
See Figure \ref{f:rnn_sru_unit}(a) for the graphical representation of this SRU structure.
The moving average structure of the state $h_t$ allows the RNN network with SRU units to enjoy some advantages compared to other RNN models. 
The current state $h_t$ is related to the previous state $h_{t-1}$ both {\it directly} and {\it indirectly} and hence mitigate the problem of multiplying the same quantities multiple times as in the Simple RNN model. 
The novel architecture of the SRU allows the model to capture long term dependencies in data via simple moving averages.

\subsection{The SR-SV model}\label{sec:lstm-sv}
This section proposes the SR-SV model that combines SV and SRU for financial volatility modelling. 
The key idea is that we use the SRU structure to capture the complicated effects such as long-term memory and non-linear auto-dependence, in the volatility dynamics that are overlooked by the basic SV models.
This leads to a prior distribution for the log volatility process $z$ that is much more flexible than the AR(1) prior (c.f. Section \ref{sec:SVmodel}). 
Our proposed SR-SV model is as follows
\bea
h_t&=&\text{SRU}(x_{t},h_{t-1}),\;\;t=2,...,T,\;\; \text{with} \; h_1:=0, \; x_t = (\eta_{t-1},z_{t-1})^\top, \label{eq:SR-SVmodel1}\\
\eta_t&=&\beta_0+\beta_1h_t+\eps_t^\eta,\;\;\eps_t^\eta\stackrel{iid}{\sim}\N(0,\sigma^2),\;\;t=1,...,T,\label{eq:SR-SVmodel2}\\
z_t &=& \eta_t+\phi z_{t-1},\;\;t=1,...,T,\label{eq:SR-SVmodel3}\\
y_t &=& e^{\frac12z_t}\eps^y_t,\;\;\eps^y_t\stackrel{iid}{\sim}\N(0,1),\;\;t=1,2,...,T,\label{eq:SR-SVmodel4}
\eea 
that is, we use a SRU to model the dynamics of the hidden states $h_t$.
Here, $z_0$ is the initial value of the log volatility process and a convenient choice of $z_0$ is the log of the unconditional variance of the observed series $y$, i.e., $z_0 = \log(\text{var}(y))$. We follow the literature to initialize $h_1 = 0$ as the recurrent units initially have no memory. Figure \ref{f:sr_sv_full} plots the graphical representation of the SR-SV model. See Appendix \ref{sec:app_2} for the fully-written version of the SR-SV model. 
\begin{figure}[ht]
	\centering
	\includegraphics[width=1\columnwidth]{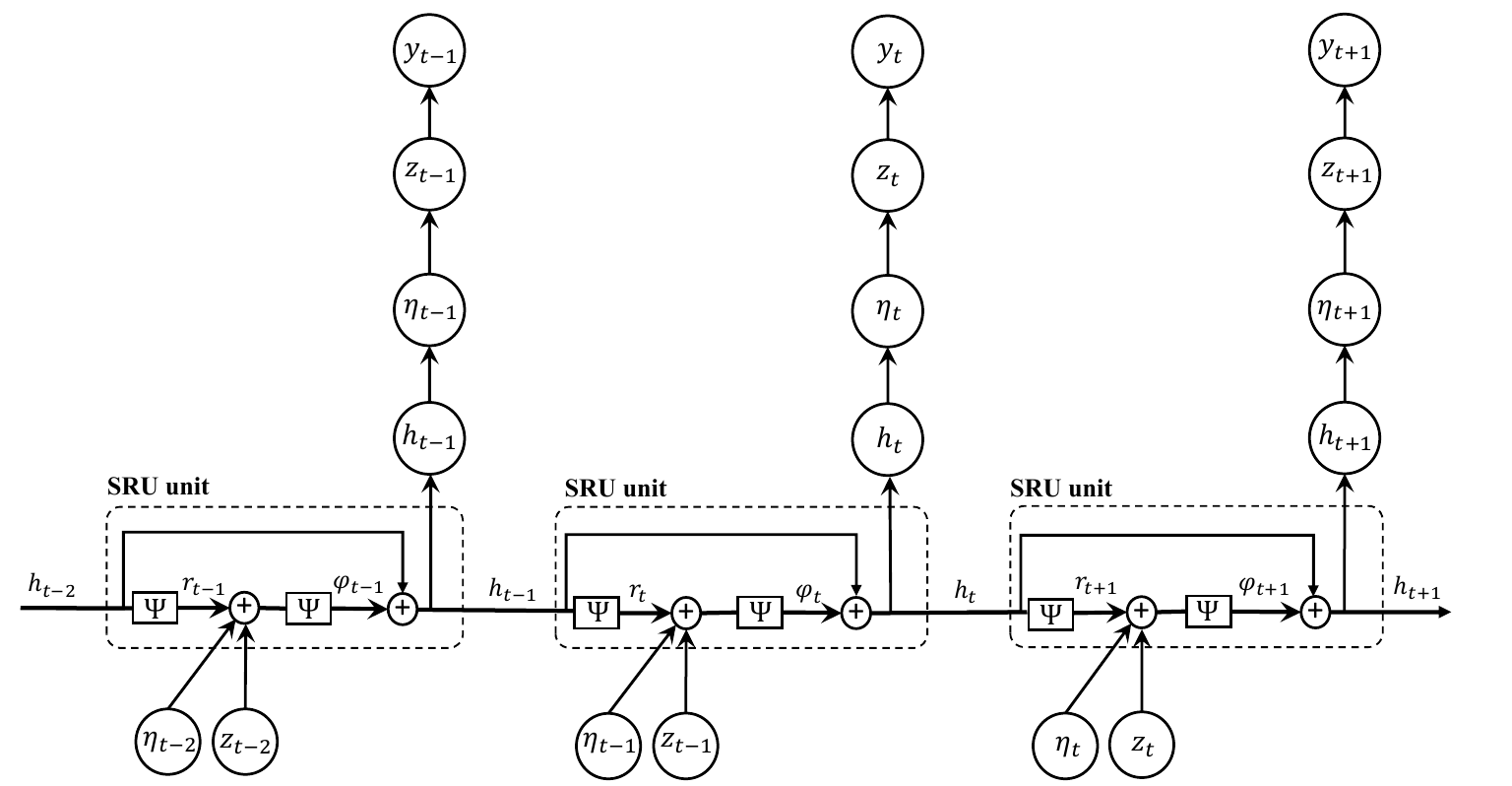}
	\caption{Graphical representation of the SR-SV model. The $\oplus$ symbol represents the addition operation.}
	\label{f:sr_sv_full}
\end{figure}

We note the following important properties of the SR-SV model. 
First, the SR-SV model in \eqref{eq:SR-SVmodel1}-\eqref{eq:SR-SVmodel4} retains the measurement equation \eqref{eq:SR-SVmodel4} and the linear part $\phi z_{t-1}$ of the AR(1) process from the standard SV model, and captures the volatility effects not captured by the AR(1) process, e.g. non-linear and long-memory auto-dependence, via the latent state $h_t$ of the SRU structure. 
The log volatility at time $t$ in \eqref{eq:SR-SVmodel3} can be written as 
\beq
z_t = \beta_0+\beta_1\text{SRU}(\eta_{t-1},z_{t-1},h_{t-1})+\phi z_{t-1} + \epsilon^\eta_t\label{eqn:zt_2}.
\eeq
Therefore, the parameter $\beta_1$ characterizes all the effects in the underlying log volatility process $z$ rather than the short-term linear effect captured by the AR(1) process.
We refer to $\beta_1$ as the {\it non-linearity long-memory} coefficient.
If $\beta_1=0$ and $\eps_1^\eta{\sim}\N(\beta_0/(1-\phi),\sigma^2/(1-\phi^2))$, the SR-SV model becomes the SV model \eqref{SVmodel1}-\eqref{SVmodel2} and hence the SV model is a special case of the SR-SV model. 
We therefore follow the SV literature and assume that $|\phi|<1$. 
The $z$ process, and thus the $y$ process of the SR-SV model, is not guaranteed to be stationary unless $\beta_1=0$ and 
$\eps_1^\eta{\sim}\N(\beta_0/(1-\phi),\sigma^2/(1-\phi^2))$. 
Non-stationarity for volatility is often argued to be more realistic in practice (e.g. \cite{Bellegem:2012}), although it might be mathematically less appealing.
The equation in \eqref{eqn:zt_2} can be further written out as
\beq
z_t = \beta_0+\beta_1\mathfrak{N}(\eta_{t-1},w_{z}z_{t-1},h_{t-1})+\phi z_{t-1} + \epsilon^\eta_t\label{eqn:zt_3},
\eeq
where $\mathfrak{N}(\cdot)$ is a non-linear function and $w_{z}$ is the weight corresponding to $z_{t-1}$; see the full version of the SR-SV model in the Appendix.
If $w_{z}=0$ in \eqref{eqn:zt_3}, then $z_t$ only depends linearly on $z_{t-1}$, therefore this equation indicates that the parameter $w_{z}$ characterizes the serial dependence rather than linearity that the previous log volatility $z_{t-1}$ has on $z_t$.
We will analyse $w_{z}$ in more details in Section \ref{sec:simulation and applications}.

Second, \cite{Oliva:2017} set the scales $\alpha$ of the SRU model to several pre-specified values to obtain a vector of summary statistics $h_t$
at different moving average weights. 
We, however, treat $\alpha$ as a model parameter and learn it from the data.
We note that a higher $\alpha$ weighs more on the historical information while a smaller $\alpha$ puts a more weight on the current information.
We show later in the empirical study that this parameter $\alpha$ is able to quantify the existence of the long-memory auto-dependence commonly exhibited in the volatility dynamics of the financial time series.  

Third, neural networks are highly flexible but often subject to overfitting, i.e., they have over-confident in-sample fit and bad out-of-sample forecasts.
Regularization is often needed to avoid overfitting.
Injecting noise into the layers of the network has been found an effective regularization approach in the Machine Learning literature,
and seen as a form of data augmentation at multiple levels of abstraction \citep{Sietsma:1991,poole:2014,Goodfellow:2016,dieng:2018}.   
In the SR-SV model, by allowing $z_{t-1}$ and $\eta_{t-1}$ to be the inputs of the SRU structure at time $t$, we inject the noise $\epsilon^\eta_{t-1}$ of the volatility process to the input and hidden layers of the SRU. 
This noise-injecting regularization approach makes the SR-SV model perform well on both in-sample fitting and out-of-sample forecast, even with the simplest specification of the SRU structure where all the $r_t$, $\varphi_t$ and $h_t$ are scalars. Our SR-SV model can be categorized as a {\it parametric} model with 
the vector of model parameters $\theta$ consisting of eleven parameters: four main parameters $\beta_0$, $\beta_1$, $\phi$, $\sigma^2$ and the parameters in the SRU including $\alpha, w_h,b_r,w_r,w_\eta,w_z$ and $b_\varphi$.     

Finally, $\beta_0$ plays the role of the scale factor $\tau=e^{\beta_0/2}$ for the variance of $y_t$.
One could set $\beta_0=0$ and modify \eqref{eq:SR-SVmodel4} to $y_t = \tau e^{\frac12z_t}\eps^y_t$; however, this parameterization might be less statistically efficient in terms of Bayesian estimation, especially for the parameter $\tau$ \cite[see][]{Kim:1998}. 

It is straightforward to extend the SR-SV model in \eqref{eq:SR-SVmodel1}-\eqref{eq:SR-SVmodel4} by incorporating other advances in the SV literature.
For example, we can use a Student's $t$ distribution instead of a Gaussian for the measurement shock $\eps^y_t$ and take into account the leverage effect by correlating   
$\eps^y_t$ with the volatility shock $\eps_t^\eta$.
We do not consider these extensions here, however, because using the most basic version makes it easier to understand the strengths and weaknesses of the new model.   

\section{Bayesian inference}\label{Sec:Bayes}
This section discusses Bayesian estimation and inference for the SR-SV model. For a generic sequence $\{x_t\}$ we use $x_{i:j}$ to denote the series $(x_i,...,x_j)$. 
The SR-SV model is a state-space model with the measurement equation
\beq
y_t|z_t	\sim \N(0,e^{z_t}) \label{eqn:lstmsv_g},
\eeq
and the state transition equation
\beq
\label{eqn:transition_density}
z_t|z_{1:t-1},h_t\sim \N(\phi z_{t-1}+\beta_0 + \beta_1 h_t,\sigma^2),\;\;t\geq 2,\;\;\;z_1\sim\N(\beta_0,\sigma^2).
\eeq
We are interested in sampling from the posterior distribution of $\theta$
\bea
\label{eqn:Bayes_SSM}
\pi(\theta) = p(\theta|y_{1:T}) = \frac{p(y_{1:T}|\theta)p(\theta)}{p(y_{1:T})},
\eea
where $p(y_{1:T}|\theta)$ is the likelihood function, $p(\theta)$ is the prior and $p(y_{1:T}) = \int_{\Theta}p(y_{1:T}|\theta)p(\theta)d\theta$ is the marginal likelihood. Recall that the vector of model parameters $\theta$ consists of $\beta_0$, $\beta_1$, $\phi$, $\sigma^2$ and the 7 parameters within the SRU model \eqref{eq:SRU1}-\eqref{eq:SRU3}.

The likelihood function in  (\ref{eqn:Bayes_SSM}) is
\bea
\label{eqn:likelihood}
p(y_{1:T}|\theta) = \int p(y_{1:T}|z_{1:T},\theta)p(z_{1:T}|\theta)dz_{1:T},
\eea
which is computationally intractable for non-linear non-Gaussian state space models like the SV and SR-SV models,
but can be estimated unbiasedly by a particle filter \citep{DelMoral:2004}. 
Bayesian inference for SR-SV can be performed using recent advances in the Sequential Monte Carlo literature that we present next. 

\subsection{The Density Tempered  Sequential Monte Carlo for the SR-SV model}\label{sec:SMC}
\cite{Duan:2015} propose the Density Tempered  Sequential Monte Carlo (DT-SMC) approach to Bayesian inference for state space models where the likelihood is intractable. 
The DT-SMC sampler generalizes the SMC method of \cite{Neal:2001} and \cite{DelMoral:2006} when the likelihood can be computed analytically. In order to sample from the posterior $\pi(\theta)$, the DT-SMC method first samples a set of $M$ weighted particles $\{W^j_0,\theta_0^j\}^M_{j=1}$ from an easy-to-sample distribution $\pi_0(\theta)$, such as the prior $p(\theta)$, and then traverses these particles through intermediate distributions $\pi_t(\theta), \;\; t=1,...,K$, which target the posterior distribution $\pi(\theta)$ eventually, i.e. $\pi_K(\theta)=\pi(\theta)$. The DT-SMC method uses the following intermediate distributions
\bea
\pi_t(\theta):=\pi_t(\theta|y_{1:T}) \propto \widehat{p}(y_{1:T}|\theta,u)^{\gamma_t}p(\theta),
\label{eqn:lik_anneal}
\eea
where the $\gamma_t$ is referred to as the level temperature and $0 = \gamma_0 < \gamma_1 < \gamma_2 < ... < \gamma_K=1$, $\widehat{p}(y_{1:T}|\theta,u)$ is the unbiased estimator of the likelihood $p(y_{1:T}|\theta)$ and $u$ is the set of pseudo random numbers used within a particle filter to estimate the likelihood $p(y_{1:T}|\theta)$. 
For the purpose of this paper where it is possible to sample from the prior $p(\theta)$, we set $\pi_0(\theta)=p(\theta)$.  Algorithm \ref{alg:lik_annealing} summarizes the DT-SMC method for the SR-SV model.

The DT-SMC method consists of three main steps: reweighting, resampling and Markov move.
At the begining of SMC iteration $t$, the set of weighted particles $\{W_{t-1}^j,\theta_{t-1}^j\}^M_{j=1}$ that approximate the intermediate distribution $\pi_{t-1}(\theta)$ is reweighted to approximate the target $\pi_{t}(\theta)$.
The efficiency of these weighted particles is often measured by the effective sample size (ESS) \citep{Robert:1998,LiuChen:1998} defined in \eqref{eq:ESS}. 
If the ESS is below a prespecified threshold, the particles are resampled; the resulting equally-weighted resamples, which are now approximate samples from $\pi_{t}(\theta)$, are then refreshed by a Markov kernel whose invariant distribution is $\pi_{t}(\theta)$. For example, \cite{Duan:2015} uses the pseudo marginal Metropolis-Hastings (PMMH) kernal of \cite{Andrieu:2010} with the likelihood estimated unbiasedly by the particle filter in the Markov move step. However,  \cite{Pitt:2012} suggest that the PMMH approach works efficiently when the variance of the log of the
estimated likelihood is around 1. For some state space models like the SR-SV model, a large number of particles might be required to obtain a likelihood estimator with log variance to be around 1, which is computationally inefficient. 
To tackle this problem, we incorporate the Correlated Pseudo Marginal (CPM) approach of \cite{Deligiannidis:2018} into the Markov move step.
The CPM method makes the current set of random numbers $u$ and proposal $u^\prime$ correlated, and helps reduce the variance of the ratio $\widehat{p}(y_{1:T}|\theta^{\prime},u^{\prime}) / \widehat{p}(y_{1:T}|\theta,u)$ in \eqref{eqn:proposal}, thus leading to a better mixing Markov chain while using less number of particles in the particle filter. Similar to the SMC methods of \cite{DelMoral:2006} and \cite{Neal:2001}, the DT-SMC method is parallelizable as the particles move independently in the Markov move step, and provides an estimate of the marginal likelihood as a by-product.
  
In Algorithm \ref{alg:lik_annealing}, we use a random walk proposal for $q(\theta^\prime|\theta)$.
We follow \cite{Dang:2018} and choose the tempering sequence $\gamma_t$ adaptively to ensure a sufficient level of particle efficiency by selecting the next value of $\gamma_t$ such that ESS stays above a threshold. 

\begin{algorithm}
	\caption{The Density Tempered  Sequential Monte Carlo for the SR-SV model}  
	\label{alg:lik_annealing}
	1. Sample  $\theta^j_0 \sim p(\theta)$, $u^j_0 \sim p(u)$ and set  $W_0^j=1/M$ for $j=1...M$ \\
	2. \textbf{For} $t=1,...,K$, 
	\begin{itemize}
		\item[] \textbf{Step 1: }  \bol{Reweighting}: Compute the unnormalized weights
		\beq\label{eq: SMC llh re-weight}
		w_t^j = W_{t-1}^j\frac{\widehat{p}(y_{1:T}|\theta_{t-1}^j,u_{t-1}^j)^{\gamma_t}p(\theta_{t-1}^j)}{\widehat{p}(y_{1:T}|\theta_{t-1}^j,u_{t-1}^j)^{\gamma_{t-1}}p(\theta_{t-1}^j)} = W_{t-1}^j\widehat{p}(y_{1:T}|\theta_{t-1}^j,u_{t-1}^j)^{\gamma_t - \gamma_{t-1}}, \; \; j=1,...,M
		\eeq
		and set the new normalized weights
		\bea
		W^j_t = \frac{w^j_t}{\sum_{s=1}^{M}w^s_t}, \; \; j=1,...,M.
		\eea			
		\item[] \textbf{Step 2: } Compute the effective sample size (ESS): 
		\bea
		\text{ESS} = \frac{1}{\sum_{j=1}^{M} \left(W_t^j\right)^2}.
		\label{eq:ESS}
		\eea
		\begin{itemize}
			\item[] \textbf{if} $\text{ESS} < c M$ for some $0<c<1$, \textbf{then}   
			\begin{itemize}
				\item[(i)] \bol{Resampling}: Resampling from $\{\theta_{t-1}^j,u_{t-1}^j\}_{j=1}^M$ using the weights $\{W_{t}^j\}^M_{j=1}$, and then set $W_t^j=1/M$ for $j=1...M$, to obtain the new equally-weighted particles $\{\theta_{t}^j,u_{t}^j,W_{t}^j\}^M_{j=1}$.
				\item[(ii)] \bol{Markov move}: For each $j=1,...,M$, move the samples $\theta_t^j$, $u_t^j$ according to $N_{\text{CPM}}$ CPM steps:
				\begin{itemize}
					\item[(a)] Sample $\theta_t^{j\prime}$ from the proposal density $q(\theta_t^{j\prime}|\theta_t^j)$.
					\item[(b)] Sample $\epsilon^j \sim \N(0_D,I_D)$ and set $u_t^{j\prime}=\rho u_t^j + \sqrt{1-\rho^2}\epsilon^j$ with $\rho \in (-1,1)$ is a correlation factor. 
					\item[(c)] Compute the estimated likelihood $\widehat{p}(y_{1:T}|\theta_t^{j \prime},u_t^{j \prime})$ using a particle filter (see Algorithm \ref{alg:PF} in Appendix \ref{sec:app_3})					
					\item[(d)] Set $\theta_t^j = \theta_t^{j \prime}$ and $u_t^j = u_t^{j \prime}$ with the probability 
					\bea
					\text{min}\left(1,\frac{\widehat{p}(y_{1:T}|\theta_t^{j \prime},u_t^{j \prime})^{\gamma_t}p(\theta_t^{j \prime})}{\widehat{p}(y_{1:T}|\theta_t^{j},u_t^{j})^{\gamma_t}p(\theta_t^{j})} \frac{q(\theta_t^j|\theta_t^{j\prime})}{q(\theta_t^{j\prime}|\theta_t^j)}\right), \label{eqn:proposal}
					\eea
					otherwise keep $\theta_t^j$, $u_t^j$ unchanged.
				\end{itemize}
			\end{itemize}
			\textbf{end}
		\end{itemize}
		3. The log of marginal likelihood estimate is
		\bea
		\text{log}\widehat{p}(y_{1:T}) = \sum^K_{t=1} \text{log}\left(\sum_{j=1}^M w_t^j\right).
		\label{eq:marllh}
		\eea
	\end{itemize}
\end{algorithm}
	
\subsection{Model choice by marginal likelihood}\label{sec:IS2}
The marginal likelihood is often used to choose between models via the Bayes factor \citep{jeffreys:1935,Kass:1995}. In order to compare the relative performance between two models $M_1$ and $M_2$ on a given data $y_{1:T}$, we can use the Bayes factor defined by  
\beq
F_{M_1,M_2}=\frac{p(y_{1:T}|M_1)}{p(y_{1:T}|M_2)},
\label{eq:bayesfactor}
\eeq
providing a Bayesian alternative to hypothesis testing.
The larger the Bayes factor $F_{M_1,M_2}$, the stronger $M_1$ is supported by the data than $M_2$. \cite{Jeffreys:1961} suggests a scale of interpretation of the Bayes factor $F_{M_1,M_2}$ as listed in Table \ref{tab:bayes_factor}.
We note that the DT-SMC sampler in the previous section provides an efficient way to compute the marginal likelihood.

\begin{table}[h]
	\begin{center}
		\begin{tabular}{ccccc}
			\hline\hline
			\rule{0pt}{3ex}
			\bol{Grade} &\boldsymbol{$F_{M_1,M_2}}$      &\boldsymbol{$\textbf{log}_{10}F_{M_1,M_2}$}  &\boldsymbol{$\textbf{ln}F_{M_1,M_2}$}  &\bol{Strength of evidence}\\
			\hline
			\rule{0pt}{3ex}
			0&$< 10^0$            &$<0$         &$<0$       &Negative (supports $M_2$)\\
			1 & $10^0 -10^{1/2}$  &$0.0-0.5$    &$0.0-1.2$  &Barely worth mentioning\\
			2 & $10^{1/2}-10^1$   &$0.5-1.0$    &$1.2-2.3$  &Substantial\\
			3 & $10^1-10^{3/2}$   &$1.0-1.5$    &$2.3-3.5$  &Strong\\
			4 & $10^{3/2}-10^2$   &$1.5-2.0$    &$3.5-4.6$  &Very strong\\
			5 & $> 10^2$          &$>2.0$       &$>4.6$     &Decisive\\
			\hline\hline
		\end{tabular}
	\end{center}
	\caption{Jeffreys' scale of interpretation of the Bayes Factor $F_{M_1,M_2}$.}
	\label{tab:bayes_factor}
\end{table}

\section{Simulation studies and applications}\label{sec:simulation and applications}
This section evaluates the performance of the SR-SV model relative to the SV, N-SV and LMSV models using a simulation study and real data applications. 
We do not report the results for GARCH as it performs similarly to SV. 
We use the DT-SMC sampler for the Bayesian inference in the SV, N-SV and SR-SV models. 
As the LMSV model does not have an explicit state-space representation and its likelihood function is analytically intractable,
we follow \cite{Breidt:1998} and estimate the LMSV model on the frequency domain. 
The implementation of Bayesian inference for the LMSV model is presented in Appendix \ref{sec:app_1}.
Table \ref{tab:implementation} in Appendix \ref{sec:app_3} lists our implementation details of the DT-SMC sampler.
All the computations are run on a High-Performance Computing (HPC) machine with 16-core CPU and 16 Gigabytes of RAM. 
The DT-SMC sampler was initialized by sampling from the priors in Table \ref{table:prior}.

\begin{table}[ht]
 	\begin{center}
 		\begin{tabular}{c c c c c c c c}
 			\hline\hline
 			\multicolumn{3}{c}{SR-SV} &\multicolumn{3}{c}{SV} &\multicolumn{2}{c}{N-SV}\\
 			\cline{1-2}
 			\cline{4-5}
 			\cline{7-8}
 			\rule{0pt}{3ex}
 			Parameter       &Prior                    &   &Parameter       &Prior &         &Parameter       &Prior \\
 			\hline
 			\rule{0pt}{3ex}
 			$\beta_0$   & $\N(0,0.1)$               &   & $\mu$    & $\N(0,25)$          & &$\mu$     & $\N(0,25)$ \\
 			$\frac{\phi + 1}{2}$      & Beta(20,1.5)            &   & $\frac{\phi + 1}{2}$     & Beta(20,1.5) & &$\frac{\phi + 1}{2}$      & Beta(20,1.5) \\
 			$\sigma^2$  & $IG(2.5,0.25)$  &   & $\sigma^2$ & $IG(2.5,0.25)$    & &$\sigma^2$  & $IG(2.5,0.25)$ \\
 			$\beta_1$   & $IG(2.5,1)$ &   &            &              & &$\delta$    & $\N(0,0.1)$ \\
 			$\alpha$    & Beta(2,2)     &&&&&&  \\
 			$w_h,w_{\phi}$,$w_{\eta}$   & $\N(0,0.1)$   &&&&&&  \\
 			$b_r,b_{\phi}$   & $\N(0,0.1)$       &&&&&&  \\
 			$w_{z}$   & $IG(2.5,1)$       &&&&&&  \\
 			\hline\hline
 		\end{tabular}
 	\end{center}
 	\caption{Prior distributions for the parameters in the SR-SV, SV and N-SV models. The notation $\N$, $IG$ and $\text{Beta}$ denote the Gaussian, inverse-Gamma and Beta distributions, respectively.}
 	\label{table:prior}
\end{table}
	 
We now motivate the choice of the priors in Table \ref{table:prior}. 
We follow \cite{YuJun:2006} and \cite{Kim:1998} to set the same prior, which is a Beta distribution, for the persistence parameters $\phi$ of the three models SV, N-SV and SR-SV. We also use an inverse-Gamma prior for the parameters $\sigma^2$ in all models but make it more flat than the priors used in \cite{YuJun:2006} and \cite{Kim:1998}. We follow \cite{YuJun:2006} to use an informative but reasonably flat prior distribution for the intercept $\mu$ in the SV and N-SV models.
For the SR-SV model, we found that the posterior distributions of $\beta_1$ and $w_{z}$ to be unimodal under inverse-Gamma priors. 
We use a normal prior with a zero mean and a small variance for the SRU parameters, except $w_{z}$, because empirical results from the SRU literature show that the values of the SRU parameters are often small.
Finally, we set a normal prior with a zero mean and a small variance for the intercept $\beta_0$ in the SR-SV model as the empirical results often show small values of $\beta_0$.

Table \ref{tab:forecast_score} lists the predictive scores used to measure the out-of-sample performance. 
The smaller the predictive scores, the better.

\begin{table}[h]
	\begin{center}
		\begin{tabular}{ccc|ccc}
			\hline\hline
			\rule{0pt}{3ex}
			\bol{Score}         &&\bol{Definition}   &    \bol{Score}         &&\bol{Definition}    \\
			\hline
			\rule{0pt}{3ex}
			$\text{PPS}$   &&$-T_{test}^{-1}\sum_{D_{test}} \log p(y_t|y_{1:t-1},\wh\theta)$  &$\text{MSE}_1$   &&$T_{test}^{-1}\sum_{D_{test}} (\sigma_t -\widehat{\sigma}_t)^2$\\
			&&&&&\\
			$\text{QLIKE}$  &&$T_{test}^{-1}\sum_{D_{test}} \left(\text{log}(\widehat{\sigma}_t^2) + \sigma_t^2 \widehat{\sigma}_t^{-2}\right)$&$\text{MSE}_2$   &&$T_{test}^{-1}\sum_{D_{test}} (\sigma_t^2 -\widehat{\sigma}_t^2)^2$  \\
			&&&&&\\
			$\text{R}^2\text{LOG}$   &&$T_{test}^{-1}\sum_{D_{test}} \left[\text{log}(\sigma_t^2 \widehat{\sigma}_t^{-2})\right]^2$  &$\text{MAE}_1$   &&$T_{test}^{-1}\sum_{D_{test}} |\sigma_t -\widehat{\sigma}_t|$  \\
			&&&&&\\
			&& &$\text{MAE}_2$   &&$T_{test}^{-1}\sum_{D_{test}} |\sigma_t^2 -\widehat{\sigma}_t^2|$ \\
			\hline\hline
		\end{tabular}
	\end{center}
	\caption{Definition of the predictive scores to measure the out-of-sample performance on simulation and real index data. Here, $\widehat{\sigma}_t$ is an estimate of the volatility $\sigma_t$,
	$T_{test}$ is the number of observations in test data $D_{test}$ and $\wh\theta$ is a posterior mean estimate of $\theta$.}
	\label{tab:forecast_score}
\end{table}

\subsection{Simulation studies}
We consider three volatility models specified in Table \ref{tab:sim_data}. Model 1 is a GARCH(1,1). Model 2 is an extension of the GARCH(1,1) model by applying a Box-Cox power transformation \citep{Box:1964} to both the conditional variance equation and the volatility dynamics. Model 2 is similar to the non-linear ARCH models of \cite{Higgins:1992} but they use lagged innovations to construct the conditional variance. Model 3 is a FIGARCH(1,$d$,1) model of \cite{Baillie:1996} which uses a long-memory process AFRIMA(1,$d$,1) to simulate the long-memory auto-dependence.  
\begin{table}[h]
	\begin{center}
		\begin{tabular}{c|c|c}
			\hline\hline
			\rule{0pt}{3ex}
			\textbf{Data}          &\textbf{Model}     &\textbf{Parameters} \\
			\hline
			\rule{0pt}{3ex}
			\multirow{3}{*}{SIM I}      &\multirow{3}{*}{$
				\begin{aligned}
				\sigma^2_t &= \mu + \alpha y_{t-1}^2 + \beta \sigma_{t-1}^2,\;\;\;t=2,...,T\\
				y_t &= \sigma_t\eps_t,\;\;\;\eps_t\sim\N(0,1),\;\;\;t=1,...,T\\
				\end{aligned}$}                    &$\sigma^2_1 = 0.1$, $\mu=0.1$\\
			&&$\alpha=0.07$, $\beta=0.92$ \\
			&&\\
			\hline
			\rule{0pt}{3ex}
			\multirow{4}{*}{SIM II}      &\multirow{4}{*}{$
			\begin{aligned}
			h_t &= \mu + \alpha \frac{(y_{t-1}^2)^\delta-1}{\delta} + \beta h_{t-1},\;\;\;t=2,...,T\\
			y_t &= \Big(1+\delta h_t\Big)^{1 \slash 2\delta}\eps_t,\;\;\;\eps_t\sim\N(0,1),\;\;\;t=1,...,T\\
			\end{aligned}$}                    &$h_1 = 0.1$, $\mu=0.1$\\
			&&$\alpha=0.15$, $\beta=0.82$ \\
			&&$\delta=0.9$\\
			&&\\
			\hline
			\rule{0pt}{3ex}
			\multirow{4}{*}{SIM III}      &\multirow{4}{*}{$
				\begin{aligned}
				\sigma^2_t &= \mu + \Big[1-\beta B- (1-\phi B)(1-B)^d\Big]y_t^2 +\beta \sigma^2_{t-1},\;\;\;t=2,...,T\\
				y_t &= \sigma_t\eps_t,\;\;\;\eps_t\sim\N(0,1),\;\;\;t=1,...,T\\
				\end{aligned}$}                    &$\sigma^2_1 = 0.1$, $\mu=0.01$\\
			&&$\phi=0.01$, $\beta=0.5$ \\
			&&$d=0.62$\\
			&&\\
			\hline\hline
		\end{tabular}
	\end{center}
	\caption{Simulation: Data generating process.}
	\label{tab:sim_data}
\end{table}

\begin{figure}[H]
	\centering
	\includegraphics[width=1\columnwidth]{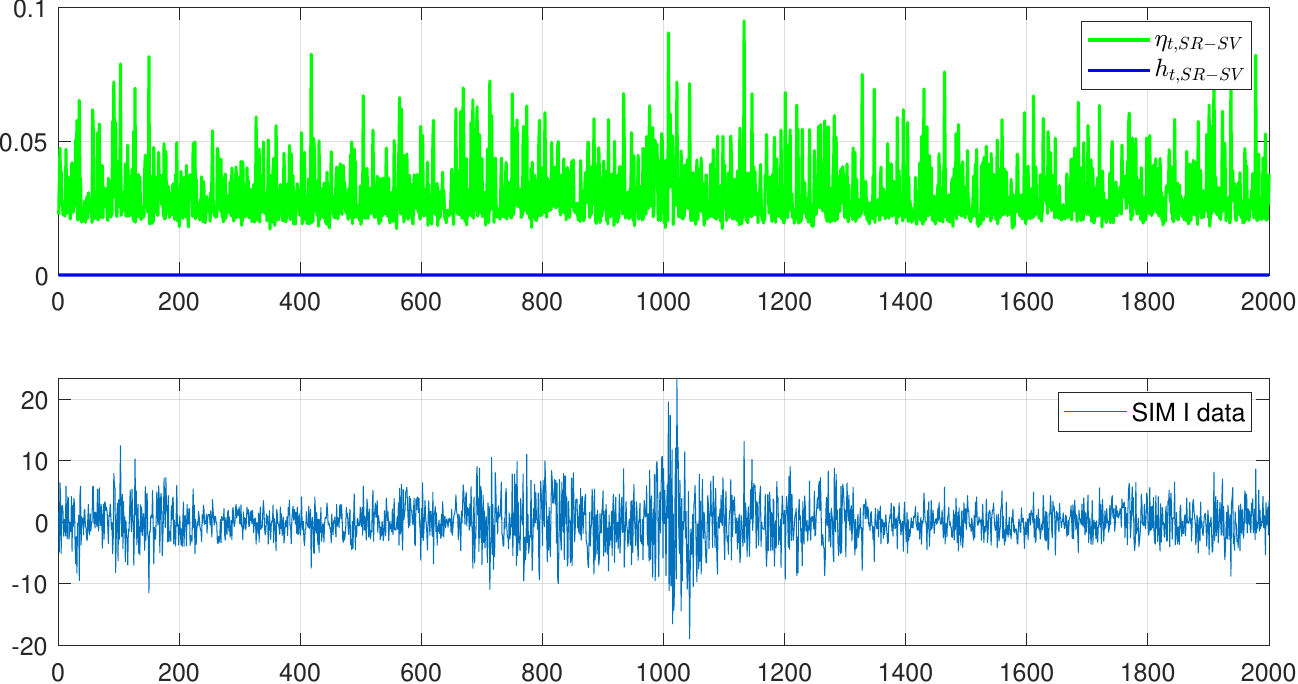}
	\caption{SIM I: The filtered values of $\eta_t$ and $h_t$ of the SR-SV model, together with the SIM I in-sample data. (This is better viewed in colour).}
	\label{f:sim_1_in_sample}
\end{figure}

\begin{figure}[H]
	\centering
	\includegraphics[width=1\columnwidth]{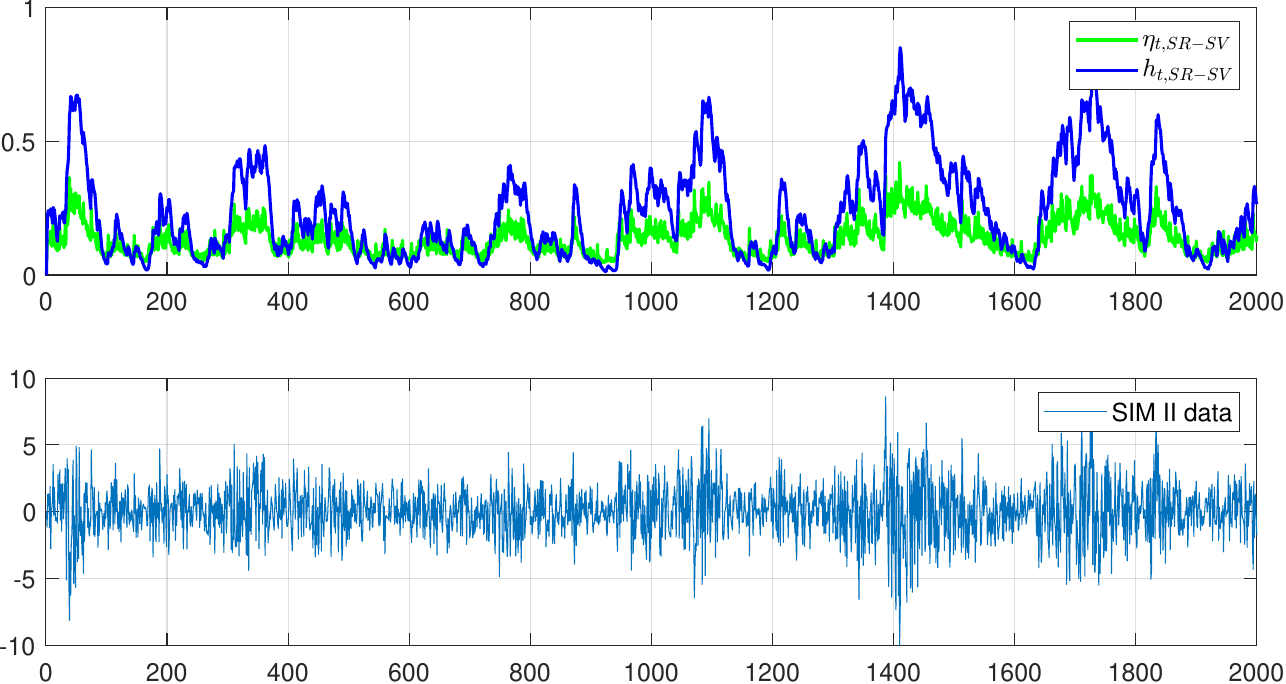}
	\caption{SIM II: The filtered values of $\eta_t$ and $h_t$ of the SR-SV model, together with the SIM II in-sample data. (This is better viewed in colour).}
	\label{f:sim_2_in_sample}
\end{figure}

\begin{figure}[h]
	\centering
	\includegraphics[width=1\columnwidth]{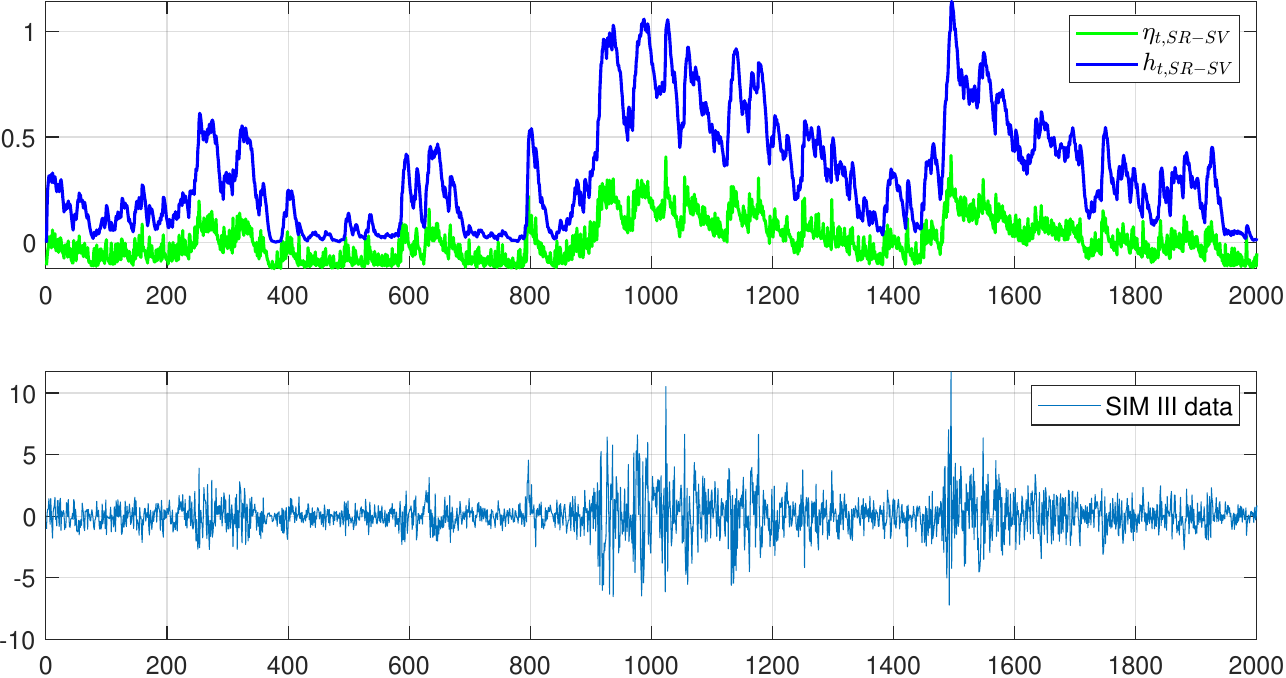}
	\caption{SIM III: The filtered values of $\eta_t$ and $h_t$ of the SR-SV model, together with the SIM III in-sample data. (This is better viewed in colour).}	
	\label{f:sim_3_in_sample}
\end{figure}

We generate time series of $T=3000$ observations from these three models and name the simulation datasets as SIM I, SIM II, SIM III, accordingly. The parameters are set so that $y_t$ somewhat resembles real financial time series data exhibiting volatility clustering with non-linearity (SIM II) and long-memory (SIM III) auto-dependence in the underlying volatility dynamics. For each dataset, the first $T_{in} = 2000$ observations are used for model estimation and the last $T_{out} = 1000$ are for out-of-sample analysis. Table \ref{tab:Simulation_params} shows the posterior mean estimates for the parameters of the SV and SR-SV models, with the posterior standard deviations in brackets; for the SR-SV model we only show the results for the main parameters. The last column in the table shows the marginal likelihood estimates, averaged over 10 different runs of the DT-SMC sampler, together with the Monte Carlo standard errors in the brackets.  Figures \ref{f:sim_1_in_sample}, \ref{f:sim_2_in_sample} and \ref{f:sim_3_in_sample} plot the filtered values of the $\eta_t$ and $h_t$ components of the SR-SV model in SIM I, SIM II and SIM III datasets, respectively.  
Figure \ref{f:sim_filtered_vol} in Appendix \ref{sec:app_4} plots the true volatility together with the filtered volatility produced by the SV and SR-SV models, for all the three simulation datasets.

\begin{table}[H]
	\centering
	\small
	\begin{tabular}{ccccccccl}
		\hline\hline
		\rule{0pt}{3ex}
		& $\mu$ & $\phi$ & $\sigma^2$ & $\alpha$  & $\beta_0$& $\beta_1$ &$w_{z}$& Mar.llh \\
		\hline
		\rule{0pt}{3ex}
		\textbf{SIM I} &&&&&&&& \\
		\rule{0pt}{3ex}
		\multirow{1}{*}{SV}   &\bol{2.145} &\bol{0.985}   &\bol{0.019}   & & & & &$-5100.8$\\
							  &(0.237)     &(0.004)       &(0.003)        &&&&& (0.131)\\
		\rule{0pt}{3ex}
		\multirow{1}{*}{SR-SV} & &\bol{0.974} &\bol{0.020} &\bol{0.534} &\bol{0.027}&\bol{0.388}&$-\bol{0.205}$ &$-5099.9$\\
							   & &(0.023)     &(0.005)     &(0.166)     &(0.031)       &(0.235)    &(0.261)     &(0.300)\\
		\hline
		\rule{0pt}{3ex}
		\textbf{SIM II} &&&&&&&& \\
		\rule{0pt}{3ex}
		\multirow{1}{*}{SV}   &\bol{1.050} & \bol{0.967}  & \bol{0.032}  &  & & & &$-4060.3$\\
		                      &(0.125)     &(0.008)       &(0.006)&&&&& (0.164)\\
		\rule{0pt}{3ex}
		\multirow{1}{*}{SR-SV} & &\bol{0.792} &\bol{0.041} &\bol{0.515} &\bol{0.043}&\bol{0.423}&\bol{0.530} &$-4057.7^\ast$\\
		                       & &(0.106)     &(0.010)     &(0.156)     &(0.044)    &(0.207)    &(0.256)     &(0.306)\\
		\hline
		\rule{0pt}{3ex}
		\textbf{SIM III} &&&&&&&& \\
		\rule{0pt}{3ex}
		\multirow{1}{*}{SV}   &\bol{0.134} & \bol{0.984}  & \bol{0.041}  &  & &  & &$-3146.9$\\
		&(0.329) &(0.005)&(0.007)&&&&& (0.195)\\
		\rule{0pt}{3ex}
		\multirow{1}{*}{SR-SV} & &\bol{0.896} &\bol{0.056} &\bol{0.645} &$-\bol{0.093}$&\bol{0.325}&\bol{0.290} &$-3144.2^*$\\
	                           & &(0.035)     &(0.013)     &(0.240)     &(0.045)       &(0.132)    &(0.115)     &(0.316)\\	      
		\hline\hline
	\end{tabular}
	\caption{Simulation: Posterior means of the parameters with the posterior standard deviations in brackets. The last column shows the estimated log marginal likelihood with the Monte Carlo standard errors in brackets, averaged over 10 different runs of the DT-SMC sampler. The asterisks indicate the cases when the Bayes factors strongly support the SR-SV model over the SV model. The marginal likelihood are reported in natural log scale.}
	\label{tab:Simulation_params}
\end{table} 

The estimation results suggest the following conclusions. 
First, for the SIM I data, the difference of marginal likelihood estimates in Table \ref{tab:Simulation_params} between the SV and SR-SV models is insignificant,
the coefficient $\beta_1$ is insignificant, and the filtered volatilities from these two models in Figure \ref{f:sim_filtered_vol} are identical and close to the true volatility.
This implies that the SV and SR-SV models fit equally well to the SIM I data and that the SR-SV model is close to the SV model if the true data generating process, which is GARCH(1,1) in this example, exhibits no other effects rather than short-memory linear auto-dependence within the volatility dynamics.  

Second, the estimation results on the SIM II and SIM III data show that the additional neural network structure of the SR-SV model is able to efficiently capture the volatility effects overlooked by the basic SV model. This is supported by the Bayes factors of the SR-SV model compared to the SV model of more than $e^{2.3}$, which, according to the interpretation in Table \ref{tab:bayes_factor}, strongly support the SR-SV model. The plots of the $h_t$ and $\eta_t$ components of the SR-SV model in Figures \ref{f:sim_1_in_sample},  \ref{f:sim_2_in_sample} and \ref{f:sim_3_in_sample} clearly show that $h_t$, and hence $\eta_t$, is well responsive to volatility effects rather than the linear short-memory effects.
For example, in the SIM I data when the volatility exhibits no non-linear and long-memory effects, the $h_t$ shown in Figure \ref{f:sim_1_in_sample} is significantly small at all time steps and hence the $\eta_t$ component simply fluctuates around $\beta_0$ during both low and high volatility periods. Figures \ref{f:sim_2_in_sample} and \ref{f:sim_3_in_sample}, on the other hand, show that the $h_t$ response adaptively to the changes in the volatility dynamics. As the result, $\eta_t$ is small during the low volatility periods and large in the high volatility periods. The non-linear (SIM II) and long-memory (SIM III) auto-dependence of the simulated volatility are well captured by the SRU structure of the SR-SV model. The plots of filtered volatility in Figure \ref{f:sim_filtered_vol} show that the filtered volatility of the SR-SV model are generally closer to the true volatility than those of the SV model.

\begin{figure}[H]
	\centering
	\includegraphics[width=0.9\columnwidth]{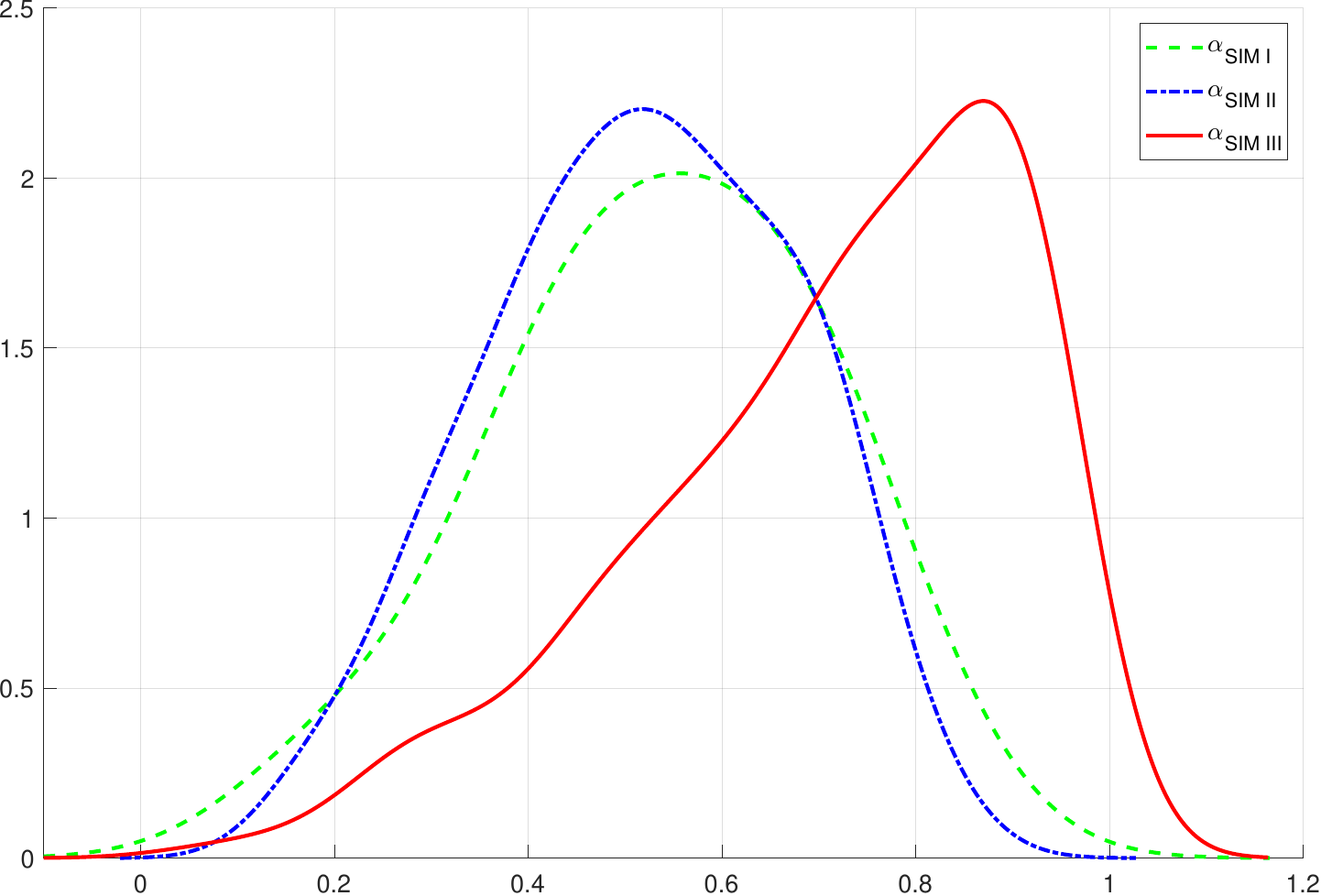}
	\caption{Simulation: Posterior densities of the moving avarage weight $\alpha$ on the three simulation datasets.}
	\label{f:alpha_density_sim}
\end{figure}

Third, the parameters of the SR-SV model are able to characterize well the existence of the various volatility effects in these simulation data. 
The estimated posterior means of parameter $\beta_1$ are more than two standard deviations from zero in the SIM II and SIM III data, suggesting the existence of volatility effects rather than linearity in the volatility dynamics of these two datasets; while $\beta_1$ is less than two standard deviations from zero in the SIM I data, suggesting that only simple 
linear effects are detectable in the volatility dynamics of this dataset.
Similarly, the non-linear coefficient $w_{z}$ of the SR-SV model (c.f. equation \eqref{eqn:zt_3}) is more than two standard deviations from zero in the SIM II and SIM III data but not in the SIM I data, indicating that the SR-SV model is able to detect the serial dependence rather than the linear dependence that the past log volatility $z_{t-1}$ has on the current log volatility $z_t$. The estimated posterior mean of the moving average weight parameter $\alpha$ in the SIM III data is higher than those in the SIM I and SIM II data, supporting further the evidence of the long-memory auto-dependence exhibited in the volatility dynamic of the SIM III data, which is generated from a FIGARCH(1,$d$,1) model. 
Figure \ref{f:alpha_density_sim} shows that the posterior mode of $\alpha$ in the SIM III data is much closer to 1 than those in the SIM I and SIM II data. Finally, it is worth noting that the persistence parameter $\phi$ of the SR-SV is close to the persistence parameter $\phi$ of the SV model in the SIM I data but much smaller than that of the SV model in the SIM II and SIM III model. This is probably because the non-linear coefficient $w_z$ with respect to the past log volatility $z_{t-1}$ is significant in the SIM II and SIM III data, and hence the historical information has been also well stored in the $\eta_t$ process.

\begin{table}[H]
	\centering
	\small
	\begin{tabular}{rcccccccc}
		\hline\hline
		\rule{0pt}{3ex}
		&PPS& $\text{MSE}_1$ &$\text{MSE}_2$ & $\text{MAE}_1$&$\text{MAE}_2$&$\text{QLIKE}$&$\text{R}^2\text{LOG}$&Count\\
		\hline
		\rule{0pt}{3ex}
		\bol{SIM I} \\
		\rule{0pt}{3ex}
		SV             &\bol{2.355}      &0.480      &0.485      &0.506      &0.540      &0.481      &0.482      &1\\
		&(0.001)    &(0.003)    &(0.003)    &(0.002)    &(0.004)    &(0.001)    &(0.004)&     \\
		\rule{0pt}{3ex}
		SR-SV          &2.357      &\bol{0.435}&\bol{0.381}&\bol{0.342}&\bol{0.319}&\bol{0.450}&\bol{0.395}&6\\
		&(0.000)    &(0.002)    &(0.003)    &(0.001)    &(0.002)    &(0.001)    &(0.003)    & \\
		\hline
		\rule{0pt}{3ex}
		\bol{SIM II} \\
		\rule{0pt}{3ex}
		SV             &1.881     &0.189      &0.282      &0.309      &0.383      &0.359      &0.512      &0\\
		&(0.000)   &(0.000)    &(0.001)    &(0.001)    &(0.001)    &(0.001)    &(0.001)    & \\
		\rule{0pt}{3ex}
		SR-SV          &\bol{1.878}     &\bol{0.076}&\bol{0.122}&\bol{0.139}&\bol{0.189}&\bol{0.172}&\bol{0.284}&7\\
		&(0.000)   &(0.002)    &(0.003)    &(0.002)    &(0.000)    &(0.003)    &(0.003)    & \\
		\hline
		\rule{0pt}{3ex}
		\bol{SIM III} \\
		\rule{0pt}{3ex}
		SV             &1.722     &0.919      &0.967      &0.707      &0.796      &0.754      &0.529      &0\\
		&(0.001)   &(0.004)    &(0.004)    &(0.003)    &(0.003)    &(0.004)    &(0.004)    & \\
		\rule{0pt}{3ex}
		SR-SV          &\bol{1.720}     &\bol{0.733}&\bol{0.775}&\bol{0.548}&\bol{0.625}&\bol{0.588}&\bol{0.399}&7\\
		&(0.000)   &(0.004)    &(0.003)    &(0.002)    &(0.003)    &(0.003)    &(0.002)    & \\
		\hline\hline
	\end{tabular}
	\caption{Simulation: Forecast performance of the SR-SV and SV models. In each panel, the bold numbers indicate the best predictive scores and the count indicates the number of times a model has better forecast scores than the other one. Monte Carlo standard errors in brackets, averaged over 10 different runs.}
	\label{tab:Sim_forecast_score_compare}
\end{table}

Table \ref{tab:Sim_forecast_score_compare} reports the predictive performance scores of the SV and SR-SV models with the Monte Carlo standard errors in brackets. 
For the SIM II and SIM III data, the SR-SV model outperforms the SV model for all the predictive scores, which is consistent with the in-sample analysis showing that the SR-SV model fits these simulation datasets better than the SV model. 
For the SIM I data, the SR-SV model also outperforms the SV model for all scores except the PPS score.  
These results illustrate the impressive out-of-sample forecast ability of the SR-SV model. The results for the real data applications in the next section further support this claim.

\subsection{Applications}\label{sec:applications}
This section evaluates the SR-SV model using five popular daily stock indexes from different international markets: The German stock index DAX30 (DAX), the Hong Kong stock index HS50 (HSI), the France market index CAC40 (FCHI), the US stock market index SP500 (SPX) and the Canada market index TSX250 (TSX).  
\subsubsection{The datasets and exploratory data analysis}\label{sec: datasets and EDA}
The datasets were downloaded from the Realized Library of The Oxford-Man Institute\footnote{https://realized.oxford-man.ox.ac.uk/}. 
We used the adjusted closing prices $\{P_t, \ t=1,...,T_P\}$ and calculated the demeaned return process as 
\bea
\label{eqn:log-return}
y_{t}=100\left(\log\frac{P_{t+1}}{P_{t}}-\frac1{T_P-1}\sum_{i=1}^{T_P-1}\log\frac{P_{i+1}}{P_{i}}\right),\;\;\;t=1,2,...,T_P-1,
\eea
and using the first $T_{\text{in}}=2000$ returns for in-sample analysis and the rest $T_{\text{out}}=1000$ for out-of-sample analysis. Table \ref{tab:data summaries} describes the relevant aspects of the datasets.

\begin{table}[h]
	\begin{center}
		\begin{tabular}{ccccc}
			\hline\hline
			\rule{0pt}{3ex}
			&\bol{In-sample Period}         &\bol{Out-of-sample Period}           &$T_{\text{in}}$   & $T_{\text{out}}$\\
			\hline
			\rule{0pt}{3ex}
			DAX    &23 Apr 2004 -- 21 Feb 2012    &22 Feb 2012 -- 05 Feb 2016  &2000    &1000 \\
			HSI    &27 Oct 2003 -- 28 Nov 2011    &29 Nov 2011 -- 21 Dec 2015  &2000    &1000 \\ 
			FCHI   &09 Jun 2004 -- 22 Mar 2012    &23 Mar 2012 -- 23 Feb 2016  &2000    &1000 \\ 
			SPX    &27 Feb 2004 -- 06 Feb 2012    &07 Feb 2012 -- 28 Jan 2016  &2000    &1000 \\
			TSX    &03 Feb 2004 -- 01 Feb 2012    &02 Feb 2012 -- 27 Jan 2016  &2000    &1000 \\ 
			\hline\hline
		\end{tabular}
	\end{center}
	\caption{Descriptions of the five index datasets. }
	\label{tab:data summaries}
\end{table}

\begin{table}[H]
	\begin{center}
		\begin{tabular}{cccccclll}
			\hline\hline
			\rule{0pt}{3ex}
			&Min         &Max      &Std         & Skew    & Kurtosis &$V_n(10)$& $V_n(20)$&$V_n(30)$\\
			\hline
			\rule{0pt}{3ex}
			\multirow{2}{*}{}DAX    &$-7.437$    &9.993     &1.267       &0.115     & 10.960 &$3.226^\ast$ &$2.501^\ast$ &$2.146^\ast$\\
			                                                                          &&&&&  &$2.456^\ast$ &$1.926^\ast$ &1.670 \\
			\rule{0pt}{3ex}
			\multirow{2}{*}{}HSI    &$-11.616$   &12.155    &1.186       &0.307     & 17.551 &$3.934^*$&$3.030^*$&$2.587^*$\\
			                                                                           &&&&& &$2.564^*$&$2.088^*$&$1.844^*$\\
			\rule{0pt}{3ex}
			\multirow{2}{*}{}FCHI   &$-7.215$    &6.663     &1.132       &$-0.320$  & 7.383  &$3.782^*$&$2.976^*$&$2.575^*$\\
		                                                                               &&&&& &$3.018^*$&$2.453^*$&$2.165^*$\\
			\rule{0pt}{3ex}
			\multirow{2}{*}{}SPX    &$-9.351$   &10.220   &1.307       &$-0.256$     & 12.502&$3.188^*$&$2.412^*$&$2.047^*$\\
																					   &&&&&  &$2.664^*$&$2.040^*$&$1.748^*$\\
			\rule{0pt}{3ex}
			\multirow{2}{*}{}TSX    &$-9.879$   &9.194     &1.262       &$-0.727$     &12.202 &$3.558^*$&$2.692^*$&$2.277^*$\\
			                                                                   &&&&&          &$2.877^*$&$2.199^*$&$1.875^*$\\
			\hline\hline
		\end{tabular}
	\end{center}
	\caption{Descriptive statistics for the demeaned returns of the DAX, HSI, FCHI, SPX and TSX datasets. $V_n(q), \ q=10, \ 20 \ \text{and} \ 30$, shows the test statistics of Lo's modified R/S test of long memory with lag $q$. Upper and lower values of the 3 last columns are the Lo's test statistics for absolute and squared returns, respectively. The asterisks indicate significance at the 5\% level.}
	\label{tab:data statistics}
\end{table}

Figure \ref{f:plot_3_data} in Appendix \ref{sec:app_4} plots the time series data and shows the existence of the volatility clustering effect commonly seen in financial data.
Table \ref{tab:data statistics} reports some descriptive statistics together with Lo's modified R/S test \citep{Lo:1991} for long-range memory in the absolute and squared returns. Lo's modified R/S test is widely used in the financial time series literature; see, e.g., \cite{Lo:1991}, \cite{Giraitis:2003}, \cite{Breidt:1998}. 
All the index data exhibit some negative skewness, high excess kurtosis and high variation.
The result of Lo's modified R/S test for long-memory dependence with several different lags $q$ indicates that there is significant evidence of long-memory dependence in the stock indices. 

The Realized Library provides different realized measures\footnote{See https://realized.oxford-man.ox.ac.uk/documentation/estimators for the list of the available realized measures} that can be used in financial econometrics as a proxy to the latent $\sigma_t^2$. We use the following four common realized measures including Realized Variance (RV) \citep{Andersen:1998}, Bipower Variation (BV) \citep{Nielsen:2004}, Median Realized Volatility (MedRV) \citep{Andersen:2012}, Realized Kernel Variance \citep{Barndorff&Nielsen:2008} with the Non-Flat Parzen kernal ($\text{RKV}$) to evaluate the forecast performance of the volatility models using the predictive scores in Table \ref{tab:forecast_score}.  See \cite{Shephard:2010} for more details about the Realized Library.

Denote by $RV_t$ the realized measure of $\sigma_t^2$ at time $t$. As the realized measures ignore the variation of the prices overnight and sometimes the variation in the first few minutes of the trading day when recorded prices may contain large errors \citep{Shephard:2010}, we follow \cite{Hansen:2005} to scale the realized measure $RV_t$ as 
\beq
\label{eqn:realized_measure_scale}
\widetilde\sigma^2_t = \widehat{c} \cdot RV_t \;\; \text{where} \;\;\ \widehat{c} = \dfrac {T_\text{out}^{-1} \sum_{t=T_\text{in}+1}^{T}\left(y_t-\E(y_t)\right)^2}{T_\text{out}^{-1}\sum_{t=T_{in}+1}^{T} RV_t},\;\;\;t=T_\text{in}+1,2,...,T,
\eeq
and use $\widetilde\sigma^2_t$ as the estimate of the latent conditional variance $\sigma_t^2$. See Table \ref{tab:data summaries} for $T_\text{in}$ and $T_\text{out}$ used in our datasets. See \cite{Martens:2002} and \cite{Fleming:2003} for the similar scaling estimator of the daily volatility. 

\subsubsection{In-sample analysis}
\begin{figure}[H]
	\centering
	\includegraphics[width=0.9\columnwidth]{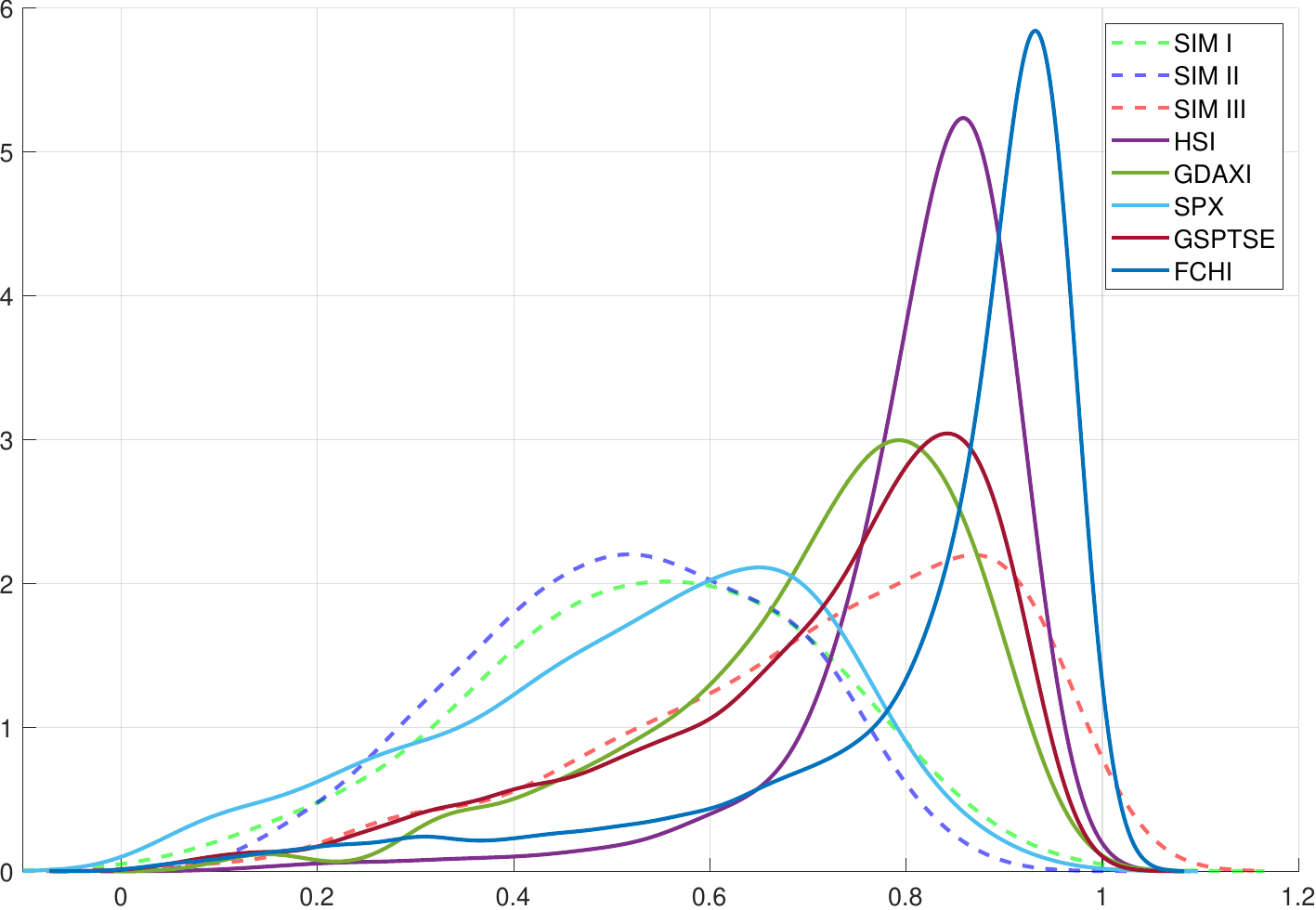}
	\caption{Posterior densities of the moving average weight $\alpha$ on simulation and real datasets. (This is better viewed in colour).}
	\label{f:alpha_density}
\end{figure}

Table \ref{tab:Real_params} summarizes the estimation results of fitting the SV, N-SV and SR-SV models to the five datasets. Table \ref{tab:LMSV_Real_params} in Appendix \ref{sec:app_4} shows the estimation results for the LMSV model.
For the SR-SV model, we only show the results of the key parameters. 
Figure \ref{f:alpha_density} shows the posterior densities of the moving average weight parameter $\alpha$ of the SR-SV model in all simulation and real datasets.  
We draw some conclusions from Table \ref{tab:Real_params} and the listed figures. 

\begin{table}[H]
	\centering
	\footnotesize
	\begin{tabular}{cccccccccl}
		\hline\hline
		\rule{0pt}{3ex}
		& $\mu$ & $\phi$ & $\sigma^2$ &$\delta$ & $\alpha$  & $\beta_0$& $\beta_1$ &$w_{z}$& Mar.llh \\
		\hline
		\rule{0pt}{3ex}
		\textbf{DAX} &&&&&&&&& \\
		\rule{0pt}{3ex}
		\multirow{1}{*}{SV}   &$-\bol{0.098}$ & \bol{0.979}  & \bol{0.038}  &  & & & & &$-2871.3$\\
		&(0.233) &(0.006)&(0.008)&&&&&& (0.171)\\
		\rule{0pt}{3ex}
		\multirow{1}{*}{N-SV} &$-\bol{0.138}$&\bol{0.977}&\bol{0.037}&$-\bol{0.198}$&&&&&$-2872.4$\\
		&(0.212)&(0.006)&(0.008)&(0.086)&&&&&(0.224) \\
		\rule{0pt}{3ex}
		\multirow{1}{*}{SR-SV} & &\bol{0.863} &\bol{0.064} &&\bol{0.605} &$-\bol{0.117}$&\bol{0.410}&\bol{0.397} &$-2868.8^\ast$\\
		& &(0.052)     &(0.021)     &&(0.204)     &(0.061)       &(0.201)    &(0.153)     &(0.301)\\
		\hline
		\rule{0pt}{3ex}
		\textbf{HSI} &&&&&&&&& \\
		\rule{0pt}{3ex}
		\multirow{1}{*}{SV}   &$-\bol{0.205}$ & \bol{0.987}  & \bol{0.022}  &  & & & & &$-2692.0$\\
		&(0.320) &(0.004)&(0.008)&&&&&& (0.184)\\
		\rule{0pt}{3ex}
		\multirow{1}{*}{N-SV} &$-\bol{0.366}$&\bol{0.987}&\bol{0.021}&$-\bol{0.242}$&&&&&$-2691.0$\\
		&(0.270)&(0.004)&(0.004)&(0.081)&&&&&(0.214) \\
		\rule{0pt}{3ex}
		\multirow{1}{*}{SR-SV} & &\bol{0.824} &\bol{0.054} &&\bol{0.784} &$-\bol{0.196}$&\bol{0.536}&\bol{0.387} &$-2687.8^{**}$\\
		& &(0.061)     &(0.021)     &&(0.137)     &(0.083)       &(0.262)    &(0.139)     &(0.337)\\
		\hline
		\rule{0pt}{3ex}
		\textbf{FCHI} &&&&&&&&& \\
		\rule{0pt}{3ex}
		\multirow{1}{*}{SV}   &$-\bol{0.213}$ & \bol{0.977}  & \bol{0.047}  &  & & & & &$-2787.1$\\
		 &(0.230) &(0.007)&(0.010)&&&&&& (0.225)\\
		\rule{0pt}{3ex}
		\multirow{1}{*}{N-SV} &$-\bol{0.217}$&\bol{0.979}&\bol{0.041}&$-\bol{0.198}$&&&&&$-2787.3$\\
		                      &(0.257)&(0.006)&(0.009)&(0.089)&&&&&(0.234) \\
		\rule{0pt}{3ex}
		\multirow{1}{*}{SR-SV} & &\bol{0.843} &\bol{0.093} &&\bol{0.780} &$-\bol{0.179}$&\bol{0.449}&\bol{0.363} &$-2784.2^{*}$\\
		                        & &(0.049)     &(0.027)     &&(0.197)     &(0.070)       &(0.199)    &(0.134)     &(0.326)\\
		\hline
		\rule{0pt}{3ex}
		\textbf{SPX} &&&&&&&&& \\
		\rule{0pt}{3ex}
		\multirow{1}{*}{SV}   &$-\bol{0.228}$ & \bol{0.985}  & \bol{0.034}  &  & & & & &$-2748.3$\\
	                          &(0.344) &(0.005)&(0.006)&&&&&& (0.201)\\
		\rule{0pt}{3ex}
		\multirow{1}{*}{N-SV} &$-\bol{0.267}$&\bol{0.9837}&\bol{0.036}&$-\bol{0.121}$&&&&&$-2749.4$\\
		                      &(0.268)        &(0.004)     &(0.007)    &(0.080)&&&&&(0.211) \\
		\rule{0pt}{3ex}
		\multirow{1}{*}{SR-SV} & &\bol{0.844} &\bol{0.056} &&\bol{0.527} &$-\bol{0.180}$&\bol{0.481}&\bol{0.373} &$-2745.6^\ast$\\
		                        & &(0.060)     &(0.017)     &&(0.186)     &(0.186)       &(0.241)    &(0.132)     &(0.311)\\		            
		\hline
		\rule{0pt}{3ex}
		\textbf{TSX} &&&&&&&&& \\
		\rule{0pt}{3ex}
		\multirow{1}{*}{SV}   &$-\bol{0.200}$ & \bol{0.985}  & \bol{0.028}  &  & & & & &$-2770.1$\\
		&(0.323) &(0.004)&(0.006)&&&&&& (0.231)\\
		\rule{0pt}{3ex}
		\multirow{1}{*}{N-SV} &$-\bol{0.249}$&\bol{0.984}&\bol{0.029}&-$\bol{0.141}$&&&&&$-2769.9$\\
		                      &(0.298)    &(0.005)    &(0.006)    &(0.077)    &&&&&(0.245)\\
		\rule{0pt}{3ex}
		\multirow{1}{*}{SR-SV} & &\bol{0.868} &\bol{0.051} &&\bol{0.697} &$-\bol{0.129}$&\bol{0.414}&\bol{0.355} &$-2767.2^\ast$\\
		                        & &(0.056)     &(0.015)     &&(0.195)     &(0.071)       &(0.201)    &(0.141)     &(0.347)\\      	      
		\hline\hline
	\end{tabular}
	\caption{Applications: Posterior means of the parameters with the posterior standard deviations in brackets. The last column shows the estimated log marginal likelihood with the Monte Carlo standard errors in brackets, averaged over 10 different runs of the DT-SMC sampler. The single and double asterisks indicate the cases when the Bayes factors strongly and very strongly support the SR-SV model over the SV model, respectively. The marginal likelihood are reported in natural log scale. }
	\label{tab:Real_params}
\end{table} 

First, the marginal likelihood estimates show that the SR-SV model fits the five index data better than the SV and N-SV models. The Bayes factors of the SR-SV model compared to the SV and N-SV models are more than $e^{2.3}$, which strongly support the SR-SV in all cases. 
There are no significant differences between the SV and N-SV models in terms of marginal likelihood estimates across the five panels. 

Second, the evidence of the volatility effects rather than linearity, e.g. non-linearity and long-memory auto-dependence, in the volatility dynamics of the index datasets is clear as the posterior means of the non-linearity long-memory parameter $\beta_1$ of the SR-SV model are more than two standard deviations from zero in all cases. 
The estimation results of the LMSV in Table \ref{tab:LMSV_Real_params} show that the posterior means of the fractional integration parameter $d$ are also more than two standard deviations from zero and close to $0.5$ in all cases, suggesting a strong evidence of the long-memory dependence in the volatility process of these five index datasets.    
The posterior means of the non-linear parameter $w_{z}$ with respect to the past log volatility $z_{t-1}$ are more than two standard deviations from zero, indicating the existence of the serial dependence rather than linearity that the past log volatility $z_{t-1}$ has on the current log volatility $z_t$, and that the SR-SV model is able to detect this serial dependence. 
The posterior density plots of parameter $\alpha$ in Figure \ref{f:alpha_density} suggest the existence of the long-memory auto-dependence in the volatility processes of the index datasets as the posterior densities of $\alpha$ are highly skewed for all cases and the posterior modes are close to 1, which is similar to the results in the SIM III data. 

Third, it is worth noting that, in all cases, the persistence parameter $\phi$ in the SR-SV model is smaller than the persistence parameters in the SV and N-SV models as the parameter $w_{z}$ is significant and hence the linear effect that $z_{t-1}$ has on $z_t$ is reduced.  
As illustrated in the SIM I data, if the volatility process exhibits no volatility effects rather than a short-memory linear auto-dependence, the persistence parameters $\phi$ in the SV and SR-SV model are similar and the parameter $w_{z}$ is insignificant. 

Using the posterior mean estimates in Table \ref{tab:Real_params}, the {\it filtered} values of $z_t$ of the SR-SV, SV and N-SV models can be computed using the particle filter. 
We discuss how to obtain the filtered values of $z_t$ of the LMSV model in Appendix \ref{sec:app_1}.
Figure \ref{f:eta_h_SPX} plots the filtered log volatility of the SV and SR-SV model, together with the filtered values of the components $h_t$ and $\eta_t$ of the SR-SV model in all time steps, for the SPX data. Figure \ref{f:eta_h_DAX}, \ref{f:eta_h_HSI}, \ref{f:eta_h_FCHI} and \ref{f:eta_h_TSX} in Appendix \ref{sec:app_4} show the similar plots for the HSI, DAX, FCHI and TSX data, respectively. Figure \ref{f:eta_h_SPX} shows that the component $h_t$, and hence $\eta_t$, of the SR-SV model is well responsive to changes in the volatility dynamics, e.g. being small during the low volatility periods and large in the high volatility periods of the SPX data. The SRU structure of the SR-SV model is able to capture these distinct behaviors of financial time series. We observe the similar behaviors of the $h_t$ and $\eta_t$ components for the other datasets as shown in the Figures \ref{f:eta_h_HSI}-\ref{f:eta_h_TSX}.

\begin{figure}[H]
	\centering
	\includegraphics[width=1\columnwidth]{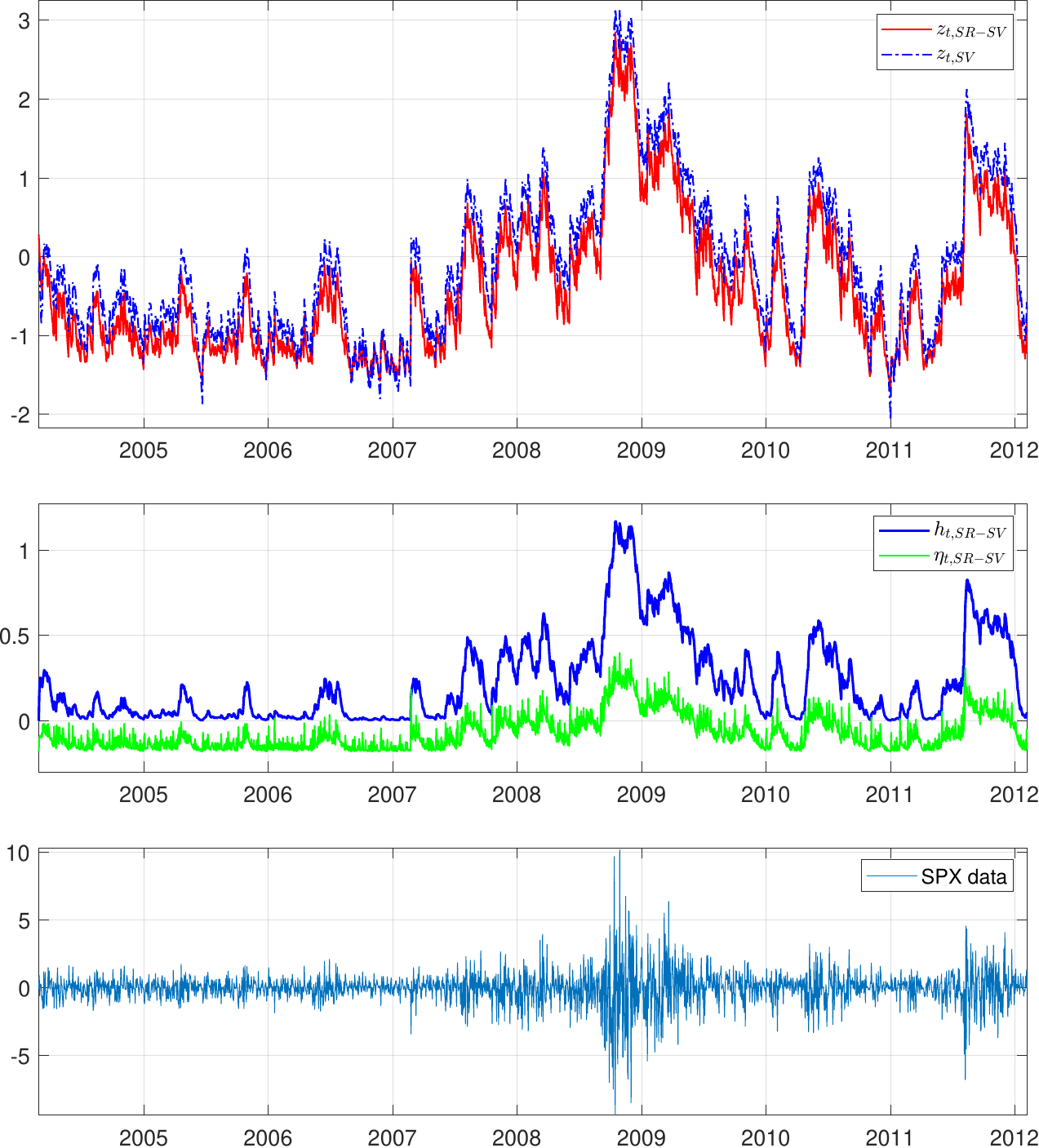}
	\caption{SPX: (\textit{Top}) The filtered log volatility of the SR-SV and SV models. (\textit{Middle}) The filtered values of $\eta_t$ and $h_t$ of the SR-SV model. (\textit{Bottom}) The SPX in-sample data. (This is better viewed in colour).}
	\label{f:eta_h_SPX}
\end{figure}

Table \ref{tab:Real Model diagnostics} provides summary statistics on the in-sample filtered volatilities and residuals $\wh\eps_t^y$ of the SR-SV, SV, N-SV and LMSV models. We note that if the assumption about the normality of returns of the three models is justified then the residuals $\wh\eps_t^y$ should have a zero skewness and a kurtosis of 3. Table \ref{tab:Real Model diagnostics} shows that all the residual distributions produced from the four models are close to the standard normal distribution, but still slightly skewed and leptokurtic. 
The $p$-values of the Ljung-Box (LB) autocorrelation test of the residuals are high in all index datasets, suggesting that there is no evidence of autocorrelation in the residuals. We conjecture that extending the SR-SV model, for example, by using a Student's $t$ distribution instead of a Gaussian for the measurement shock $\eps^y_t$ and taking into account the leverage effect by correlating $\eps^y_t$ with the volatility shock $\eps_t^\eta$, is likely to lead to better diagnostics for the residuals. However, we do not consider these extensions here.

\begin{table}[h!]
	\begin{center}
		\begin{tabular}{rccccccccc}
			\hline\hline
			\rule{0pt}{3ex}
			&\multicolumn{4}{l}{Filtered volatility } &&\multicolumn{4}{l}{Residual $\wh\eps_t^y$}\\
			\cline{2-5}\cline{7-10}
			\rule{0pt}{3ex}
			&Mean &Std &Kurtosis &Skew & & Std      & Kurtosis           &Skew      &LB-$\wh\eps_t$  \\
			\hline
			\rule{0pt}{3ex}
			\textbf{DAX}      &&&&&&&&&\\
			\rule{0pt}{3ex}
			SV         &1.531&1.930&24.465&4.075   & &0.985&2.817&$-0.215$&0.978\\
			\rule{0pt}{3ex}
			N-SV       &1.591&2.290&38.239&5.112   & &0.982&2.742&$-0.213$&0.978 \\
			\rule{0pt}{3ex}
			LMSV       &1.423&1.670&13.734&2.885   & &0.999&3.584&$-0.163$&0.887 \\
			\rule{0pt}{3ex}
			SR-SV     &1.325&1.559&26.276&4.182   & &0.991&2.819&$-0.207$&0.983 \\
			\hline
			\rule{0pt}{3ex}
			\textbf{HSI}      &&&&&&&&&\\
			\rule{0pt}{3ex}
			SV         &1.343&2.084&55.696&6.187   & &0.966&2.801&$-0.040$&0.162\\
			\rule{0pt}{3ex}
			N-SV       &1.428&3.133&179.820&11.399 & &0.965&2.776&$-0.040$&0.226 \\
			\rule{0pt}{3ex}
			LMSV       &1.271&1.632&22.216&3.720   & &0.999&3.946&$-0.028$&0.326 \\
			\rule{0pt}{3ex}
			SR-SV     &1.101&1.651&50.909&5.821   & &0.978&2.768&$-0.052$&0.132 \\
			\hline
			\rule{0pt}{3ex}
			\textbf{FCHI}      &&&&&&&&&\\
			\rule{0pt}{3ex}
			SV         &1.416&1.563&13.561&2.851 & &0.982&2.724&$-0.123$&0.101 \\
			\rule{0pt}{3ex}
			N-SV       &1.483&1.867&22.140&3.705 & &0.981&2.690&$-0.117$&0.103 \\
			\rule{0pt}{3ex}
			LMSV       &1.287&1.451&11.815&2.645 & &1.000&3.548&$-0.058$&0.108 \\
			\rule{0pt}{3ex}
			SR-SV     &1.167&1.159&13.285&2.756 & &0.985&2.722&$-0.136$&0.105 \\
			\hline
			\rule{0pt}{3ex}
			\textbf{SPX}      &&&&&&&&&\\
			\rule{0pt}{3ex}
			SV        &1.648&2.712&25.696&4.395 & &0.994&2.769&$-0.254$&0.136 \\
			\rule{0pt}{3ex}
			N-SV      &1.729&3.139&33.443&5.010 & &0.995&2.737&$-0.251$&0.143 \\
			\rule{0pt}{3ex}
			LMSV      &1.592&2.810&25.516&4.381 & &0.999&3.623&$-0.103$&0.219 \\
			\rule{0pt}{3ex}
			SR-SV    &1.892&3.862&29.532&4.856 & &1.002&2.746&$-0.252$&0.149 \\
			\hline
			\rule{0pt}{3ex}
			\textbf{TSX}      &&&&&&&&&\\
			\rule{0pt}{3ex}
			SV       &1.543&2.516&27.582&4.641 & &0.971&2.705&$-0.311$&0.977 \\
			\rule{0pt}{3ex}
			N-SV     &1.590&2.846&35.343&5.256 & &0.971&2.683&$-0.307$&0.969 \\
			\rule{0pt}{3ex}
			LMSV     &1.395&2.124&20.325&3.893 & &0.999&3.353&$-0.301$&0.915 \\
			\rule{0pt}{3ex}
			SR-SV   &1.389&2.347&28.167&4.702 & &0.973&2.659&$-0.310$&0.961 \\
			\hline\hline
		\end{tabular}
	\end{center}
	\caption{Applications: Model diagnostics of the filtered log volatility and residual $\wh\eps^y_t$. The LB p-values denote the p-value from the Ljung-Box test with 10 lags.}
	\label{tab:Real Model diagnostics}
\end{table} 

\subsubsection*{Out-of-sample analysis}
The marginal likelihood estimates in Table \ref{tab:Real_params} suggest that the SR-SV model fits the in-sample data of the five index datasets better than the SV and N-SV models. 
We now examine if this in-sample performance is consistent with the out-of-sample performance.
Table \ref{tab:Real_Model_forecast_diagnostics} provides summary statistics on the one-step-ahead forecasts of volatility and standardized residuals of the SR-SV, SV, N-SV and LMSV models. We note two conclusions from Table \ref{tab:Real_Model_forecast_diagnostics}.

First, the SR-SV model does not suffer from overfiting as often observed in neural network based volatility models \citep{Pagan:1990, Donaldson:1997},
as the one-step-ahead forecast volatilities and the forecast residuals appear to be well behaved, compared to those from the more parsimonious SV and N-SV models.
We emphasize that, as discussed in Section \ref{sec:lstm-sv}, the use of noise-injecting regularization in the novel structure of the SR-SV model helps prevent it from
the well-known overfitting problem.

\begin{figure}[h]
	\centering
	\includegraphics[width=1\columnwidth]{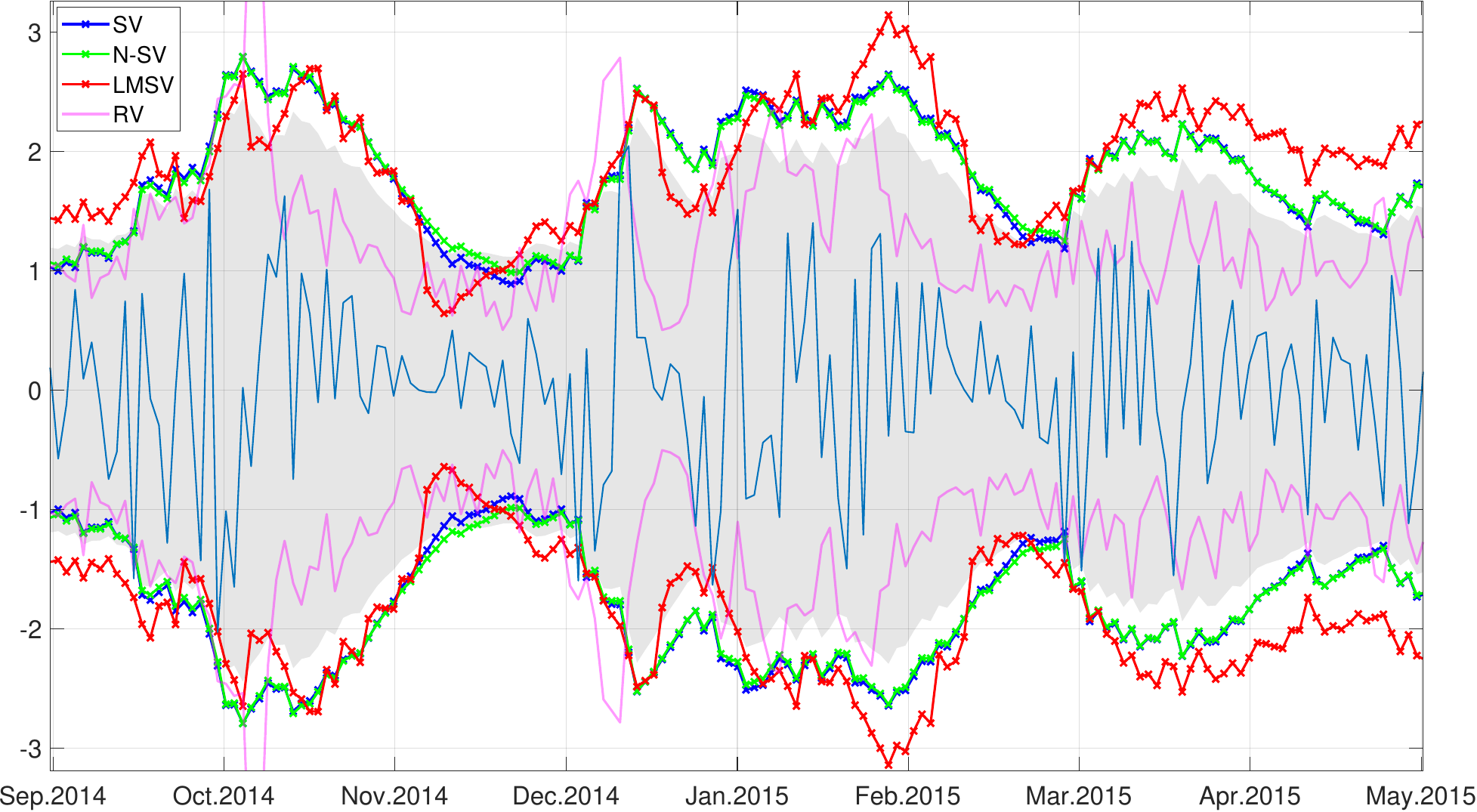}
	\caption{SPX: 99\% One-step-ahead forecast intervals of the SR-SV, SV, N-SV and LMSV models, together with the 99\% interval of index data estimated by the realized variance (RV) during Sep 2014 - May 2015. The shaded area is the 99\% one-step-ahead forecast interval produced by the SR-SV model.}
	\label{f:SP500_forcast_interval}
\end{figure}

Second, the \text{means} and standard deviations of one-step-ahead forecast volatility of the SR-SV model are smaller than those of the SV, N-SV and LMSV in all five index datasets.
The SR-SV forecasts are generally more conservative in low volatility periods, in the sense that the forecast intervals often have a smaller band compared to the forecasts produced by the other models. The comparison of 99\% one-step-ahead forecast intervals during the period Sep 2014 - May 2015 of the SPX data in Figure \ref{f:SP500_forcast_interval} shows that the SR-SV model gives a safe buffer against abrupt changes in low volatility regions, e.g. Nov-Dec 2014, because it maintains a wider forecast band, while it does not produce overly large forecast intervals in high volatility regions, e.g. Oct 2014, Dec 2014 - Jan 2015. 
Therefore, the SR-SV model is less sensitive to the data values in the shorter time periods, and maintains a good trade-off between the information in recent observations and the information in the long-term memory.    
The SV, N-SV and LMSV models, compared to the SR-SV model, produce a smaller forecast volatility in low volatility regions and a higher volatility forecast in high volatility regions. 
The figure also shows that the SV and N-SV forecasts depend mainly on the return at the previous step, as the persistence parameters $\phi$ in the SV and N-SV models are larger than the persistence parameter of the SR-SV model. The SR-SV intervals are closer to the intervals made by the realized variance and hence seem to track the out-of-sample returns better than the SV, N-SV and LMSV models.   

\begin{table}[h!]
	\begin{center}
		\begin{tabular}{rccccccccc}
			\hline\hline
			\rule{0pt}{3ex}
			&\multicolumn{4}{l}{Forecast Volatility } &&\multicolumn{4}{l}{Forecast Residual $\wh\eps_t^y$}\\
			\cline{2-5}\cline{7-10}
			\rule{0pt}{3ex}
			&Mean &Std &Kurtosis &Skew & & Std      & Kurtosis           &Skew      &LB-$\wh\eps_t$  \\
			\hline
			\rule{0pt}{3ex}
			\textbf{DAX}      &&&&&&&&&\\
			\rule{0pt}{3ex}
			SV         &1.069&0.580&2.992&0.711   & &0.997&3.964&$-0.334$&0.652\\
			\rule{0pt}{3ex}
			N-SV       &1.072&0.588&3.464&0.893   & &0.992&3.937&$-0.342$&0.659 \\
			\rule{0pt}{3ex}
			LMSV       &1.083&0.616&3.467&0.879   & &1.001&3.942&$-0.310$&0.577 \\
			\rule{0pt}{3ex}
			SR-SV     &0.943&0.458&3.261&0.879   & &1.034&3.835&$-0.328$&0.597 \\
			\hline
			\rule{0pt}{3ex}
			\textbf{HSI}      &&&&&&&&&\\
			\rule{0pt}{3ex}
			SV         &0.655&0.380&14.099&2.823   & &0.982&4.353&0.036&0.390\\
			\rule{0pt}{3ex}
			N-SV       &0.649&0.408&23.440&3.827   & &0.981&4.298&0.021&0.379 \\
			\rule{0pt}{3ex}
			LMSV       &0.740&0.386&9.633 &2.065   & &0.928&4.465&0.057&0.283 \\
			\rule{0pt}{3ex}
			SR-SV     &0.491&0.215&15.963&3.102   & &1.091&4.135&$-0.008$&0.368 \\
			\hline
			\rule{0pt}{3ex}
			\textbf{FCHI}      &&&&&&&&&\\
			\rule{0pt}{3ex}
			SV         &1.087&0.620&3.796&0.989 & &0.963&4.590&$-0.436$&0.734 \\
			\rule{0pt}{3ex}
			N-SV       &1.089&0.646&4.629&1.249 & &0.965&4.415&$-0.432$&0.686 \\
			\rule{0pt}{3ex}
			LMSV       &1.144&0.633&5.126&1.400 & &0.971&4.656&$-0.371$&0.657 \\
			\rule{0pt}{3ex}
			SR-SV     &0.894&0.434&3.626&0.918 & &0.995&4.174&$-0.375$&0.609 \\
			\hline
			\rule{0pt}{3ex}
			\textbf{SPX}      &&&&&&&&&\\
			\rule{0pt}{3ex}
			SV        &0.675&0.438&9.938&2.135  & &0.983&3.970&$-0.456$&0.352 \\
			\rule{0pt}{3ex}
			N-SV      &0.679&0.453&12.583&2.503 & &0.978&3.970&$-0.458$&0.365 \\
			\rule{0pt}{3ex}
			LMSV       &0.763&0.422&4.993&1.291 & &0.920&4.075&$-0.416$&0.547 \\
			\rule{0pt}{3ex}
			SR-SV    &0.523&0.287&9.978&2.242  & &1.074&3.797&$-0.406$&0.450 \\
			\hline
			\rule{0pt}{3ex}
			\textbf{TSX}      &&&&&&&&&\\
			\rule{0pt}{3ex}
			SV       &0.551&0.314&3.231&0.976 & &0.960&3.859&$-0.537$&0.123 \\
			\rule{0pt}{3ex}
			N-SV     &0.551&0.304&3.412&1.035 & &0.952&3.811&$-0.533$&0.105 \\
			\rule{0pt}{3ex}
			LMSV     &0.541&0.272&3.156&0.798 & &0.983&4.060&$-0.583$&0.098 \\
			\rule{0pt}{3ex}
			SR-SV   &0.500&0.229&3.948&1.234 & &0.973&3.734&$-0.502$&0.077 \\
			\hline\hline
		\end{tabular}
	\end{center}
	\caption{Applications: Summary statistics on the one-step-ahead out-of-sample forecast conditional variances $\widehat\sigma^2_t$ and residual $\wh\eps_t$. The LB p-values denote the p-value from the Ljung-Box test with 10 lags.}
	\label{tab:Real_Model_forecast_diagnostics}
\end{table} 

\begin{table}[h!]
	\centering
	\footnotesize
	\begin{tabular}{crcccccccc}
		\hline\hline
		\rule{0pt}{3ex}
		Measure&&PPS& $\text{MSE}_1$ &$\text{MSE}_2$ & $\text{MAE}_1$&$\text{MAE}_2$&$\text{QLIKE}$&$\text{R}^2\text{LOG}$&Count\\
		\hline
		\rule{0pt}{3ex}
		&SV             &1.122      &0.103      &1.333      &0.229      &0.392      &0.347      &0.667      &0\\
		&               &(0.001)    &(0.001)    &(0.002)    &(0.001)    &(0.002)    &(0.001)    &(0.002)&     \\
		\rule{0pt}{3ex}
		BV&N-SV         &1.122      &0.103      &1.327      &0.229      &0.392      &0.346      &0.669      &0\\
		&               &(0.000)    &(0.001)    &(0.002)    &(0.001)    &(0.002)    &(0.001)    &(0.003)    & \\
		\rule{0pt}{3ex}
		&LMSV           &           &0.125      &1.406      &0.265      &0.451      &0.402      &0.837  &0 \\
		&               &&&&&&&& \\
		\rule{0pt}{3ex}
		&SR-SV         &\bol{1.113}&\bol{0.091}&\bol{1.321}&\bol{0.209}&\bol{0.354}&\bol{0.330}&\bol{0.572}&7\\
		&               &(0.001)    &(0.000)    &(0.002)    &(0.000)    &(0.001)    &(0.002)    &(0.000)    & \\
		\hline
		\rule{0pt}{3ex}
		&SV             &     &0.114      &0.830      &0.256      &0.428      &0.323      &0.862      &0\\
		&               &     &(0.000)    &(0.002)    &(0.001)    &(0.001)    &(0.001)    &(0.003)    & \\
		\rule{0pt}{3ex}
		MedRV&N-SV      &     &0.114      &0.826      &0.257      &0.428      &0.323      &0.866      &0\\
		&               &     &(0.000)    &(0.002)    &(0.001)    &(0.000)    &(0.001)    &(0.004)    & \\
		\rule{0pt}{3ex}
		&LMSV           &     &0.140      &0.901      &0.294      &0.488      &0.385      &1.064  &0 \\
		&               &&&&&&&& \\
		\rule{0pt}{3ex}
		&SR-SV         &     &\bol{0.102}&\bol{0.821}&\bol{0.235}&\bol{0.389}&\bol{0.308}&\bol{0.757}&6\\
		&              &     &(0.000)    &(0.002)    &(0.000)    &(0.000)    &(0.001)    &(0.001)    & \\
		\hline
		\rule{0pt}{3ex}
		&SV             &     &0.114      &0.834      &0.256      &0.419      &0.363      &0.915      &0\\
		&               &     &(0.000)    &(0.001)    &(0.000)    &(0.001)    &(0.000)    &(0.002)    & \\
		\rule{0pt}{3ex}
		RKV  &N-SV      &     &0.114      &0.829      &0.256      &0.420      &0.361      &0.918      &0\\
		&               &     &(0.000)    &(0.002)    &(0.000)    &(0.000)    &(0.000)    &(0.001)    & \\
		\rule{0pt}{3ex}
		&LMSV           &     &0.137      &0.904      &0.290      &0.476      &0.405      &1.100  &0 \\
		&               &&&&&&&& \\
		\rule{0pt}{3ex}
		&SR-SV         &     &\bol{0.101}&\bol{0.822}&\bol{0.237}&\bol{0.384}&\bol{0.345}&\bol{0.808}&6\\
		&               &     &(0.000)    &(0.001)    &(0.000)    &(0.000)    &(0.000)    &(0.001)    & \\
		\hline
		\rule{0pt}{3ex}
		&SV             &     &0.121      &1.864      &0.245      &0.421      &0.331      &0.796      &0\\
		&               &     &(0.000)    &(0.002)    &(0.001)    &(0.002)    &(0.000)    &(0.002)    & \\
		\rule{0pt}{3ex}
		RV  &N-SV       &     &0.120      &1.863      &0.246      &0.422      &0.329      &0.799&0\\
		&               &     &(0.000)    &(0.002)    &(0.001)    &(0.001)    &(0.001)    &(0.003)    & \\
		\rule{0pt}{3ex}
		&LMSV           &     &0.145      &1.95       &0.283      &0.484      &0.385      &0.983  &0 \\
		&               &     &&&&&&& \\
		\rule{0pt}{3ex}
		&SR-SV         &     &\bol{0.108}&\bol{1.861}&\bol{0.224}&\bol{0.382}&\bol{0.316}&\bol{0.692}&6\\
		&               &     &(0.000)    &(0.001)    &(0.000)    &(0.000)    &(0.001)    &(0.000)    & \\
		\hline\hline
		\end{tabular}
		\caption{SPX data: Forecast performance of the SR-SV and benchmark models using different realized measures. In each panel, the bold numbers indicate the best predictive scores.}
	\label{tab:SPX_forecast_score_compare}
\end{table} 

Table \ref{tab:SPX_forecast_score_compare} shows the out-of-sample performances of the four models on the SPX data, with the Monte Carlo standard errors in brackets. 
Tables \ref{tab:DAX_forecast_score_compare}, \ref{tab:HSI_forecast_score_compare}, \ref{tab:FCHI_forecast_score_compare} and \ref{tab:TSX_forecast_score_compare} 
in Appendix \ref{sec:app_4} show the results for the other four datasets.
Each table provides predictive scores separately in four panels, corresponding to the four realized measures BV, MedRV, RKV and RV, as discussed Section \ref{sec: datasets and EDA}. In each panel, we count the number of times a particular model has lowest (best) predictive scores and list these numbers in the last column; the model with the highest count is preferred. We note that the PPS predictive score is independent of the realized measures of volatility, and that the PPS predictive score is not available for the LMSV model.  

As shown in these tables, the SR-SV model consistently has the best out-of-sample performance in all the five index datasets. 
The superior predictive performance of the SR-SV model is consistent with its in-sample performance discussed earlier and provides further evidence to support the conclusion that the SR-SV model does not overfit the five index datasets. We note that the forecast performances of the SV and N-SV models are mixed with no model consistently outperforming the other across the predictive scores and the datasets. The LMSV model consistently makes the least accurate forecasts in all cases .

\section{Conclusions}
\label{Sec:conclusion}
This paper proposes a statistical recurrent stochastic volatility (SR-SV) model, by combining the statistical recurrent unit architecture from Machine Learning and the stochastic volatility model from Financial Econometrics.
These two techniques are combined in a principled and non trivial way to form a new approach 
that is, as carefully illustrated through the extensive simulation and empirical studies, highly efficient for volatility modelling and forecasting.
It is easy to carry Bayesian inference in the SR-SV model using standard Bayesian computation methods,
such as the Density Tempered  Sequential Monte Carlo method as used in this paper.
The simulation and empirical studies suggest that the SR-SV model is able to capture various volatility effects overlooked by the SV benchmark models, and is able to produce highly accurate forecast volatilities. 

Extending the SR-SV model by incorporating features such as the leverage effect is an interesting research question.
Another interesting research question is extending the present SR-SV model to multivariate financial time series.
We conjecture that the RNN architectures will be even more powerful in this case as they can naturally capture the interaction between the inputs.
This research is in progress. 

 
\appendix
\section{Appendix}
\subsection{Bayesian inference and forecast for the LMSV model}\label{sec:app_1}
Denote by  $x=\{x_t=\text{log}\; y_t^2,\ t=1,...,T\}$ the series of log squared returns, the LMSV model in \eqref{eq:LMSVmodel2}-\eqref{eq:LMSVmodel1} can be transformed to a stationary process with respect to $x_t$ as
\bea
(1-B)^d \Phi(B) z_t &=& \Theta (B)\eta_t,\;\;\eta_t\sim\N(0,\sigma^2_\eta),\;\;t=2,...,T,\label{LMSVmodel2_logsquared} \\
x_t &=& c + z_t + \xi_t,\;\; \xi_t \sim (0,\sigma^2_\xi)\;\; t=1,2,...,T,\label{LMSVmodel1_logsquared}
\eea
where $\xi_t = \text{log}\; \eps_t^2 - \E[\text{log}\;\eps_t^2]$ is i.i.d with mean zero and variance $\sigma_\xi^2$, $c = \text{log}(\kappa^2) + \E[\text{log}\;\eps_t^2]$. The process $x_t$ is the sum of the long-memory ARFIMA$(p,d,q)$ process $z_t$ and a non-Gaussian noise, with $\E[x_t] = c$ and the auto-covariance function (ACVF)
\beq
\label{eq:lmsv_acvf}
\gamma_x(h) = \text{Cov}(x_t,x_{t+h}) = \gamma(h) + \sigma_\xi^2 \mathbb{1}_{h=0},
\eeq  
where $h$ is the lag number, $\gamma(h)$ is the ACVF of the $z_t$ process and $\mathbb{1}_{h=0}$ is an indicator function which equals to 1 if $h=0$ and $0$ otherwise. 
\cite{Breidt:1998} estimate the LMSV model by maximizing the Whittle log-likelihood \citep{whittle:1953}, defined as  
\beq
\label{eq:whittle}
\ell_W(\beta_x) = 2\pi T^{-1} \sum^{[T/2]}_{k=1}\left\{\text{log}\;f_{\beta_x}(\omega_k)+\frac{J(\omega_k)}{f_{\beta_x}(\omega_k)}\right\},
\eeq
where $[\cdot]$ denotes the integer part, $\beta_x = (d,\phi_1,...,\phi_p, \theta_1,...,\theta_q, \sigma_\eta^2, \sigma_\xi^2,c)$ is the vector of model parameters, $\omega_k=2\pi k T^{-1}$ is the $k$th Fourier frequency, $J(\omega_k)$ is the $k$th normalized periodogram ordinate 
\beq
J(\omega_k) = \frac{1}{2 \pi T}\left( \sum^T_{t=1}x_t \text{cos}\; \omega_k t\right)^2 + \frac{1}{2 \pi T}\left( \sum^T_{t=1}x_t \text{sin}\; \omega_k t\right)^2,
\eeq
and $f_{\beta_x}(\omega_k)$ is the spectral density of the LMSV model in \eqref{LMSVmodel2_logsquared}-\eqref{LMSVmodel1_logsquared}
\beq
f_{\beta_x}(\omega_k) = \frac{\sigma^2_\eta \left|\Theta(e^{-i\omega_k})\right|^2}{2\pi \left|1-e^{-i\omega_k}\right|^{2d}\left|\Phi(e^{-i\omega_k})\right|^{2}} + \frac{\sigma^2_\xi}{2\pi}.
\eeq
The Whittle likelihood is an approximation of the time-domain likelihood and is exact if the data are i.i.d. Gaussian. 

Let $\pi_W(\beta_x) \propto L_W(\beta_x) \; p(\beta_x)$ be the posterior density based on the Whittle likelihood $L_W(\beta_x) = \text{exp}(\ell_W(\beta_x))$, given the log of squared return series $x=\{x_t,\ t=1,...,T\}$. 
To sample from $\pi_W(\beta_x)$, we use an adaptive random walk MCMC method summarized in Algorithm \ref{alg:MCMC},
with the covariance matrix in the random walk proposal adaptively scaled to target an overall acceptance probability of $25\%$ \cite{Garthwaite2010}.
We note that the vector of parameters $\beta_x$ in Algorithm \ref{alg:MCMC} does not include the constant $c$ in \eqref{LMSVmodel1_logsquared}, which is simply estimated by the sample mean, i.e., $c = \frac{1}{T}\sum_{t=1}^T x_t$ \citep{HARVEY:2007}. 

\begin{algorithm}
\caption{Markov Chain Monte Carlo with random walk proposal}  
\label{alg:MCMC}
Sample $\beta_x \sim p(\beta_x)$ 

For each MCMC iteration: 
\begin{enumerate}
	\item Sample $\beta_x^\prime$ from the proposal density $q(\beta^\prime_x|\beta_x)$.
	\item Compute the Whittle likelihood $L_W(\beta_x^\prime) = \text{exp}(\mathcal{L}_w(\beta_x^\prime))$ with $\mathcal{L}_w(\cdot)$ defined in \eqref{eq:whittle}.
	\item Accept the proposal $\beta_x^\prime$ with the probability
	\bean
	\text{min} \left\{ 1,\frac{L_W(\beta_x^\prime)}{L_W(\beta_x)}\frac{p(\beta_x^\prime)}{p(\beta_x)}\frac{q(\beta_x|\beta_x^{\prime})}{q(\beta_x^{\prime}|\beta_x)}\right\}.
	\eean
\end{enumerate}
\end{algorithm}
	
Given the samples of model parameters from the posterior density $\pi_W(\beta_x)$, \cite{HARVEY:2007} suggests a convenient way to obtain the estimated values of conditional variance $\sigma_t^2$ for the LMSV model as follows. Suppose that $z_t$ is a stationary process and denote by $\Sigma_z$ and $\Sigma_\xi$ the covariance matrices of $z_t$ and $\xi_t$, respectively, then the covariance matrix $\Sigma$ of the log squared returns series $x$ is $\Sigma = \Sigma_z + \Sigma_\xi$. The minimum mean square linear estimator of the log volatility $\tilde{z}=\{\tilde{z}_t,\ t=1,...,T\}$ is calculated as
\beq
\label{eq:z_t}
\tilde{z} = (\textbf{I}_T - \sigma_\xi^2 \Sigma^{-1})x' + \sigma_\xi^2 \Sigma^{-1} \boldsymbol{\iota},
\eeq   
where $\textbf{I}_T$ is the identity matrix of size $T$, $\boldsymbol{\iota}$ is a $T \times 1$ column vector of ones. 
As $\xi_t$ are i.i.d and serially uncorrelated, the covariance matrix $\Sigma_\xi$ is $\Sigma_\xi = \sigma_\xi^2 \bol{I}_T$. We note that for a general ARFIMA$(p,d,q)$ process, there is no closed form for the ACVF $\gamma(h)$, and hence covariance matrix $\Sigma_z$, so approximations of $\Sigma_z$ are needed, e.g. see \cite{Sowell:1992, Doornik:2003}. However, \cite{Hosking:1981} suggests exact ACVF for some simple cases of $(p,d,q)$ such as ARFIMA$(0,d,0)$, ARFIMA$(1,d,0)$ and ARFIMA$(0,d,1)$. 

Given the estimates of $z_t$ in \eqref{eq:z_t}, the conditional variance $\sigma_t^2$ is computed as 
\[
\widetilde{\sigma_t}^2 = \tilde{\kappa}^2 \text{exp}(\widetilde{z_t}),
\]
where the scale factor $\tilde{\kappa}^2$ is estimated as
\beq
\label{eq:scale}
\tilde{\kappa}^2 = \frac{1}{T} \sum_{t=1}^T \widetilde{y_t},
\eeq
with $\widetilde{y_t} = y_t \;\text{exp}(-\widetilde{z_t}/2)$ the heteroscredasticity corrected observations. 

Denote by $R$ the $1 \times T$ the vector of covariance between ${x}_{T+1}$ and $x$, e.g. $R = [\gamma_x(1),...,\gamma_x(T)]$, the one-step-ahead forecast value of the log squared return is calculated as \citep{HARVEY:2007}
\[\widehat{x}_{T+1} = c + R \Sigma^{-1}(x-c \textbf{1}_T)\]
The one-step-ahead forecast of the conditional variance is $\widehat{\sigma}_{T+1} = \tilde{\kappa}^2 \text{exp}(\widehat{x}_{T+1}-c)$.

Table \ref{tab:LMSV_Real_params} shows the estimation results of the LMSV model using a ARFIMA$(1,d,0)$ process to model the log volatility $z_t$, with the vector of parameters $\beta_x = [d,\phi_1,\sigma_\eta^2,\sigma_\xi^2]$. 
We also report the estimation of the scale factor $\kappa$ in \eqref{eq:LMSVmodel1} and the constant $c$ in \eqref{LMSVmodel1_logsquared}.
For the parameters $\phi_1$ and $\sigma_\eta^2$, we choose the priors to be similar to those of the SV model as shown in Table \ref{table:prior}. We use the same inverse-Gamma prior, as that of $\sigma_\eta^2$, for the parameter $\sigma_\xi^2$. 
For the fractional integration parameter $d$, we set the prior $2d \sim \text{Beta(20,5)}$. 
We run $N_{\text{MCMC}}=100000$ MCMC iterations of Algorithm \ref{alg:MCMC} and then discard the first $10,000$ iterations as burn-ins.

\begin{table}[h]
	\centering
	\footnotesize
	\begin{tabular}{ccccccc}
		\hline\hline
		\rule{0pt}{3ex}
		& $d$ & $\phi$ & $\sigma^2_\eta$ & $\sigma^2_\xi$ &$\kappa$ & $c$  \\
		\hline
		\rule{0pt}{3ex}
		\multirow{1}{*}{\text{DAX}}   &${0.442}$ & {0.708}  & {0.054} &{4.933} &{1.974}  &$-{1.616}$ \\
		&(0.026) &(0.082)&(0.024)&(0.170)&&\\  
		\rule{0pt}{3ex}
		\multirow{1}{*}{\text{HSI}}    &${0.431}$ & {0.747}  & {0.052} &{5.469} &{2.047}  &$-{1.526}$ \\ 
		&(0.030) &(0.069)&(0.021)&(0.183)&&\\   	
		\rule{0pt}{3ex}
		\multirow{1}{*}{\text{FCHI}}   &${0.434}$ & {0.716}  & {0.062} &{5.241}  &{1.984}  &$-{1.562}$ \\ 
		&(0.029) &(0.083)&(0.029)&(0.182)&&\\  
		\rule{0pt}{3ex}
		\multirow{1}{*}{\text{SPX}}  &${0.428}$ & {0.809}  & {0.040} &{5.488} &{2.017}  &$-{1.650}$ \\ 
		&(0.035) &(0.060)&(0.016)&(0.177)&&\\  
		\rule{0pt}{3ex}
		\multirow{1}{*}{\text{TSX}}   &${0.445}$ & {0.714}  & {0.047} &{4.964} &{1.918}  &$-{1.493}$ \\
		&(0.026) &(0.084)&(0.021)&(0.160)&&\\  	      	      	            	      
		\hline\hline
	\end{tabular}
	\caption{Applications: Posterior means of the parameters of the LMSV model with the posterior standard deviations in brackets. We also report the estimation of the scale factor $\kappa$ and the constant $c$.}
	\label{tab:LMSV_Real_params}
\end{table} 

\subsection{The SR-SV model}
\label{sec:app_2}
The SR-SV model in Section \ref{sec:lstm-sv} is fully written as
\label{Appendix}
\bean
r_t &=&\Psi(w_h h_{t-1}+b_r)\\
\varphi_t &=& \Psi(w_{r}r_t+w_{\eta}\eta_{t-1}+w_{z}z_{t-1} +b_{\varphi})\\
h_t &=& \alpha h_{t-1} + (1-\alpha)\varphi_t\\
\eta_t&=&\beta_0+\beta_1h_t+\eps_t^\eta,\;\;\eps_t^\eta\stackrel{iid}{\sim}\N(0,\sigma^2),\;\;t=1,2,...,T\\
z_t &=& \eta_t+\phi z_{t-1},\;\;t=1,...,T,\\
y_t &=& e^{\frac12z_t}\eps^y_t,\;\;\eps^y_t\stackrel{iid}{\sim}\N(0,1),\;\;t=1,2,...,T,
\eean 
where $\Psi(\cdot)$ is the ReLU activation function, $\Psi(x)=\max(0,x)$.
The model parameter vector is $\theta = (\beta_0,\beta_1, \phi, \sigma^2, \alpha, w_h, b_r, w_{r}, b_\varphi, w_{\eta}, w_{z})$.  

\subsection{Particle filter and implementation details of the DT-SMC sampler}
\label{sec:app_3}
Algorithm \ref{alg:PF} describes the particle filter for the SR-SV model where $\textbf{Z}_t=(Z^1_t,...,Z^N_t)$ denotes the vector of particles at time $t$. 
The set of standard normal random numbers $U$ includes two sources of randomness: the set of random numbers $\{U^P_{t,k}, t=1,...,T; k=1,...,N\}$ used to propose new particles in each time step, and the set of random numbers $\{U^R_{t,k}, t=1,...,T-1;k=1,...,N\}$ used in the resampling step.  
For the resampling step, we use multinomial resampling, with sorting, to obtain the vector ancestor indexes $\{A^k_{t-1},k=1,...,N\}$ used to propose particles at time $t$. The sorting step helps eliminate the discontinuity issues of the selected particles in the ordinary multinomial resampling scheme \citep{Gerber2014}. This sorted resampling scheme allows the selected particles to still be close after being resampled and hence helps to reduce the variability of the likelihood ratio estimator $\widehat{p}(y_{1:T}|\theta^{\prime},u^{\prime})/\widehat{p}(y_{1:T}|\theta,u)$ shown in the Algorithm \ref{alg:lik_annealing} \citep{Deligiannidis:2018}.

\begin{table}[h!]
	\begin{center}
		\begin{tabular}{clc}
			\hline\hline
			\rule{0pt}{3ex}
			\bol{Variable}      & \bol{Description}   &\bol{Value}\\
			\hline
			\rule{0pt}{3ex}
			$K$&    Number of annealing levels    & 10000 \\
			$M$&    Number of particles & 10000 \\
			$N$&    Number of particles in the particle filter& 200 \\
			$\rho$& Correlation factor in the CPM algorithm  & 0.999 \\
			$c$&    Constant of the ESS threshold & 0.800 \\
			$N_{\text{CPM}}$ & Number of CPM moves & 20\\
			\hline\hline
		\end{tabular}
	\end{center}
	\caption{Implementation settings of the DT-SMC sampler.}
	\label{tab:implementation}
\end{table}

The multinomial resampling scheme in step 2a and 2b generates the ancestor index $A^k_{t-1}, k=1,...,N,$ from the multinomial distribution denoted as $\F(\cdot|\textbf{p},\textbf{u})$ with \textbf{p} the vector of parameters of the multinomial distribution and \textbf{u} the uniform random numbers used within a multinomial random number generator. 
We use the standard normal cumulative distribution function $\textPhi(\cdot)$ in the resampling step to transform the normal random numbers $U^R_{t-1,k}$ to the uniform random numbers, denoted as $\overline{U}^R_{t-1,k}$. 
\begin{algorithm}
	\caption{Particle filter for the SR-SV model}  
	\label{alg:PF}
	\textbf{Input:} $T,N,y_{1:T}, \theta, U = (U^P_{1,1},...,U^P_{T,N},U^R_{1,1},...,U^R_{T-1,N})$
	\begin{itemize}
		\item[] 1. At time $t=1$, 
		\begin{itemize}
			\item[] (a) for $k=1,...N$, initialize the particles $(H_1^k,\eta^k_1,Z^k_1)$, e.g., $H_1^k=0$, as the SRU unit initially has no memory, and
			\bean
			\eta^k_1 &=& \beta_0 + \sigma U^P_{1,k} \\
			Z^k_1 &=& \eta^k_1
			\eean
			\item[] (b) compute and normalize the weights 
			\bean
			w_1(Z^k_1) &=&\frac{\mu_\theta(Z^k_1)g_\theta(y_1|Z^k_1)}{q_\theta(Z^k_1|y_1)} = g_\theta(y_1|Z^k_1)\\
			W^k_1 &=&\frac{w_1(Z^k_1)}{\sum_{m=1}^{N}w_1(Z^m_1)}
			\eean
			\item[] (c) compute the estimated likelihood $\widehat{p}(y_1|\theta)$ as
			\bean
			\widehat{p}(y_1|\theta,U) = \frac{1}{N}\sum_{k=1}^{N}w_1(Z^k_1).
			\eean
		\end{itemize}
		\item[] 2. At times $t = 2,...,T$,
		\begin{itemize}
			\item[] (a) sort the particle vector $\textbf{Z}_{t-1}$ in ascending order to obtain the vector of sorted particles $\overline{\textbf{Z}}_{t-1} = (\overline{Z}^1_{t-1},...,\overline{Z}^N_{t-1})$. The sorted index vector associated with $\overline{\textbf{Z}}_{t-1}$ is denoted as $\textbf{I}_{t-1}=(I^1_{t-1},...,I^N_{t-1})$. In this setting, we have the relation $\overline{Z}^k_{t-1} = Z^{I^k_{t-1}}_{t-1}$ with $k=1,...,N$.	
			Use the sorted index vector $\textbf{I}_{t-1}$ to define the vector of sorted weights $(\overline{W}^1_{t-1},...,\overline{W}^N_{t-1})$ such that 
			\bean
			\overline{W}^k_{t-1} &=& W^{I^k_{t-1}}_{t-1}
			\eean
			\item[] (b) sample $A^k_{t-1} \sim \F(\cdot|\overline{W}^k_{t-1},\overline{U}^R_{t-1,k})$ where $\overline{U}^R_{t-1,k} = \textPhi(U^R_{t-1,k})$ for $k=1,...,N$.
		\end{itemize}
	\end{itemize} 
\end{algorithm}
\begin{algorithm}[h]
	\textcolor{white}{}
	\begin{itemize}
		\item[]
		\begin{itemize}
			\item[] (c) for $k=1,...N$, generate particles $Z^k_t$ by
			\bean
			x_{t-1} &=& [\eta^{A^k_{t-1}}_{t-1}, Z^{A^k_{t-1}}_{t-1}]\\
			H^k_t &=& \text{SRU}(x_{t-1},H^{A^k_{t-1}}_{t-1}) \\
			\eta^k_t &=& \beta_0 + \beta_1 H^k_t  + \sigma U^P_{t,k} \\ 
			Z^k_t &=& \eta^k_t + \phi Z^{A^k_{t-1}}_{t-1}
			\eean
			and set $Z^k_{1:t} = (Z^{A^k_{t-1}}_{1:t-1},Z^k_t)$.
			\item[] (d) compute and normalize the weights 
			\bean
			w_t(Z^k_{1:t}) &=&\frac{f_\theta(Z^k_t|Z^{A^k_{t-1}}_{t-1})g_\theta(y_t|Z^k_t)}{q_\theta(Z^k_t|y_t,Z^{A^k_{t-1}}_{t-1})} = g_\theta(y_1|Z^k_1)\\
			W^k_t &=&\frac{w_t(Z^k_{1:t})}{\sum_{m=1}^{N}w_t(Z^m_{1:t})}
			\eean
			\item[] (e) compute the estimated likelihood $\widehat{p}(y_t|y_{1:t-1},\theta)$ as
			\bean
			\widehat{p}(y_t|y_{1:t-1},\theta,U) = \frac{1}{N}\sum_{k=1}^{N}w_t(Z^k_{1:t}).
			\eean
		\end{itemize}
	\end{itemize}
	\textbf{Output:} Estimate of the likelihood 
	\bean
	\widehat{p}(y_{1:T}|\theta,U) = \widehat{p}(y_1|\theta,U)\prod_{t=2}^T \widehat{p}(y_t|y_{1:t-1},\theta,U).
	\eean
\end{algorithm}

\subsection{Additional results}
\label{sec:app_4}
This section provides the additional tables and figures discussed in Section \ref{sec:applications}.

\begin{figure}[h]
	\centering
	\includegraphics[width=1\columnwidth]{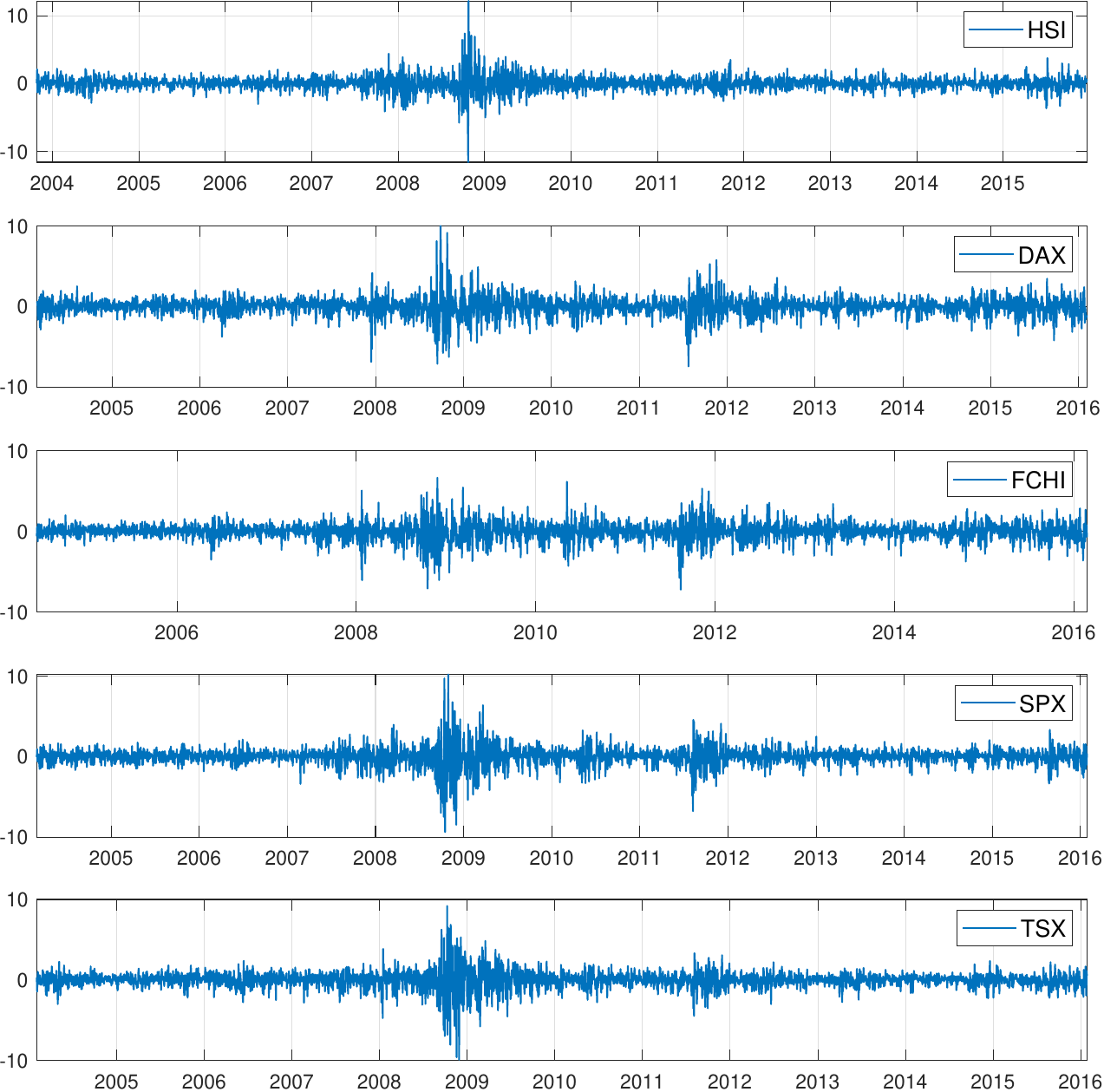}
	\caption{Applications: Time series plots for the HSI, DAX, FCHI, SPX and TSX datasets.}
	\label{f:plot_3_data}
\end{figure}

\begin{figure}[h]
	\centering
	\includegraphics[width=1\columnwidth]{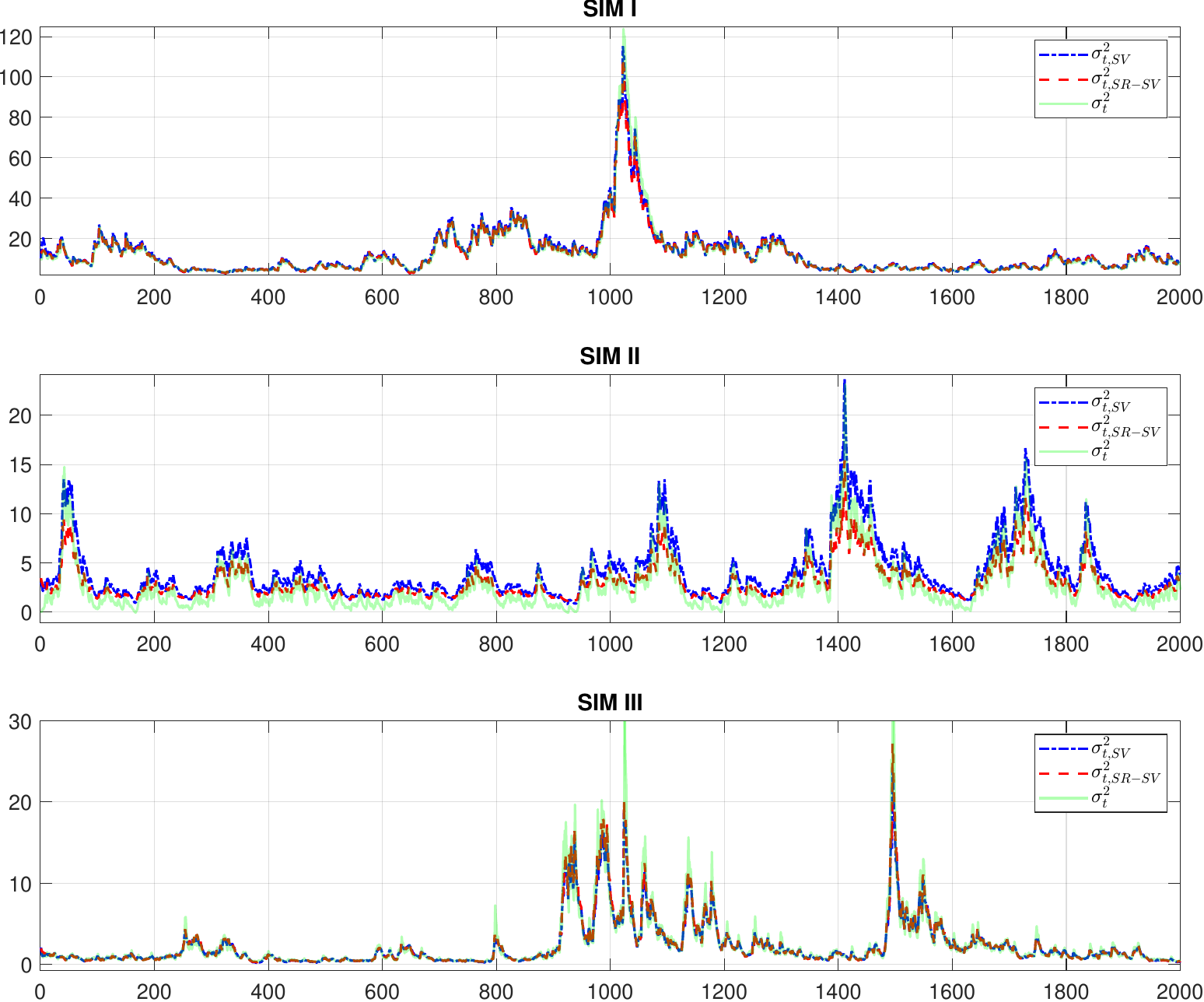}
	\caption{Simulation: Filtered volatility of the SV and SR-SV models, together with the true volatility, on three simulation datasets. (This is better viewed in colour).}
	\label{f:sim_filtered_vol}
\end{figure}

\begin{figure}[h]
	\centering
	\includegraphics[width=1\columnwidth]{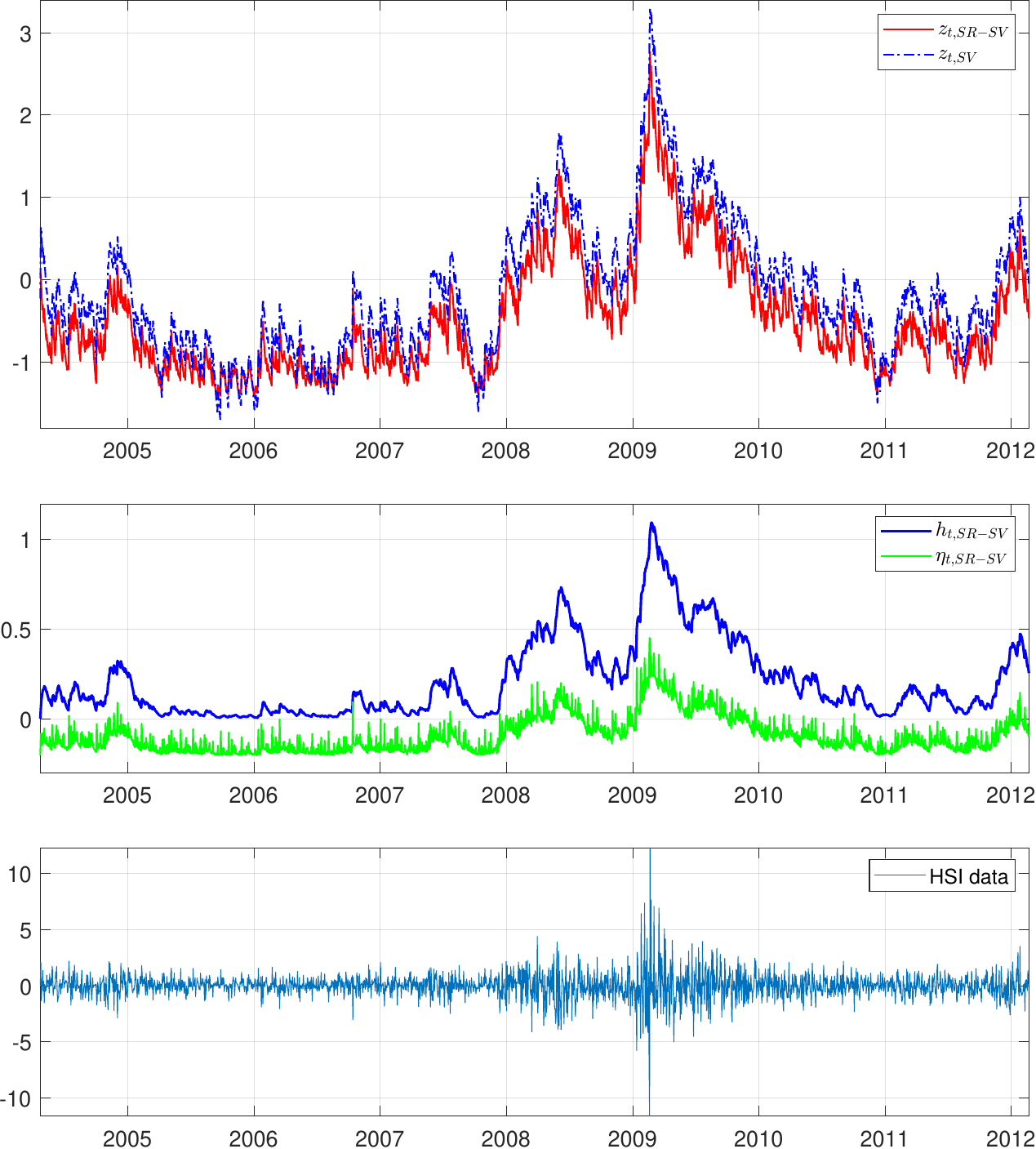}
	\caption{HSI: (\textit{Top}) The filtered log conditional variance of the SR-SV and SV models. (\textit{Middle}) The filtered values of $\eta_t$ and $h_t$ of the SR-SV model. (\textit{Bottom}) The in-sample data. (This is better viewed in colour).}
	\label{f:eta_h_HSI}
\end{figure}

\begin{figure}[h]
	\centering
	\includegraphics[width=1\columnwidth]{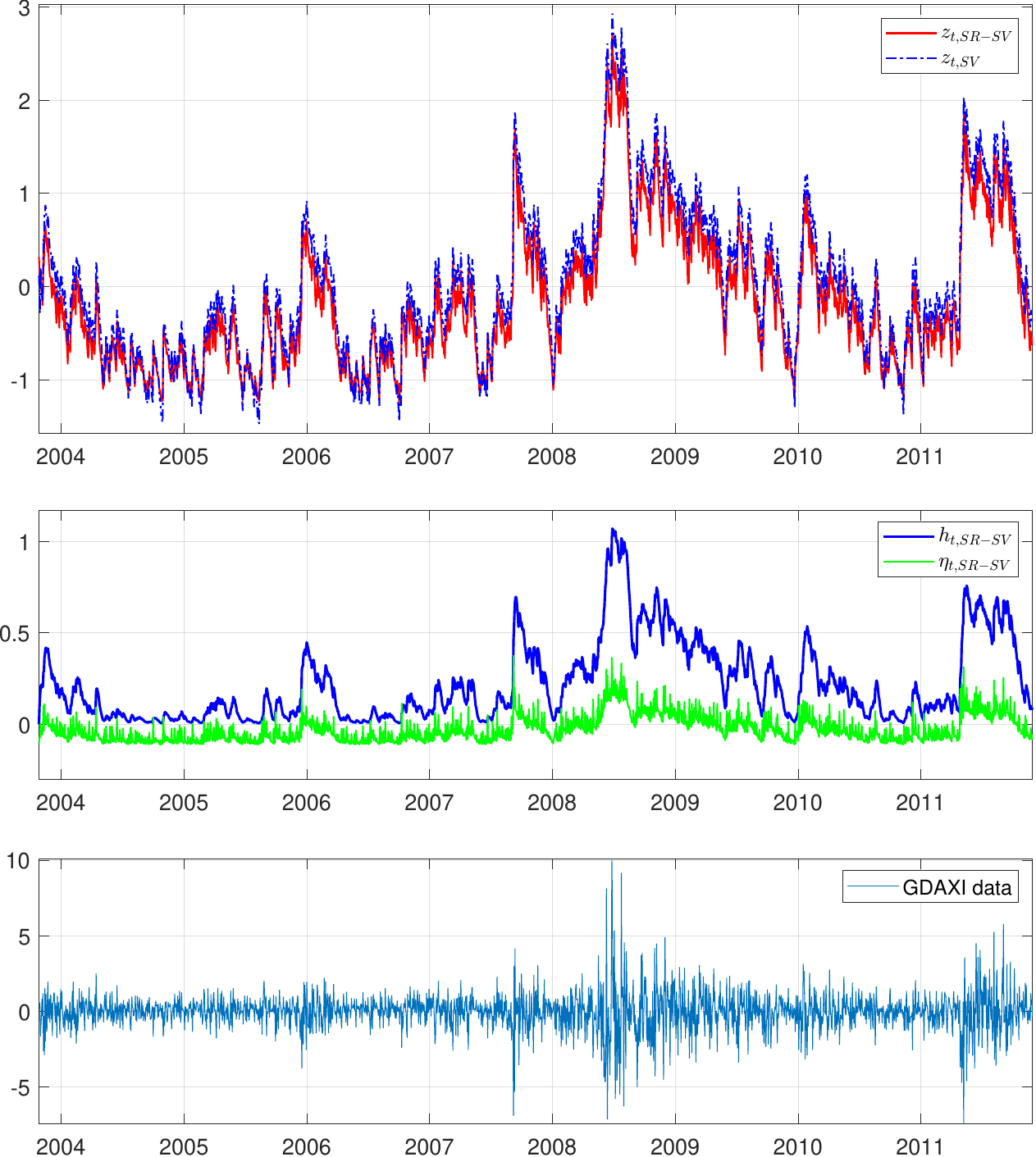}
	\caption{DAX: (\textit{Top}) The filtered log conditional variance of the SR-SV and SV models. (\textit{Middle}) The filtered values of $\eta_t$ and $h_t$ of the SR-SV model. (\textit{Bottom}) The in-sample data. (This is better viewed in colour).}
	\label{f:eta_h_DAX}
\end{figure}

\begin{figure}[h]
	\centering
	\includegraphics[width=1\columnwidth]{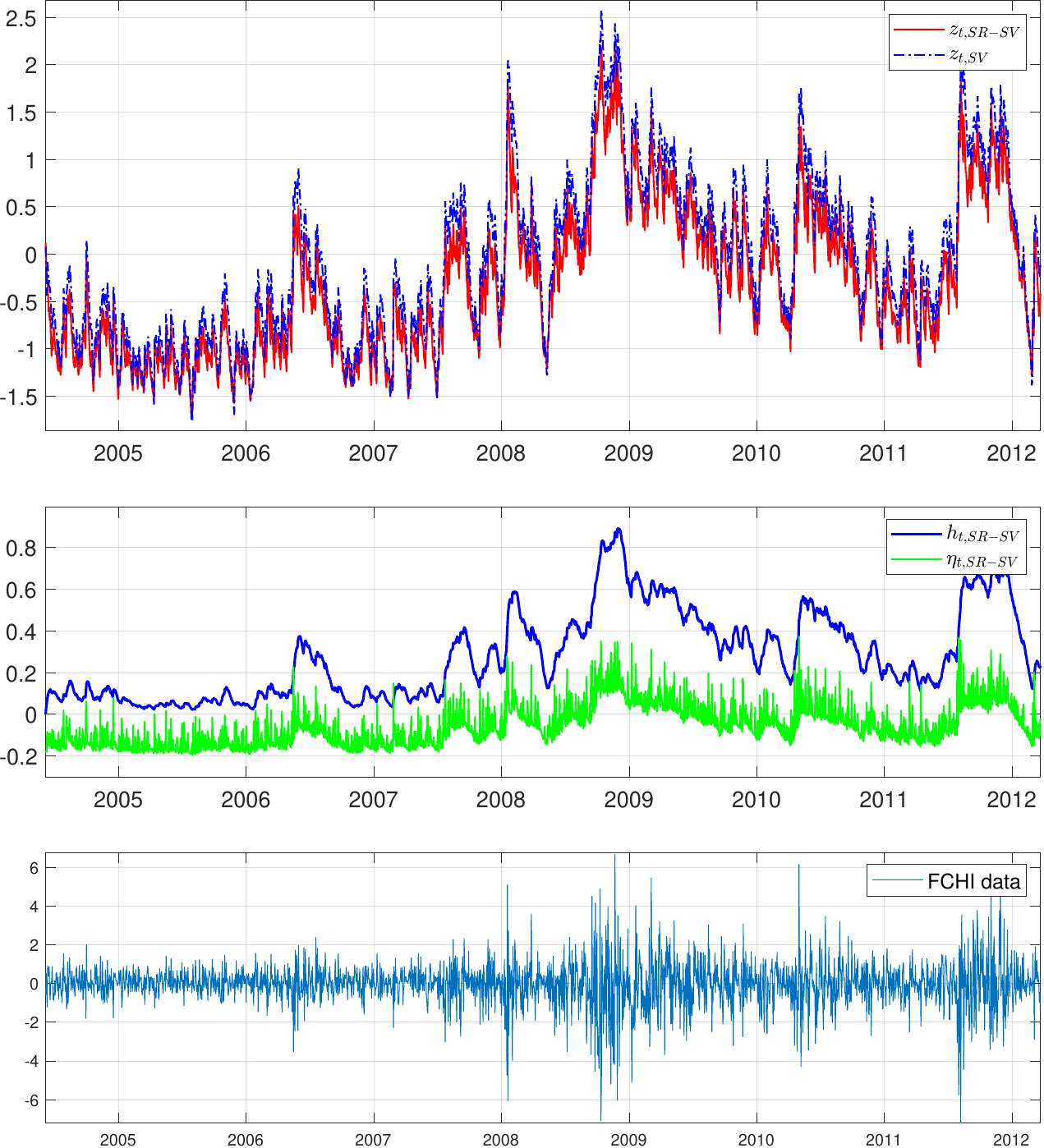}
	\caption{FCHI: (\textit{Top}) The filtered log conditional variance of the SR-SV and SV models. (\textit{Middle}) The filtered values of $\eta_t$ and $h_t$ of the SR-SV model. (\textit{Bottom}) The in-sample data. (This is better viewed in colour).}
	\label{f:eta_h_FCHI}
\end{figure}

\begin{figure}[h]
	\centering
	\includegraphics[width=1\columnwidth]{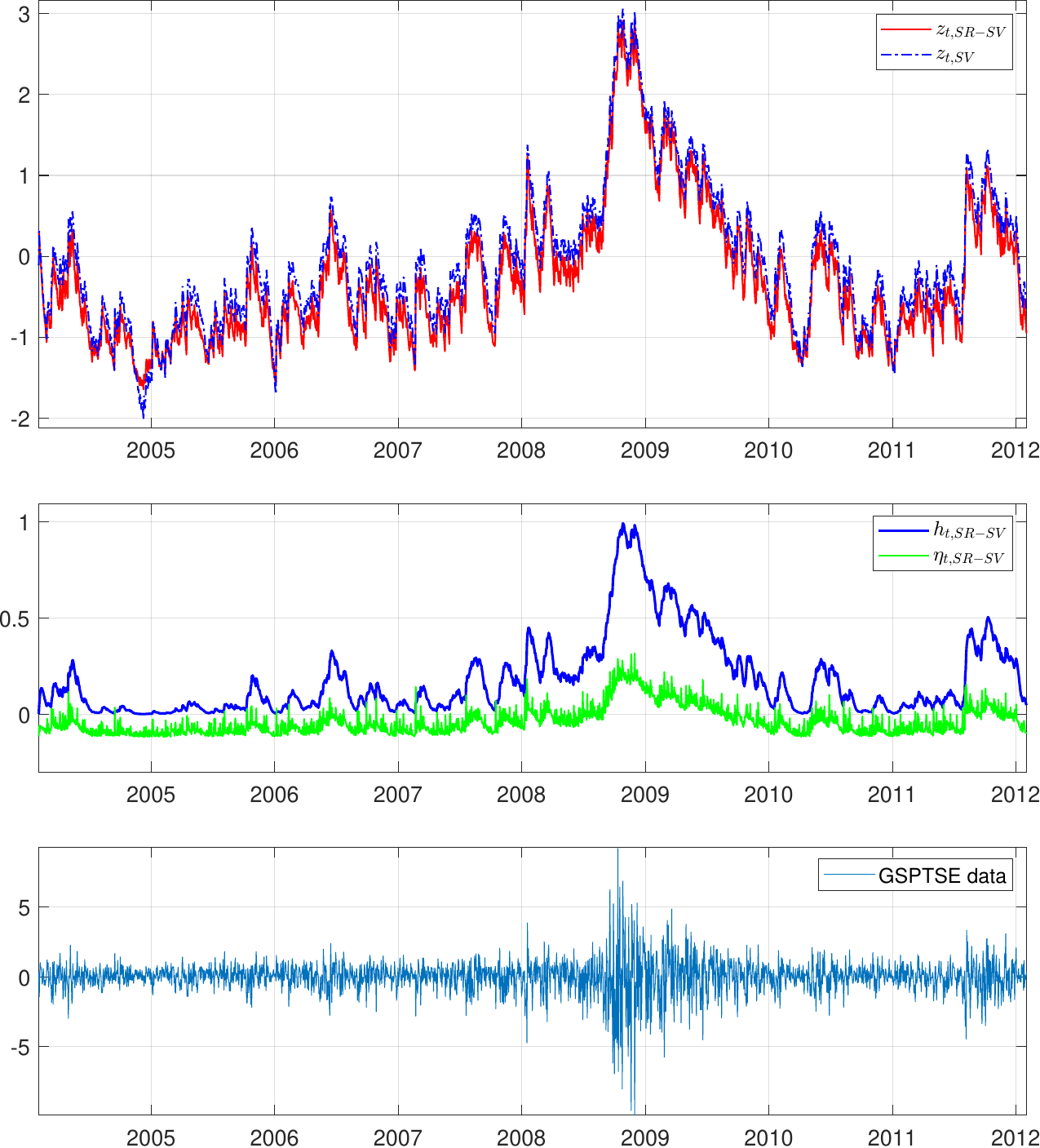}
	\caption{TSX: (\textit{Top}) The filtered log conditional variance of the SR-SV and SV models. (\textit{Middle}) The filtered values of $\eta_t$ and $h_t$ of the SR-SV model. (\textit{Bottom}) The in-sample data. (This is better viewed in colour).}
	\label{f:eta_h_TSX}
\end{figure}

\begin{table}[h]
	\centering
	\footnotesize
	\begin{tabular}{crcccccccc}
		\hline\hline
		\rule{0pt}{3ex}
		Measure&&PPS& $\text{MSE}_1$ &$\text{MSE}_2$ & $\text{MAE}_1$&$\text{MAE}_2$&$\text{QLIKE}$&$\text{R}^2\text{LOG}$&Count\\
		\hline
		\rule{0pt}{3ex}
		&SV             &1.368  &0.099  &0.763  &0.234  &0.485  &0.853  &0.423  &0 \\
		&               &(0.000)&(0.000)&(0.001)&(0.000)&(0.000)&(0.000)&(0.000)& \\
		\rule{0pt}{3ex}
		BV&N-SV         &1.368  &0.099  &0.769  &0.234  &0.487  &0.850  &0.422  &0\\
		&               &(0.000)&(0.001)&(0.003)&(0.001)&(0.002)&(0.001)&(0.002)& \\
		\rule{0pt}{3ex}
		&LMSV           &       &0.108  &0.805  &0.245  &0.501  &0.862  &0.435  &0 \\
		&               &&&&&&&& \\
		\rule{0pt}{3ex}
		&SR-SV         &\bol{1.365}&\bol{0.09	}&\bol{0.745}&\bol{0.220}&\bol{0.452}&\bol{0.847}&\bol{0.386}&7\\
		&               &(0.000)    &(0.000)    &(0.001)    &(0.000)    &(0.001)    &(0.000)    &(0.001)    & \\
		\hline
		\rule{0pt}{3ex}
		&SV             &     &0.094  &0.582  &0.237  &0.486  &0.810  &0.443  &0 \\
		&               &     &(0.000)&(0.002)&(0.000)&(0.000)&(0.000)&(0.001)& \\
		\rule{0pt}{3ex}
		MedRV&N-SV      &     &0.094  &0.585  &0.238  &0.489  &0.810  &0.447  &0\\
		&               &     &(0.001)&(0.004)&(0.001)&(0.002)&(0.001)&(0.002)& \\
		\rule{0pt}{3ex}
		&LMSV           &     &0.099  &0.612  &0.245  &0.504  &0.825  &0.467&0 \\
		&               &&&&&&&& \\
		\rule{0pt}{3ex}
		&SR-SV          &    &\bol{0.088}&\bol{0.580}&\bol{0.227}&\bol{0.462}&\bol{0.807}&\bol{0.417}&6 \\
		&               &    &(0.000)    &(0.001)    &(0.000)    &(0.001)    &(0.000)    &(0.001)    & \\
		\hline
		\rule{0pt}{3ex}
		&SV             &     &0.126&0.994&0.268&0.558&0.858&0.581&0 \\
		&               &     &(0.000)&(0.001)&(0.000)&(0.000)&(0.000)&(0.001)& \\
		\rule{0pt}{3ex}
		RKV  &N-SV      &     &0.126&1.001&0.269&0.559&0.854&0.581&0\\
		&               &     &(0.001)&(0.003)&(0.001)&(0.001)&(0.001)&(0.002)& \\
		\rule{0pt}{3ex}
		&LMSV           &     &0.135  &1.023  &0.279  &0.578  &0.878  &0.603&0 \\
		&               &&&&&&&& \\
		\rule{0pt}{3ex}
		&SR-SV          &    &\bol{0.116}&\bol{0.974}&\bol{0.254}&\bol{0.516}&\bol{0.851}&\bol{0.538}&6 \\
		&               &    &(0.000)    &(0.001)    &(0.000)    &(0.000)    &(0.000)    &(0.001)    & \\
		\hline
		\rule{0pt}{3ex}
		&SV             &     &0.107&0.909&0.239&0.501&0.859&0.442&0 \\
		&               &     &(0.000)&(0.001)&(0.000)&(0.000)&(0.000)&(0.001)& \\
		\rule{0pt}{3ex}
		 RV  &N-SV      &     &0.108  &0.915  &0.239  &0.503  &0.855  &0.441  &0\\
		 &              &     &(0.001)&(0.003)&(0.001)&(0.002)&(0.001)&(0.002)& \\
		 \rule{0pt}{3ex}
		 &LMSV          &     &0.115  &0.936  &0.250  &0.527  &0.875  &0.464  &0 \\
		 &              &&&&&&&& \\
		\rule{0pt}{3ex}
		&SR-SV          &    &\bol{0.098}&\bol{0.889}&\bol{0.223}&\bol{0.462}&\bol{0.851}&\bol{0.400}&6 \\
		&               &    &(0.000)    &(0.001)    &(0.000)    &(0.001)    &(0.000)    &(0.001)    & \\
		\hline\hline
	\end{tabular}
	\caption{DAX data: Forecast performance of the SR-SV and benchmark models using different realized measures. In each panel, the bold numbers indicate the best predictive scores. }
	\label{tab:DAX_forecast_score_compare}
\end{table} 

\begin{table}[h]
	\centering
	\footnotesize
	\begin{tabular}{crcccccccc}
		\hline\hline
		\rule{0pt}{3ex}
		Measure&&PPS& $\text{MSE}_1$ &$\text{MSE}_2$ & $\text{MAE}_1$&$\text{MAE}_2$&$\text{QLIKE}$&$\text{R}^2\text{LOG}$&Count\\
		\hline
		\rule{0pt}{3ex}
		&SV             &1.131      &0.069      &\bol{0.497}      &0.186      &0.319      &0.359      &0.390      &1 \\
		&               &(0.000)    &(0.000)    &(0.001)    &(0.000)    &(0.001)    &(0.001)    &(0.001)    & \\
		\rule{0pt}{3ex}
		BV&N-SV         &1.130      &0.067      &0.499      &0.182      &0.313      &0.357      &0.376      &0\\
		&               &(0.000)    &(0.000)    &(0.001)    &(0.000)    &(0.001)    &(0.001)    &(0.002)    & \\
		 \rule{0pt}{3ex}
		&LMSV          &            &0.076      &0.516      &0.205      &0.343      &0.370      &0.423      &0 \\
		&              &&&&&&&& \\
		\rule{0pt}{3ex}
		&SR-SV         &\bol{1.127}&\bol{0.060}&0.504&\bol{0.152}&\bol{0.261}&\bol{0.355}&\bol{0.294}&5\\
		&               &(0.000)    &(0.00)     &(0.002)    &(0.000)    &(0.000)    &(0.000)    &(0.001)    & \\
		\hline
		\rule{0pt}{3ex}
		&SV             &     &0.066  &0.371  &0.191  &0.317  &0.347  &0.435  &0 \\
		&               &     &(0.000)&(0.001)&(0.000)&(0.001)&(0.000)&(0.001)& \\
		\rule{0pt}{3ex}
		MedRV&N-SV      &     &0.065  &\bol{0.370}  &0.188  &0.312  &0.344  &0.423  &1\\
		&               &     &(0.001)&(0.000)&(0.001)&(0.000)&(0.001)&(0.002)& \\
		\rule{0pt}{3ex}
		&LMSV          &     &0.073   &0.396  &0.207  &0.335  &0.356  &0.469  &0 \\
		&              &&&&&&&& \\
		\rule{0pt}{3ex}
		&SR-SV          &    &\bol{0.059}&0.389&\bol{0.164}&\bol{0.272}&\bol{0.341}&\bol{0.338}&5 \\
		&                &    &(0.000)    &(0.001)    &(0.000)    &(0.001)    &(0.000)    &(0.001)    & \\
		\hline
		\rule{0pt}{3ex}
		&SV             &     &0.100&\bol{0.740}&0.230&0.385&0.366&0.665&1 \\
		&               &     &(0.000)&(0.001)&(0.000)&(0.001)&(0.001)&(0.001)& \\
		\rule{0pt}{3ex}
		RKV  &N-SV      &     &0.098&0.741&0.226&0.380&0.364&0.648&0\\
		&               &     &(0.000)&(0.002)&(0.001)&(0.001)&(0.000)&(0.002)& \\
		\rule{0pt}{3ex}
		&LMSV          &     &0.112   &0.764  &0.246  &0.405  &0.380  &0.744  &0 \\
		&              &&&&&&&& \\
		\rule{0pt}{3ex}
		&SR-SV          &    &\bol{0.087}&0.748&\bol{0.194}&\bol{0.323}&\bol{0.360}&\bol{0.519}&5 \\
		&                &    &(0.000)    &(0.002)    &(0.000)    &(0.001)    &(0.000)    &(0.001)    & \\
		\hline
		\rule{0pt}{3ex}
		&SV             &     &0.069  &\bol{0.522}  &0.186  &0.318  &0.367  &0.390  &1 \\
		&               &     &(0.000)&(0.001)&(0.000)&(0.001)&(0.001)&(0.001)& \\
		\rule{0pt}{3ex}
		RV  &N-SV       &     &0.068  &0.524  &0.181  &0.311  &0.365  &0.374  &0\\
		&               &     &(0.000)&(0.001)&(0.000)&(0.001)&(0.001)&(0.002)& \\
		\rule{0pt}{3ex}
		&LMSV          &      &0.077  &0.552  &0.204  &0.353  &0.386  &0.419  &0 \\
		&              &&&&&&&& \\
		\rule{0pt}{3ex}
		&SR-SV         &     &\bol{0.060}&0.530&\bol{0.150}&\bol{0.258}&\bol{0.361}&\bol{0.291}&5 \\
		&               &     &(0.000)    &(0.002)    &(0.000)    &(0.000)    &(0.001)    &(0.001)    & \\
		\hline\hline
	\end{tabular}
	\caption{HSI data: Forecast performance of the SR-SV and benchmark models using different realized measures. In each panel, the bold numbers indicate the best predictive scores.}
	\label{tab:HSI_forecast_score_compare}
\end{table} 

\begin{table}[h!]
	\centering
	\footnotesize
	\begin{tabular}{crcccccccc}
		\hline\hline
		\rule{0pt}{3ex}
		Measure&&PPS& $\text{MSE}_1$ &$\text{MSE}_2$ & $\text{MAE}_1$&$\text{MAE}_2$&$\text{QLIKE}$&$\text{R}^2\text{LOG}$&Count\\
		\hline
		\rule{0pt}{3ex}
		&SV             &1.384      &0.108      &1.076      &0.235      &0.504      &0.863      &0.426      &0\\
		&               &(0.000)    &(0.000)    &(0.001)    &(0.000)    &(0.001)    &(0.000)    &(0.001)&     \\
		\rule{0pt}{3ex}
		BV&N-SV         &1.383      &0.108      &1.086      &0.234      &0.507      &0.860      &0.420      &0\\
		&               &(0.000)    &(0.000)    &(0.002)    &(0.000)    &(0.001)    &(0.000)    &(0.001)    & \\
		\rule{0pt}{3ex}
		&LMSV          &            &0.118      &1.104      &0.246      &0.527      &0.872      &0.469  &0 \\
		&              &&&&&&&& \\
		\rule{0pt}{3ex}
		&SR-SV         &\bol{1.381}&\bol{0.095}&\bol{1.057}&\bol{0.210}&\bol{0.448}&\bol{0.856}&\bol{0.354}&7\\
		&               &(0.000)    &(0.00)     &(0.001)    &(0.000)    &(0.001)    &(0.000)    &(0.001)    & \\
		\hline
		\rule{0pt}{3ex}
		&SV             &     &0.100      &0.670      &0.238      &0.500      &0.833      &0.543      &0\\
		&               &     &(0.000)    &(0.001)    &(0.000)    &(0.001)    &(0.001)    &(0.002)    & \\
		\rule{0pt}{3ex}
		MedRV&N-SV      &     &0.100      &0.672      &0.237      &0.501      &0.832      &0.538      &0\\
		&               &     &(0.000)    &(0.002)    &(0.001)    &(0.000)    &(0.000)    &(0.001)    & \\
		\rule{0pt}{3ex}
		&LMSV          &      &0.112      &0.695      &0.247      &0.513      &0.849      &0.582  &0 \\
		&              &&&&&&&& \\
		\rule{0pt}{3ex}
		&SR-SV         &     &\bol{0.090}&\bol{0.665}&\bol{0.216}&\bol{0.452}&\bol{0.828}&\bol{0.472}&6\\
		&               &     &(0.000)    &(0.001)    &(0.000)    &(0.000)    &(0.000)    &(0.001)    & \\
		\hline
		\rule{0pt}{3ex}
		&SV             &     &0.158      &1.271      &0.301      &0.624      &0.901      &0.750      &0\\
		&               &     &(0.000)    &(0.001)    &(0.000)    &(0.001)    &(0.000)    &(0.002)    & \\
		\rule{0pt}{3ex}
		RKV  &N-SV      &     &0.159      &1.285      &0.301      &0.628      &0.896      &0.745      &0\\
		&               &     &(0.000)    &(0.002)    &(0.000)    &(0.000)    &(0.000)    &(0.001)    & \\
		\rule{0pt}{3ex}
		&LMSV          &      &0.168      &1.332      &0.315      &0.656      &0.908      &0.815  &0 \\
		&              &&&&&&&& \\
		\rule{0pt}{3ex}
		&SR-SV         &     &\bol{0.139}&\bol{1.229}&\bol{0.275}&\bol{0.562}&\bol{0.890}&\bol{0.645}&6\\
		&               &     &(0.000)    &(0.001)    &(0.000)    &(0.000)    &(0.000)    &(0.001)    & \\
		\hline
		\rule{0pt}{3ex}
		&SV             &     &0.103      &0.908      &0.232      &0.495      &0.877      &0.411      &0\\
		&               &     &(0.000)    &(0.001)    &(0.000)    &(0.000)    &(0.000)    &(0.001)    & \\
		\rule{0pt}{3ex}
		RV  &N-SV       &     &0.103      &0.920      &0.232      &0.498      &0.873      &0.404      &0\\
		&               &     &(0.000)    &(0.002)    &(0.000)    &(0.001)    &(0.000)    &(0.001)    & \\
		\rule{0pt}{3ex}
		&LMSV           &     &0.113      &0.945      &0.245      &0.521      &0.884      &0.449  &0 \\
		&               &&&&&&&& \\
		\rule{0pt}{3ex}
		&SR-SV          &     &\bol{0.090}&\bol{0.880}&\bol{0.209}&\bol{0.440}&\bol{0.869}&\bol{0.340}&6\\
		&               &     &(0.000)    &(0.001)    &(0.000)    &(0.000)    &(0.000)    &(0.001)    & \\
		\hline\hline
	\end{tabular}
	\caption{FCHI data: Forecast performance of the SR-SV and benchmark models using different realized measures. In each panel, the bold numbers indicate the best predictive scores and the model with highest count of best predictive scores is preferred. }
	\label{tab:FCHI_forecast_score_compare}
\end{table}

\begin{table}[h!]
	\centering
	\footnotesize
	\begin{tabular}{crcccccccc}
		\hline\hline
		\rule{0pt}{3ex}
		Measure&&PPS& $\text{MSE}_1$ &$\text{MSE}_2$ & $\text{MAE}_1$&$\text{MAE}_2$&$\text{QLIKE}$&$\text{R}^2\text{LOG}$&Count\\
		\hline
		\rule{0pt}{3ex}
		&SV             &1.004      &0.074      &0.904      &0.192      &0.295      &0.106      &0.542      &0\\
		&               &(0.001)    &(0.000)    &(0.002)    &(0.001)    &(0.001)    &(0.001)    &(0.003)&     \\
		\rule{0pt}{3ex}
		BV&N-SV         &1.003      &0.074      &0.902      &0.192      &0.294      &\bol{0.105}      &0.543      &1\\
		&               &(0.000)    &(0.000)    &(0.002)    &(0.001)    &(0.001)    &(0.001)    &(0.002)    & \\
		\rule{0pt}{3ex}
		&LMSV           &           &0.091      &0.965      &0.217      &0.331      &0.197      &0.676  &0 \\
		&               &     &&&&&&& \\
		\rule{0pt}{3ex}
		&SR-SV         &\bol{1.001}&\bol{0.069}&\bol{0.900}&\bol{0.181}&\bol{0.273}&0.107&\bol{0.517}&6\\
		&               &(0.000)    &(0.00)     &(0.001)    &(0.000)    &(0.001)    &(0.001)    &(0.001)    & \\
		\hline
		\rule{0pt}{3ex}
		&SV             &     &0.074      &0.289      &0.210      &0.310      &0.098      &1.052      &0\\
		&               &     &(0.000)    &(0.001)    &(0.001)    &(0.001)    &(0.001)    &(0.002)    & \\
		\rule{0pt}{3ex}
		MedRV&N-SV      &     &0.073      &0.291      &0.210      &0.309      &0.098      &1.044      &0\\
		&               &     &(0.000)    &(0.002)    &(0.001)    &(0.000)    &(0.001)    &(0.001)    & \\
		\rule{0pt}{3ex}
		&LMSV           &     &0.096      &0.360      &0.239      &0.353      &0.209      &0.839  &0 \\
		&               &     &&&&&&& \\
		\rule{0pt}{3ex}
		&SR-SV         &     &\bol{0.069}&\bol{0.287}&\bol{0.201}&\bol{0.291}&0.098&\bol{0.985}&5\\
		&               &     &(0.000)    &(0.001)    &(0.000)    &(0.000)    &(0.001)    &(0.001)    & \\
		\hline
		\rule{0pt}{3ex}
		&SV             &     &0.096      &0.349      &0.246      &0.357      &0.134      &0.987   &0\\
		&               &     &(0.000)    &(0.001)    &(0.000)    &(0.001)    &(0.000)    &(0.002) & \\
		\rule{0pt}{3ex}
		RKV  &N-SV      &     &0.096      &0.347      &0.246      &0.355      &0.134      &0.991   &0\\
		&               &     &(0.000)    &(0.001)    &(0.001)    &(0.000)    &(0.001)    &(0.001) & \\
		\rule{0pt}{3ex}
		&LMSV           &     &0.110      &0.392      &0.263      &0.378      &0.220      &1.120   &0 \\
		&               &     &&&&&&& \\
		\rule{0pt}{3ex}
		&SR-SV         &     &\bol{0.089}&\bol{0.341}&\bol{0.236}&\bol{0.336}&\bol{0.131}&\bol{0.951}&6\\
		&               &     &(0.000)    &(0.001)    &(0.000)    &(0.000)    &(0.000)    &(0.001)    & \\
		\hline
		\rule{0pt}{3ex}
		&SV             &     &0.087      &1.370      &0.206      &0.319      &0.118      &0.625      &0\\
		&               &     &(0.000)    &(0.002)    &(0.000)    &(0.000)    &(0.001)    &(0.002)    & \\
		\rule{0pt}{3ex}
		RV  &N-SV       &     &0.087      &1.368      &0.206      &0.317      &\bol{0.117}      &0.627      &1\\
		&               &     &(0.000)    &(0.002)    &(0.000)    &(0.001)    &(0.001)    &(0.002)    & \\
		\rule{0pt}{3ex}
		&LMSV           &     &0.100      &1.418      &0.224      &0.342      &0.195      &0.742  &0 \\
		&               &     &&&&&&& \\
		\rule{0pt}{3ex}
		&SR-SV         &     &\bol{0.081}&\bol{1.363}&\bol{0.195}&\bol{0.295}&0.119&\bol{0.597}&5\\
		&               &     &(0.000)    &(0.002)    &(0.000)    &(0.000)    &(0.001)    &(0.001)    & \\
		\hline\hline
	\end{tabular}
	\caption{TSX data: Forecast performance of the SR-SV and benchmark models using different realized measures. In each panel, the bold numbers indicate the best predictive scores.}
	\label{tab:TSX_forecast_score_compare}
\end{table}

\clearpage
\bibliographystyle{apalike}
\bibliography{references}

\begin{thebibliography}{}

\bibitem[Andersen and Bollerslev, 1998]{Andersen:1998}
Andersen, T.~G. and Bollerslev, T. (1998).
\newblock Answering the skeptics: Yes, standard volatility models do provide
  accurate forecasts.
\newblock {\em International Economic Review}, 39(4):885--905.

\bibitem[Andersen et~al., 2012]{Andersen:2012}
Andersen, T.~G., Dobrev, D., and Schaumburg, E. (2012).
\newblock Jump-robust volatility estimation using nearest neighbor truncation.
\newblock {\em Journal of Econometrics}, 169(1):75 -- 93.
\newblock Recent Advances in Panel Data, Nonlinear and Nonparametric Models: A
  Festschrift in Honor of Peter C.B. Phillips.

\bibitem[Andrieu et~al., 2010]{Andrieu:2010}
Andrieu, C., Doucet, A., and Holenstein, R. (2010).
\newblock Particle {Markov chain Monte Carlo} methods.
\newblock {\em Journal of the Royal Statistical Society, Series B}, 72:1--33.

\bibitem[Andrieu and Roberts, 2009]{Andrieu:2009}
Andrieu, C. and Roberts, G. (2009).
\newblock The pseudo-marginal approach for efficient {Monte Carlo}
  computations.
\newblock {\em The Annals of Statistics}, 37:697--725.

\bibitem[Baillie et~al., 1996]{Baillie:1996}
Baillie, R.~T., Bollerslev, T., and Mikkelsen, H.~O. (1996).
\newblock Fractionally integrated generalized autoregressive conditional
  heteroskedasticity.
\newblock {\em Journal of Econometrics}, 74(1):3 -- 30.

\bibitem[Barndorff-Nielsen et~al., 2008]{Barndorff&Nielsen:2008}
Barndorff-Nielsen, O., Hansen, P., Lunde, A., and Shephard, N. (2008).
\newblock Designing realized kernels to measure the ex post variation of equity
  prices in the presence of noise.
\newblock {\em Econometrica}, 76(6):1481--1536.
\newblock cited By 491.

\bibitem[Barndorff-Nielsen and Shephard, 2004]{Nielsen:2004}
Barndorff-Nielsen, O.~E. and Shephard, N. (2004).
\newblock {Power and Bipower Variation with Stochastic Volatility and Jumps}.
\newblock {\em Journal of Financial Econometrics}, 2(1):1--37.

\bibitem[Bollerslev, 1986]{Bollerslev1986}
Bollerslev, T. (1986).
\newblock Generalized autoregressive conditional heteroskedasticity.
\newblock {\em Journal of Econometrics}, 31(3):307 -- 327.

\bibitem[Bollerslev and Mikkelsen, 1996]{Bollerslev:1996}
Bollerslev, T. and Mikkelsen, H.~O. (1996).
\newblock Modeling and pricing long memory in stock market volatility.
\newblock {\em Journal of Econometrics}, 73(1):151 -- 184.

\bibitem[Box and Cox, 1964]{Box:1964}
Box, G. E.~P. and Cox, D.~R. (1964).
\newblock An analysis of transformations.
\newblock {\em Journal of the Royal Statistical Society: Series B
  (Methodological)}, 26(2):211--243.

\bibitem[Breidt et~al., 1998]{Breidt:1998}
Breidt, F., Crato, N., and de~Lima, P. (1998).
\newblock The detection and estimation of long memory in stochastic volatility.
\newblock {\em Journal of Econometrics}, 83(1):325 -- 348.

\bibitem[Cho et~al., 2014]{cho-etal-2014-learning}
Cho, K., van Merri{\"e}nboer, B., Gulcehre, C., Bahdanau, D., Bougares, F.,
  Schwenk, H., and Bengio, Y. (2014).
\newblock Learning phrase representations using {RNN} encoder{--}decoder for
  statistical machine translation.
\newblock In {\em Proceedings of the 2014 Conference on Empirical Methods in
  Natural Language Processing ({EMNLP})}, pages 1724--1734, Doha, Qatar.
  Association for Computational Linguistics.

\bibitem[Crato and de~Lima, 1994]{Crato:1994}
Crato, N. and de~Lima, P.~J. (1994).
\newblock Long-range dependence in the conditional variance of stock returns.
\newblock {\em Economics Letters}, 45(3):281 -- 285.

\bibitem[Del~Moral, 2004]{DelMoral:2004}
Del~Moral, P. (2004).
\newblock {\em Feynman-Kac Formulae: Genealogical and Interacting Particle
  Systems with Applications}.
\newblock Springer, New York.

\bibitem[{Del Moral} et~al., 2006]{DelMoral:2006}
{Del Moral}, P., Doucet, A., and Jasra, A. (2006).
\newblock Sequential {M}onte {C}arlo samplers.
\newblock {\em Journal of the Royal Statistical Society, Series B},
  68:411--436.

\bibitem[Deligiannidis et~al., 2018]{Deligiannidis:2018}
Deligiannidis, G., Doucet, A., and Pitt, M.~K. (2018).
\newblock The correlated pseudo marginal method.
\newblock {\em Journal of the Royal Statistical Society: Series B (Statistical
  Methodology)}, 80(5):839--870.

\bibitem[Dieng et~al., 2018]{dieng:2018}
Dieng, A.~B., Ranganath, R., Altosaar, J., and Blei, D.~M. (2018).
\newblock Noisin: Unbiased regularization for recurrent neural networks.

\bibitem[Ding et~al., 1993]{Ding:1993}
Ding, Z., Granger, C.~W., and Engle, R.~F. (1993).
\newblock A long memory property of stock market returns and a new model.
\newblock {\em Journal of Empirical Finance}, 1(1):83 -- 106.

\bibitem[Donaldson and Kamstra, 1997]{Donaldson:1997}
Donaldson, R.~G. and Kamstra, M. (1997).
\newblock An artificial neural network-garch model for international stock
  return volatility.
\newblock {\em Journal of Empirical Finance}, 4(1):17--46.

\bibitem[Doornik and Ooms, 2003]{Doornik:2003}
Doornik, J.~A. and Ooms, M. (2003).
\newblock Computational aspects of maximum likelihood estimation of
  autoregressive fractionally integrated moving average models.
\newblock {\em Computational Statistics and Data Analysis}, 42(3):333 -- 348.
\newblock Computational Ecomometrics.

\bibitem[Duan and Fulop, 2015]{Duan:2015}
Duan, J.-C. and Fulop, A. (2015).
\newblock Density-tempered marginalized {Sequential Monte Carlo} samplers.
\newblock {\em Journal of Business \& Economic Statistics}, 33(2):192--202.

\bibitem[Elman, 1990]{Elman:1990}
Elman, J.~L. (1990).
\newblock Finding structure in time.
\newblock {\em Cognitive Science}, 14:179--21.

\bibitem[Ferguson, 1973]{Ferguson:1973}
Ferguson, T.~S. (1973).
\newblock A {Bayesian} analysis of some nonparametric problems.
\newblock {\em The Annals of Statistics}, 1(2):209--230.

\bibitem[Fleming et~al., 2003]{Fleming:2003}
Fleming, J., Kirby, C., and Ostdiek, B. (2003).
\newblock The economic value of volatility timing using “realized”
  volatility.
\newblock {\em Journal of Financial Economics}, 67(3):473 -- 509.

\bibitem[Garthwaite et~al., 2010]{Garthwaite2010}
Garthwaite, P., Fan, Y., and Sisson, S. (2010).
\newblock Adaptive optimal scaling of {Metropolis-Hastings} algorithms using
  the {Robbins-Monro} process.
\newblock {\em Communications in Statistics - Theory and Methods}, 45.

\bibitem[Gerber and Chopin, 2014]{Gerber2014}
Gerber, M. and Chopin, N. (2014).
\newblock Sequential {Quasi-Monte Carlo}.
\newblock {\em Journal of the Royal Statistical Society: Series B (Statistical
  Methodology)}, 77.

\bibitem[Giraitis et~al., 2003]{Giraitis:2003}
Giraitis, L., Kokoszka, P., Leipus, R., and Teyssière, G. (2003).
\newblock Rescaled variance and related tests for long memory in volatility and
  levels.
\newblock {\em Journal of Econometrics}, 112(2):265 -- 294.

\bibitem[Goodfellow et~al., 2016]{Goodfellow:2016}
Goodfellow, I., Bengio, Y., and Courville, A. (2016).
\newblock {\em Deep Learning}.
\newblock MIT Press.

\bibitem[Granger and Joyeux, 1980]{Granger:1980}
Granger, C. W.~J. and Joyeux, R. (1980).
\newblock An introduction to long-memory time series models and fractional
  differencing.
\newblock {\em Journal of Time Series Analysis}, 1(1):15--29.

\bibitem[Gunawan et~al., 2018]{Dang:2018}
Gunawan, D., Dang, K., Quiroz, M., Kohn, R., and Tran, M. (2018).
\newblock Subsampling sequential {Monte Carlo} for static {Bayesian} models.
\newblock {\em arXiv:1805.03317}.

\bibitem[Hansen and Lunde, 2005]{Hansen:2005}
Hansen, P.~R. and Lunde, A. (2005).
\newblock A forecast comparison of volatility models: does anything beat a
  {GARCH}(1,1)?
\newblock {\em Journal of Applied Econometrics}, 20(7):873--889.

\bibitem[Harvey, 2007]{HARVEY:2007}
Harvey, A.~C. (2007).
\newblock 16 - long memory in stochastic volatility.
\newblock In Knight, J. and Satchell, S., editors, {\em Forecasting Volatility
  in the Financial Markets (Third Edition)}, Quantitative Finance, pages 351 --
  363. Butterworth-Heinemann, Oxford.

\bibitem[Higgins and Bera, 1992]{Higgins:1992}
Higgins, M.~L. and Bera, A.~K. (1992).
\newblock A class of nonlinear {ARCH} models.
\newblock {\em International Economic Review}, 33(1):137--158.

\bibitem[Hochreiter and Schmidhuber, 1997]{Hochreiter1997}
Hochreiter, S. and Schmidhuber, J. (1997).
\newblock Long short-term memory.
\newblock {\em Neural computation}, 9:1735--80.

\bibitem[Hosking, 1981]{Hosking:1981}
Hosking, J. R.~M. (1981).
\newblock Fractional differencing.
\newblock {\em Biometrika}, 68(1):165--176.

\bibitem[Jacquier et~al., 1994]{Jacquier:1994}
Jacquier, E., Polson, N.~G., and Rossi, P.~E. (1994).
\newblock Bayesian analysis of stochastic volatility models (with discussion).
\newblock {\em Journal of Business and Economic Statistics}, 12:371--417.

\bibitem[Jeffreys, 1935]{jeffreys:1935}
Jeffreys, H. (1935).
\newblock Some tests of significance, treated by the theory of probability.
\newblock {\em Mathematical Proceedings of the Cambridge Philosophical
  Society}, 31(2):203–222.

\bibitem[Jeffreys, 1961]{Jeffreys:1961}
Jeffreys, H. (1961).
\newblock {\em Theory of Probability, 3rd}.
\newblock Clarendon Press, Oxford, England.

\bibitem[Jensen and Maheu, 2010]{Jensen:2010}
Jensen, M.~J. and Maheu, J.~M. (2010).
\newblock Bayesian semiparametric stochastic volatility modeling.
\newblock {\em Journal of Econometrics}, 157(2):306 -- 316.

\bibitem[Kass et~al., 1998]{Robert:1998}
Kass, R.~E., Carlin, B.~P., Gelman, A., and Neal, R.~M. (1998).
\newblock {Markov Chain Monte Carlo} in practice: A roundtable discussion.
\newblock {\em The American Statistician}, 52(2):93--100.

\bibitem[Kass and Raftery, 1995]{Kass:1995}
Kass, R.~E. and Raftery, A.~E. (1995).
\newblock Bayes factors.
\newblock {\em Journal of the American Statistical Association},
  90(430):773--795.

\bibitem[Kili\c{c}, 2011]{KILIC2011368}
Kili\c{c}, R. (2011).
\newblock Long memory and nonlinearity in conditional variances: A smooth
  transition {FIGARCH} model.
\newblock {\em Journal of Empirical Finance}, 18(2):368 -- 378.

\bibitem[Kim and Won, 2018]{Kim:2018}
Kim, H.~Y. and Won, C.~H. (2018).
\newblock Forecasting the volatility of stock price index: A hybrid model
  integrating {LSTM} with multiple garch-type models.
\newblock {\em Expert Systems with Applications}, 103:25 -- 37.

\bibitem[Kim et~al., 1998]{Kim:1998}
Kim, S., Shephard, N., and Chib, S. (1998).
\newblock Stochastic volatility: likelihood inference and comparison with
  {ARCH} models.
\newblock {\em Review of Economic Studies}, 65:361--393.

\bibitem[Lipton et~al., 2015]{Lipton:2018}
Lipton, Z., Berkowitz, J., and Elkan, C. (2015).
\newblock A critical review of recurrent neural networks for sequence learning.
\newblock arXiv:1804.04359.

\bibitem[Liu and Chen, 1998]{LiuChen:1998}
Liu, J.~S. and Chen, R. (1998).
\newblock {Sequential Monte Carlo} methods for dynamic systems.
\newblock {\em Journal of the American Statistical Association},
  93(443):1032--1044.

\bibitem[Lo, 1991]{Lo:1991}
Lo, A.~W. (1991).
\newblock Long-term memory in stock market prices.
\newblock {\em Econometrica}, 59(5):1279--1313.

\bibitem[Makridakis et~al., 2018]{Makridakis:2018}
Makridakis, S., Spiliotis, E., and Assimakopoulos, V. (2018).
\newblock Statistical and machine learning forecasting methods: Concerns and
  ways forward.
\newblock {\em PLOS ONE}, 13(3):1--26.

\bibitem[Mandelbrot, 1967]{Benoit1967}
Mandelbrot, B. (1967).
\newblock The variation of some other speculative prices.
\newblock {\em The Journal of Business}, 40(4):393--413.

\bibitem[Martens, 2002]{Martens:2002}
Martens, M. (2002).
\newblock Measuring and forecasting {S\&P} 500 index-futures volatility using
  high-frequency data.
\newblock {\em Journal of Futures Markets}, 22(6):497--518.

\bibitem[Neal, 2001]{Neal:2001}
Neal, R. (2001).
\newblock Annealed importance sampling.
\newblock {\em Statistics and Computing}, 11:125--139.

\bibitem[Nelson, 1991]{Nelson:1991}
Nelson, D.~B. (1991).
\newblock Conditional heteroskedasticity in asset returns: A new approach.
\newblock {\em Econometrica}, 59(2):347--370.

\bibitem[Oliva et~al., 2017]{Oliva:2017}
Oliva, J.~B., P{\'o}czos, B., and Schneider, J.~G. (2017).
\newblock The statistical recurrent unit.
\newblock In {\em ICML2017}.

\bibitem[Pagan and Schwert, 1990]{Pagan:1990}
Pagan, A.~R. and Schwert, G. (1990).
\newblock Alternative models for conditional stock volatility.
\newblock {\em Journal of Econometrics}, 45(1):267 -- 290.

\bibitem[Pitt et~al., 2012]{Pitt:2012}
Pitt, M.~K., dos Santos~Silva, R., Giordani, P., and Kohn, R. (2012).
\newblock On some properties of {Markov chain Monte Carlo} simulation methods
  based on the particle filter.
\newblock {\em Journal of Econometrics}, 171(2):134 -- 151.
\newblock Bayesian Models, Methods and Applications.

\bibitem[Poole et~al., 2014]{poole:2014}
Poole, B., Sohl-Dickstein, J., and Ganguli, S. (2014).
\newblock Analyzing noise in autoencoders and deep networks.

\bibitem[Shephard and Sheppard, 2010]{Shephard:2010}
Shephard, N. and Sheppard, K. (2010).
\newblock Realising the future: forecasting with high-frequency-based
  volatility (heavy) models.
\newblock {\em Journal of Applied Econometrics}, 25(2):197--231.

\bibitem[Sietsma and Dow, 1991]{Sietsma:1991}
Sietsma, J. and Dow, R.~J. (1991).
\newblock Creating artificial neural networks that generalize.
\newblock {\em Neural Networks}, 4(1):67 -- 79.

\bibitem[Sowell, 1992]{Sowell:1992}
Sowell, F. (1992).
\newblock Maximum likelihood estimation of stationary univariate fractionally
  integrated time series models.
\newblock {\em Journal of Econometrics}, 53(1):165 -- 188.

\bibitem[Taylor, 1986]{Taylor:1986}
Taylor, S. (1986).
\newblock {\em Modelling Financial Time Series}.
\newblock John Wiley, Chichester.

\bibitem[Taylor, 1982]{Taylor:1982}
Taylor, S.~J. (1982).
\newblock Financial returns modelled by the product of two stochastic processes
  — a study of daily sugar prices 1961-79.
\newblock In Anderson, O.~D., editor, {\em Time Series Analysis: Theory and
  Practice}, page 203–226. Amsterdam: North-Holland.

\bibitem[van Bellegem, 2012]{Bellegem:2012}
van Bellegem, S. (2012).
\newblock Locally stationary volatility modeling.
\newblock In Bauwens, L., Hafner, C., and Laurent, S., editors, {\em Volatility
  Models and Their Applications}. Wiley \& Sons.

\bibitem[Whittle, 1953]{whittle:1953}
Whittle, P. (1953).
\newblock Estimation and information in stationary time series.
\newblock {\em Ark. Mat.}, 2(5):423--434.

\bibitem[Yu, 2002]{Yu:2002}
Yu, J. (2002).
\newblock Forecasting volatility in the {New Zealand} stock market.
\newblock {\em Applied Financial Economics}, 12(3):193--202.

\bibitem[Yu et~al., 2006]{YuJun:2006}
Yu, J., Yang, Z., and Zhang, X. (2006).
\newblock A class of nonlinear stochastic volatility models and its
  implications for pricing currency options.
\newblock {\em Computational Statistics and Data Analysis}, 51(4):2218 -- 2231.

\bibitem[Zhang, 2003]{ZHANG2003159}
Zhang, G. (2003).
\newblock Time series forecasting using a hybrid {ARIMA} and neural network
  model.
\newblock {\em Neurocomputing}, 50:159 -- 175.

\end{thebibliography}
\end{document}